\documentclass[twocolumn]{aastex631}

\graphicspath{{./}{figures/}}

\newcommand\Rgal[0]{R_{\mathrm{gal}}}
\newcommand\Reff[0]{R_{\mathrm{eff}}}
\newcommand\Tpeak[0]{T_{\mathrm{peak}}}
\newcommand\dV[0]{{\Delta}{V}}

\newcommand\kmPerS[0]{\mathrm{km\ s}^{-1}}

\newcommand\sigmax[0]{\sigma_{\mathrm{x}}}
\newcommand\sigmay[0]{\sigma_{\mathrm{y}}}
\newcommand\sigmav[0]{\sigma_{\mathrm{v}}}
\newcommand\sigmaxraw[0]{\sigma_{\mathrm{x,r}}}
\newcommand\sigmayraw[0]{\sigma_{\mathrm{y,r}}}
\newcommand\sigmavraw[0]{\sigma_{\mathrm{v,r}}}

\newcommand\SigmaBG[0]{\Sigma_{\mathrm{bg}}}
\newcommand\SigmaGas[0]{\Sigma}
\newcommand\SigmaGMC[0]{\Sigma}

\newcommand\Mvir[0]{{M}_{\mathrm{vir}}}
\newcommand\Mcl[0]{{M}_{\mathrm{cl}}}

\newcommand\alphaVir[0]{{\alpha_{\mathrm{vir}}}}
\newcommand\alphaVirCh[0]{{\alpha_{\mathrm{vir,0}}}}
\newcommand\LCO[0]{{L}_{\mathrm{CO}}}
\newcommand\MsunPerSqPC[0]{ {M}_{\odot}\ \mathrm{pc}^{-2} }
\newcommand\Msun[0]{ {M}_{\odot} }

\newcommand\alphaCO[0]{{\alpha_{\mathrm{CO}}}}

\newcommand\vLOS[0]{{v_{\mathrm{LOS}}}}
\newcommand\vSys[0]{{v_{\mathrm{sys}}}}
\newcommand\vRot[0]{{v_{\mathrm{rot}}}}
\newcommand\vPhi[0]{{v_{\mathrm{phi}}}}
\newcommand\vR[0]{{v_{\mathrm{R}}}}
\newcommand\vZ[0]{{v_{\mathrm{z}}}}
\newcommand\vResid[0]{{v_{\mathrm{res}}}}

\newcommand\gammaSch[0]{{\gamma_{\mathrm{exp}}}}
\newcommand\MuSch[0]{{M_{\mathrm{u,exp}}}}

\newcommand\gammaPL[0]{{\gamma_{\mathrm{PL}}}}
\newcommand\MuPL[0]{{M_{\mathrm{lim}}}}
\newcommand\NuPL[0]{{N_{\mathrm{u,PL}}}}
\newcommand\MuPLDeviate[0]{{M_{\mathrm{u,PL}}}}

\newcommand\Teq[0]{T_{\mathrm{eq}}}
\newcommand\TpeakCh[0]{T_{{0}}}
\newcommand\SigmaCh[0]{{\Sigma_{{0}}}}
\newcommand\RCh[0]{{R_{{0}}}}

\newcommand\fFill[0]{{{f}}}

\newcommand\Dkin[0]{D_{\mathrm{kin}}}

\begin{document}

\title{Whole-disk sampling of molecular clouds in M83}

\shorttitle{Molecular Clouds in M83}
\shortauthors{Hirota et al.}
\email{akihiko.hirota AT alma.cl}

\author[0000-0002-0465-5421]{Akihiko Hirota}
\affiliation{Joint ALMA Observatory, Alonso de C\'ordova 3107, Vitacura, Santiago 763-0355, Chile}
\affiliation{National Astronomical Observatory of Japan, 2-21-1 Osawa, Mitaka, Tokyo 181-8588, Japan}

\author[0000-0002-8762-7863]{Jin Koda}
\affiliation{Department of Physics and Astronomy, Stony Brook University, Stony Brook, NY 11794-3800, USA}

\author[0000-0002-1639-1515]{Fumi Egusa}
\affiliation{Institute of Astronomy, Graduate School of Science, The University of Tokyo, 2-21-1 Osawa, Mitaka, Tokyo 181-0015, Japan}

\author[0000-0002-0588-5595]{Tsuyoshi Sawada}
\affiliation{Joint ALMA Observatory, Alonso de C\'ordova 3107, Vitacura, Santiago 763-0355, Chile}
\affiliation{National Astronomical Observatory of Japan, 2-21-1 Osawa, Mitaka, Tokyo 181-8588, Japan}

\author[0000-0001-5187-2288]{Kazushi Sakamoto}
\affiliation{Academia Sinica, Institute of Astronomy and Astrophysics, Taipei 10617, Taiwan}

\author[0000-0002-3871-010X]{Mark Heyer}
\affiliation{Department of Astronomy, University of Massachusetts Amherst, 710 North Pleasant Street, Amherst, MA 01003, USA}

\author[0000-0001-8254-6768]{Amanda M Lee}
\affiliation{Department of Physics and Astronomy, Stony Brook University, Stony Brook, NY 11794-3800, USA}
\affiliation{Department of Astronomy, University of Massachusetts Amherst, 710 North Pleasant Street, Amherst, MA 01003, USA}

\author[0000-0002-8868-1255]{Fumiya Maeda}
\affiliation{Institute of Astronomy, Graduate School of Science, The University of Tokyo, 2-21-1 Osawa, Mitaka, Tokyo 181-0015, Japan}
\affiliation{Research Center for Physics and Mathematics, Osaka Electro-Communication University, 18-8 Hatsucho, Neyagawa, 572-8530, Osaka, Japan}

\author[0000-0002-9091-2366]{Samuel Boissier}
\affiliation{Aix Marseille Univ., CNRS, CNES, Laboratoire d'Astrophysique de Marseille, Marseille, France}

\author[0000-0002-5189-8004]{Daniela Calzetti}
\affiliation{Department of Astronomy, University of Massachusetts Amherst, 710 North Pleasant Street, Amherst, MA 01003, USA}

\author[0000-0002-1723-6330]{Bruce G. Elmegreen}
\affiliation{Katonah, NY 10536 USA}

\author[0000-0002-6824-6627]{Nanase Harada}
\affiliation{National Astronomical Observatory of Japan, 2-21-1 Osawa, Mitaka, Tokyo 181-8588, Japan}
\affiliation{Department of Astronomy, School of Science, Graduate University for Advanced Studies (SOKENDAI), 2-21-1 Osawa, Mitaka, Tokyo, 181-1855 Japan}

\author[0000-0001-6947-5846]{Luis C. Ho}
\affiliation{Kavli Institute for Astronomy and Astrophysics, Peking University, Beijing 100871, People's Republic of China}
\affiliation{Department of Astronomy, School of Physics, Peking University, Beijing 100871, People's Republic of China}

\author[0000-0003-3990-1204]{Masato I. N. Kobayashi}
\affiliation{Universit\"{a}t zu K\"{o}ln, I. Physikalisches Institut, Z\"{u}lpicher Str. 77, D-50937 K\"{o}ln, Germany}

\author[0000-0002-1234-8229]{Nario Kuno}
\affiliation{Division of Physics, Faculty of Pure and Applied Sciences, University of Tsukuba, 1-1-1 Tennodai, Tsukuba, Ibaraki 305-8571, Japan}
\affiliation{Tomonaga Center for the History of the Universe, University of Tsukuba, 1-1-1 Tennodai, Tsukuba, Ibaraki 305-8571, Japan}

\author[0000-0002-1576-1676]{Barry F. Madore}
\affiliation{The Observatories, Carnegie Institution for Science, 813 Santa Barbara Street, Pasadena CA 91101 USA}
\affiliation{Department of Astronomy and Astrophysics, University of Chicago, 5640 S. Ellis Ave., Chicago, IL 60637, USA}

\author[0000-0001-9281-2919]{Sergio Martín}
\affiliation{Joint ALMA Observatory, Alonso de C\'ordova 3107, Vitacura, Santiago 763-0355, Chile}
\affiliation{European Southern Observatory, Alonso de C\'ordova, 3107, Vitacura, Santiago 763-0355, Chile}

\author[0000-0002-3106-7676]{Jennifer Donovan Meyer}
\affiliation{National Radio Astronomy Observatory, 520 Edgemont Road, Charlottesville, VA 22903, USA}

\author[0000-0002-3373-6538]{Kazuyuki Muraoka}
\affiliation{Department of Physics, Graduate School of Science, Osaka Metropolitan University, 1-1 Gakuen-cho, Naka-ku, Sakai, Osaka 599-8531, Japan}

\author[0000-0002-9668-3592]{Yoshimasa Watanabe}
\affiliation{Materials Science and Engineering, College of Engineering, Shibaura Institute of Technology, 3-7-5 Toyosu, Koto-ku, Tokyo 135-8548, Japan}

\begin{abstract}
We present a catalog of clouds identified from the $^{12}$CO (1--0) data of M83, which was observed using Atacama Large Millimeter/submillimeter Array (ALMA) with a spatial resolution of $\sim$46 pc and a mass sensitivity of $\sim$10$^4$ $\Msun$ (3 $\sigma$).
The almost full-disk coverage and high sensitivity of the data allowed us to sample 5724 molecular clouds with a median mass of $\sim1.9$ $\times$ $10^5$ $\Msun$, which is comparable to the most frequently sampled mass of Giant Molecular Clouds (GMCs) by surveys in the Milky Way.
About 60 percent of the total CO luminosity in M83's disk arises from clouds more massive than 10$^6$ $\Msun$.
Such massive clouds comprise 16 percent of the total clouds in number and tend to concentrate toward the arm, bar, and center, while smaller clouds are more prevalent in inter-arm regions.
Most $>10^6$ $\Msun$ clouds have peak brightness temperatures $\Tpeak$ above 2 K with the current resolution.
Comparing the observed cloud properties with the scaling relations determined by Solomon et al. 1987 (S87),
$\Tpeak$$>2$ K clouds follow the relations, but $\Tpeak$$<2$ K clouds, which are dominant in number, deviate significantly.
Without considering the effect of beam dilution, the deviations would suggest modestly high virial parameters (median $\alphaVir$ $\sim2.7$) and low surface mass densities (median $\SigmaGMC$ $\sim$22 $\MsunPerSqPC$) for the entire cloud samples,
which are similar to values found for the Milky Way clouds by Rice et al. (2016) and Miville-Desch{\^e}nes et al. (2017).
However, once beam dilution is taken into account, the observed $\alphaVir$ and $\SigmaGas$ for a majority of the clouds (mostly $\Tpeak$ $<2$ K) can be potentially explained with intrinsic $\SigmaGas$ of $\sim$100 $\MsunPerSqPC$ and $\alphaVir$ of $\sim$1, which are similar to the clouds of S87.
\end{abstract}


\section{Introduction}
\label{sec:intro}

Molecular gas is directly associated with star formation in the nearby universe.
The bulk of H$_2$ mass resides in giant molecular clouds (GMCs), which are concentrations of molecular gas with a temperature of around 10 K, diameter of 10 pc to 100 pc, and mass of
10$^{5-6}$ $\Msun$
\cite[e.g.,][]{Sanders1985}
\footnote{
The definition of GMC slightly varies from literature to literature, but \cite{Sanders1985} defined molecular clouds with a diameter larger than 22 pc and mass above 10$^5$ $\Msun$ as GMCs.
Nearby Galactic clouds with a mass of a few 10$^4$ $\Msun$ are also often referred to as GMCs \citep[e.g,][]{Lada2010}
}.
The efficiency of star formation in GMCs is determined by the complex interplay among many processes, including self-gravity, turbulence, stellar feedback, and magnetic fields, although the exact balance is still not certain.
Characterizing GMCs' properties and their relations is vital in reaching a coherent understanding of star formation in galaxies.
\par

The conventional view of GMCs was mainly established by the early CO studies of Galactic molecular clouds about four decades ago.
At that time, GMCs were mainly identified as discrete emission features in the Galactic plane \citep[e.g.,][]{Sanders1985}.
The early studies empirically established the scaling relations of cloud properties \citep[][hereafter S87]{Larson1981, Solomon1987Larson} and also found a power-law GMC mass function \citep{Sanders1985, ScovilleSanders1987}.
In particular, the three relations determined by S87, namely the scaling relations of linewidth and size $\dV$ $\propto$ $S^{\sim0.5}$, virial mass and CO luminosity $\Mvir$ $\propto$ $\LCO$$^{\sim0.8}$, and volume density and size $\rho$ $\propto$ $S^{-1}$, are often used as a benchmark in most subsequent studies of GMCs.
Assuming that clouds are approximately in virial equilibrium, the near-linear $\Mvir$-$\LCO$ relation (2nd relation) yields a constant CO-to-H$_2$ conversion factor of $X_{\mathrm CO} \sim$2 $\times$ 10$^{20}$ cm$^{-2}\ (\mathrm{K}\ \kmPerS)^{-1}$ for a typical Galactic GMC, which agrees with the `standard' conversion factor often adopted today in metal-rich environments \citep[see,][]{Bolatto2013ConversionFactor}.
Using the constant conversion factor, the 3rd relation suggests that the surface mass densities of GMCs are nearly constant around 200$\MsunPerSqPC$, which is often taken as a typical surface density in GMCs.
\par

Subsequent studies explored the cloud properties in diverse environments, including the outer Galaxy \citep{Heyer2001OuterGalaxy}, Galactic center \citep{Oka2001}, and extragalactic environments \citep{Engargiola2003, Rosolowsky2005, Bolatto2008GMCs, DonovanMeyer2013, Colombo2014Env, Freeman2017, Hirota2018M83, Maeda2020NGC1300, Rosolowsky2021PhangsGMCs, Muraoka2023M33}.
Several studies revisited GMCs also in the inner Galactic disk with new $^{13}$CO survey data \citep{RomanDuval2010} and with a more complete coverage albeit at lower spatial resolution \citep{Rice2016, MivilleDeschenes2017MWGMC}.
Some confirmed the S87 relations in the inner Galactic disk \citep{RomanDuval2010} and in external galaxies \citep{Bolatto2008GMCs}, but others found deviations in two aspects.
\par
First, despite the near virial balance suggested for S87's clouds, recent studies report a variety of dynamical stability in clouds.
In the Galactic center \citep{Oka2001} and outer Galactic disk \citep{Heyer2009}, many clouds appear unbound if the $X_{\rm CO}$ in the inner disk is applied.
These analyses often adopt the virial parameter \citep{BertoldiMcKee1992}, the ratio of kinetic to gravitational energies of clouds, and find $\alphaVir \gg $1-2 for the unbound clouds.
Even in the inner Galactic disk, \citet{MivilleDeschenes2017MWGMC} suggests that only 15 percent of the clouds might be bound, although bound clouds account for 40 percent of the total molecular gas mass.
In addition, recent extragalactic studies found a large scatter in $\alphaVir$ with an average value of $\sim$2 in disk regions \citep{Sun2020, Rosolowsky2021PhangsGMCs}, which suggested a modestly gravitationally-bound state, but not in the virial equilibrium.
\citet{Evans2021BoundCloud} found that, in general, only 20 to 40 percent of the cloud masses are bound by compiling cloud catalogs both in Galactic and extragalactic environments.
\par

Secondly, the near constant $\SigmaGMC$ of around 200 $\MsunPerSqPC$ is also revisited by recent studies. $\SigmaGMC$ appears to show considerable variations, with a low median value of a few $\times$ 10 $\MsunPerSqPC$ in the recent Galactic plane study \citep{MivilleDeschenes2017MWGMC} and in the outer Galactic disk survey \citep{Heyer2001OuterGalaxy} but high $\SigmaGMC$, reaching up to 1000 $\MsunPerSqPC$ or higher, in galactic centers \citep{Oka2001, Leroy2015NGC253}.
\par

The large variations in $\alphaVir$ and $\SigmaGMC$ seen in recent studies, along with predictions from numerical simulations \citep{DobbsPringle2013, Smith2020CloudFactory}, potentially require a refinement of the view of GMCs:
GMCs could be modestly bound by self-gravity and are relatively transient objects, which might be prone to be dispersed by large-scale shear motions and stellar feedback \citep[see,][]{Dobbs2011NotBound, Chevance2023Review}.
\par

Extragalactic studies of molecular clouds provided essential contributions to this refinement of the view of molecular clouds.
However, most extragalactic studies were limited by low spatial resolutions of $>$ 50 pc to 100 pc \citep[i.e., larger than most of the Galactic GMC diameters, ][]{Colombo2014Env, Hirota2018M83}, even in the recent PHANGS survey \citep{Rosolowsky2021PhangsGMCs}.
These extragalactic 'disk' surveys
mostly achieved mass completeness limits
of around 10$^6$ $\Msun$ or more, while most of the inner Galactic disk surveys were made the completeness limit of 1--3 $\times$ 10$^5$ $\Msun$ \citep[e.g.,][]{Sanders1985, Solomon1987Larson, ScovilleSanders1987, RomanDuval2010, Rice2016, MivilleDeschenes2017MWGMC}.
To build a more accurate and comprehensive view, we need a higher sensitivity and resolution to characterize the cloud properties and distribution over a wide range of cloud masses and to elucidate their environmental dependence.
\par

Recently, the full extent of the molecular disk of the nearby galaxy M83 was observed in $^{12}$CO (1--0) using ALMA \citep[][]{Koda2023M83}.
The mass sensitivity, $\sim 10^4$ $\Msun$, is unprecedented for such a full-disk mapping of substantial spiral galaxies.
M83 closely resembles the Milky Way galaxy in morphology (including the bar, spiral arms, and even interarm structures), as well as in its size, gas and stellar masses, and near Solar metallicity \citep[][]{Koda2023M83}.
This study identifies molecular clouds from the $^{12}$CO data to examine the cloud properties, distributions, scaling relations, and mass distributions.
\par

We organize this paper as follows.
In \S\ref{SecIdentify}, we describe the data and procedure to catalog molecular clouds.
In \S\ref{SecResults}, we present their properties, scaling relations, and spatial distribution.
We examine the cloud mass function in \S\ref{SecMF} and estimate the vertical cloud-to-cloud velocity dispersions in \S\ref{SecVelDisp}.
In \S\ref{SecDiscussion}, we discuss the impact of beam-dilution (beam filling factor) on the cloud parameters, in particular on $\alphaVir$ and $\SigmaGMC$, and discuss its implications to our view of molecular clouds.
In \S\ref{SecSummary}, we present the conclusions.
\section{Molecular Cloud Identification}
\label{SecIdentify}
\subsection{Data}

We used the $^{12}$CO (J=1--0) mapping data of M83 observed with ALMA \citep[project code 2017.1.00079.S; ][]{Koda2023M83}.
The observations used the main 12-m array and Atacama Compact Array (ACA), i.e., the 7-m short-baseline interferometric array and the Total-Power (TP) single-dish array, to recover the flux.
The interferometric and single-dish data were calibrated separately.
The calibration procedure is described in \citet{Koda2023M83}.
The calibrated TP array data was converted into interferometric visibilities using the TP2VIS software \citep{Koda2019TP2VIS}. Then, the 12-m, 7-m, and TP array visibilities were inverted onto the image plane and jointly deconvolved using the MIRIAD software package \citep{SaultTeubenWright1995MIRIAD}.
We refer the readers to \cite{Koda2023M83} for further details on the data processing.
\par

Three data cubes were prepared in \cite{Koda2023M83} with velocity resolutions of 1, 2, and 5 $\kmPerS$, respectively.
We here adopt the one with 1 $\kmPerS$ resolution.
The restoring beam of the original data is a Gaussian with major and minor axes of {2$\arcsec$.09} $\times$ {1$\arcsec$.68} in full-width half maximum (FWHM), which corresponds to 45.6 pc $\times$ 36.7 pc at the assumed distance of M83 \citep[4.5 Mpc;][]{Thim2003M83Distance}.
To avoid complexities associated with the non-circular beam, we smoothed the original data cube in spatial directions to achieve a circular beam with an FWHM size of 2$\arcsec$.1, corresponding to $\sim$46 pc.

\par

Figure \ref{FigRMSMap}(a) shows the two-dimensional map of the root-mean-square (rms) noise level.
We adopted the following procedure to obtain a noise-free estimate of the noise distribution. First, we used MIRIAD's {\it mossen} command to generate a two-dimensional noise distribution expected from the integration time and system noise temperature assigned to visibilities. Next, the 'expected' noise map produced by {\it mossen} was compared with the 'measured' noise map, which we made by measuring rms noise along the velocity axis at each position in the data cube.
The comparison confirmed that both maps are consistent regarding the relative spatial variation within each map but are discrepant in the absolute scaling by a few tens of percent. Therefore, as the final step, we scaled the 'expected' map so that the median noise level matches that of the 'measured' noise map. The median noise level across the observed field is 6.5 mJy per beam.

\subsection{Cloud identification}
\label{SubsecIdentify}
We here describe our procedures to identify molecular clouds in the $^{12}$CO (1--0) data cube.
\par

As a preparatory step, we constructed a signal mask to isolate voxels with significant detection.
The mask was created by first identifying a set of regions that consist of voxels with SNR $\ge$ 4,
then discarding small regions whose spatial extents are less than 0.6 times the beam area or whose emissions are confined in a single $1\,\rm km\,s^{-1}$ channel along the velocity axis, and finally expanding the remaining regions to include the morphologically connected surrounding voxels with SNR $\ge$ 2.
\par

The masked data cube was decomposed into a set of independent closed surfaces using the {\it astrodendro}\footnote{http://www.dendrograms.org} software package.
Although {\it astrodendro} identifies hierarchically nested sets of structures, we adopted only the structures at the finest spatial scales, commonly referred to as {\it leaves}.
This is because the utilized spatial resolution, $\sim$46 pc at FWHM, is only a factor of two smaller than the typical height of the thin molecular gas disk \citep[e.g., about 100 pc in FWHM for the MW disk, see ][]{HeyerDame2015Review}.
As long as the disk height is similar between M83 and Milky Way and the size of the largest cloud is limited by the disk height, examining structures larger than individual leaves is unnecessary.
Parameters for the cloud decomposition algorithm are configured such that only voxels with SNR $>2$ are included in the analysis, and {\it leaf} candidates with SNR contrasts between their peak and edge-level $<2$ are discarded.
In addition, to avoid picking up noise fluctuations as clouds,
we discarded small {\it leaves} with the number of voxels less than $A_{\mathrm{beam}}/A_{\mathrm{voxel}}$, where $A_{\mathrm{beam}}$ and $A_{\mathrm{voxel}}$ are the area of the beam and a voxel.
\par

The identified {\it leaves } generally trace the small regions around the peaks of molecular clouds, especially in crowded areas, because by nature of the algorithm, the leaves are defined such that they do not overlap each other.
The {\it astrodendro} decomposition left about 80\% of the CO emission unassigned to any {\it leaves}.
The unassigned voxels to any seed structures are further segmented with the watershed algorithm, using the {\it scikit-image} software package \citep{vanDerWalt2014ScikitImage}.
\par

5724 leaves, which are expanded by the watershed segmentation, are accepted as molecular clouds
About 91 percent of the total CO luminosity within the mapped area was sampled as molecular clouds, as we will see later in \S\ref{SubsecCOLumFraction}.

\subsection{Derivation of cloud properties}
The physical properties of the identified clouds are calculated in a way similar to the one adopted in \cite*{Hirota2018M83}.
A brief description is provided here.

\subsubsection{Measured properties}
\label{SubsecBasicProps}

Each cloud's centroid position and velocity are measured as the intensity-weighted first moments.
We record the maximum brightness temperature within the cloud as peak temperature $\Tpeak$.
The size parameters in two dimensions and velocity dispersion of each cloud are calculated as second moments and then extrapolated to the 0 K level with the procedure of \cite*{Rosolowsky2006CPROPS}.
Hereafter, we denote the extrapolated moment values as $\sigmaxraw$, $\sigmayraw$, and $\sigmavraw$, where $\sigmaxraw$ $>$ $\sigmayraw$.
The CO luminosity for each cloud ($\LCO$) is derived as a summation in each cloud boundary and extrapolated to 0 K level.
We estimated the uncertainty in $\sigmaxraw$, $\sigmayraw$, and $\sigmavraw$ by performing a bootstrap error estimation with 1000 trials.

\subsubsection{Derived properties}
\label{SubsecDerivedProps}
We derived the physical parameters of the clouds from the directly measured parameters presented above.

\par
\emph{Effective radius:}
Effective radius of the cloud ($\Reff$) is derived as
\begin{equation}
\Reff = \frac{3.4}{\sqrt{\pi}} \sqrt{\sigmax \sigmay}
\end{equation}
where $3.4 / \sqrt\pi$ ($\sim1.91$) is an empirical factor that is determined by S87, and $\sigmax$ and $\sigmay$ are the deconvolved cloud sizes defined below.
We note that although this definition of $\Reff$ is long used as common custom, some of the recent studies also define cloud radius as half width half maximum, $R_{\mathrm{HWHM}}$ = 1.18 $\sqrt{\sigmax \sigmay}$ \citep{Rosolowsky2021PhangsGMCs}, and thus care is needed when comparing values.
\par

The deconvolved sizes are obtained by subtracting the beam size in quadrature from the extrapolated cloud sizes of $\sigmaxraw$ and $\sigmayraw$ as
\begin{eqnarray}
\sigmax &=& \sqrt{\sigmaxraw^2 - \theta_{\mathrm{b}}^2 / \left( 8 \log{2} \right)}\\
\sigmay &=& \sqrt{\sigmayraw^2 - \theta_{\mathrm{b}}^2 / \left( 8 \log{2} \right)}
\label{EqReffCorrection}
\end{eqnarray}
where $\theta_{\mathrm{b}}$ is the FWHM resolution of the data (2$\arcsec$.1 , $\sim$46 pc).
Out of the 5724 clouds identified, 225 clouds have $\sigmayraw$ smaller than $\theta_{\mathrm{b}} / \sqrt{8 \log{2}}$.
We marked them as ``deconvolution-failed clouds" and assigned $\sigmaxraw$ and $\sigmayraw$ as the upper limits on $\sigmax$ and $\sigmay$, respectively.

\emph{Velocity dispersion:}
The velocity dispersion of the cloud ($\sigmav$) is calculated by subtracting the instrumental velocity resolution from the extrapolated $\sigmavraw$ as
\begin{equation}
\sigmav = \sqrt{ \sigmavraw^2 - \left( \Delta{V}_{\mathrm{ch}} / \sqrt{12} \right)^2 },
\end{equation}
where $\Delta{V}_{\mathrm{ch}}$ is the channel width of the data (1 $\kmPerS$).
The factor of $1/\sqrt{12}$ is the ratio of the 2nd moment to the width for a boxcar profile.

\emph{Molecular gas mass:}
Molecular gas mass for each GMC ($\Mcl$) is calculated from $\LCO$ by applying a CO-to-H$_{2}$ conversion factor of 2.0 $\times$ 10$^{20}$ cm$^{-2}$
(K km s$^{-1}$)$^{-1}$ in combination with a correction factor of 1.36 that accounts for the contribution of Helium and other elements. \citep{Bolatto2013ConversionFactor}:
\begin{equation}
\label{EqLCOtoMsun}
\left(\frac{\Mcl}{\Msun}\right) =
\left(\frac{\alphaCO}{4.4 \Msun / \left( \mathrm{K}\ \kmPerS \mathrm{pc}^2 \right) }\right)
\left(\frac{\LCO}{\mathrm{K}\ \kmPerS\ \mathrm{pc}^{2}}\right),
\end{equation}
where $\alphaCO$ is the mass-to-luminosity ratio.
The $\alphaCO$ of 4.4 $\Msun$ $\left( \mathrm{K}\ \kmPerS \mathrm{pc}^2 \right)$ is often regarded as the `standard' value for GMCs in metal-rich environments, such as in M83's inner disk \citep{Bresolin2016}, and thus the use of the `standard' conversion factor is justified as the first order approximation \citep{Bolatto2013ConversionFactor}.
We note that, however, \citet{Lee2024M83} suggested a kpc-scale radial variation in $\alphaCO$, which is a factor of a few increase from the galactocentric radius of $\sim2$ kpc to $\sim$6 kpc.
\citet{Lee2024M83} noted that the radial gradient of metallicity in the disk of M83 is insufficient to explain the observed variation in $\alphaCO$ as it can only produce a factor of $\sim1.2$ to $\sim1.3$ variations in $\alphaCO$ over the corresponding radial range by adopting the metallicity measurement of \citet{Bresolin2016} and the metallicity-dependent descriptions of $\alphaCO$ of \citet{Wilson1995Xco} and \citet{Arimoto1996}.
They suggested that the weak dependence of $\alphaCO$ on $\LCO$, expected for gravitationally bound clouds with constant surface densities \citep{Solomon1987Larson, Bolatto2013ConversionFactor}, combined with the radial variation in the cloud mass function can provide a relative variation in $\alphaCO$ comparable to the observed variation.

\emph{Surface density:} Average gas surface density of a cloud ($\SigmaGMC$) is calculated as,
\begin{equation}
\SigmaGMC = \Mcl / (\pi \Reff^2).
\label{EqSigmaDef}
\end{equation}

\emph{Virial mass:} Following \cite{BertoldiMcKee1992} virial mass is calculated as
\begin{equation}
\Mvir = \frac{1}{a_1} \frac{5 R \sigmav^2}{G},
\label{EqMVirDef}
\end{equation}
where $G$ is the gravitational constant and $a_1$ is a geometrical factor that is (1 - $k$/3) / (1 - 2$k$/5) for a power law density distribution of $\rho(r)$ $\propto$ $r^{-k}$. We assumed $k$ to be 1, following S87.
Examinations of the nearby Galactic clouds suggest slightly larger values of $k$ \citep[e.g., 1.3--1.4,][]{Kauffmann2010RelationII, Lombardi2010}.
As the dependence of the geometrical factor $a_1$ on k is small, we adopt $k$=1 to keep consistency with previous studies.

\emph{Virial parameter:} The virial parameter $\alphaVir$ is a ratio between the virial mass and gas mass and is a measure of the ratio of the kinetic to gravitational energy \citep{BertoldiMcKee1992}:
\begin{equation}
\alphaVir \equiv \Mvir / \Mcl.
\label{EqAlphaVirDef}
\end{equation}
Ignoring the support from external pressure and magnetic fields, a state of simple virial equilibrium is achieved when $\alphaVir$ is 1 or a factor of few around 1, considering the uncertainty in the assumed density profile and CO-to-H$_2$ conversion factor.
Clouds are loosely bound when the equipartition between the kinetic and gravitational energy is achieved with $\alphaVir$ $\sim$2 \citep{Larson1981, BallesterosParedes2011A}.
We note a general caveat on interpreting $\alphaVir$. As isolated clouds with simple geometry and density distribution are assumed, the observationally derived $\alphaVir$ has a limitation in characterizing the dynamical status of molecular clouds. Numerical simulations suggest that clouds could be either more bound \citep{BallesterosParedes2018, RamirezGaleano2022} or more unbound \citep{Mao2020CloudProps} than what the face value of $\alphaVir$ would suggest.

\subsection{Extrapolation factors and measurement uncertainties}
\label{SubsecErrorEstimates}
The extrapolation to 0 K emission level made to basic cloud parameters \S\ref{SubsecBasicProps} is needed to correct for the underestimation of the parameters that would arise if the parameters are determined at a higher edge level \citep{Rosolowsky2006CPROPS}.
However, the extrapolation result becomes less reliable for a cloud that requires a higher correction factor. Thus, we expect clouds with higher correction factors to have higher measurement uncertainties assigned by the bootstrap estimate.
We show the amount of correction and measurement uncertainties to the clouds here.

\par
Figure \ref{FigCorrectionFactors}(a-c) shows the correction factor of the extrapolation, which is the ratio between the extrapolated to un-extrapolated values for CO luminosity, radius, and the velocity dispersion of the clouds, plotted as functions of the cloud mass.
The correction factors tend to increase toward lower cloud masses.
At a mass of 2 $\times$ 10$^5$ $\Msun$, the median of the correction factors are about 1.25, 1.16, and 1.12 for luminosity, radius, and velocity dispersion, respectively, and are modest.
At the lowest mass bin of 2 $\times$ 10$^4$ $\Msun$, the median factors are 1.96, 1.43, and 1.41, respectively.
\par

Figure \ref{FigCorrectionFactors}(d-f) shows the measurement errors in CO luminosity, radius, and velocity dispersion expressed in the fractional form.
Uncertainties in these basic three quantities are transferred to other advanced quantities following uncertainty propagation.

\subsection{GMC Catalog}
Table \ref{TableCatalog} lists the first several entries of the compiled catalog. The complete list shall be available as an electronic table.
Each line in the catalog reports the parameter of an identified cloud, including the centroid position and receding velocity, $\Reff$, the axial ratio ($\sigmay$/$\sigmax$) and position angle, $\sigmav$, $\LCO$, $\Mvir$, and the binary flag that indicates whether the deconvolution failed for the cloud or not.
\begin{deluxetable*}{rrrrrrrrrrl}
\tablecaption{Catalog of clouds in M83}
\tablewidth{0pt}
\tablehead{
\colhead{R.A.} & \colhead{Decl.} & \colhead{$V_{\mathrm{LSR}}$} & \colhead{$T_{\mathrm{peak}}$} & \colhead{$R_{\mathrm{eff}}$} & \colhead{$\sigma_{\mathrm{y}}/\sigma_{\mathrm{x}}$} & \colhead{$P.A.$} & \colhead{$\sigma_{\mathrm{v}}$} & \colhead{$L_{\mathrm{CO}}$} & \colhead{$M_{\mathrm{vir}}$} & \colhead{deconv. flag} \\
\colhead{}&\colhead{}&\colhead{(km s$^{-1}$)}&\colhead{(K)}&\colhead{(pc)}&\colhead{}&\colhead{(deg)}&\colhead{(km s$^{-1}$})&\colhead{($10^4$ K km s$^{-1}$ pc$^{2}$})&\colhead{($10^5$ M$_{\odot}$)}&\colhead{}}
\decimalcolnumbers
\startdata
13:37:00.310 &    -29:51:52.41 &                        542.6 &                         17.39 &                           66 &                                               0.51 &             -157 &                              13 &                         640 &                          110 &                      0 \\
13:37:00.755 &    -29:52:07.75 &                        510.9 &                         16.68 &                           86 &                                               0.78 &             -122 &                              22 &                        1300 &                          420 &                      0 \\
13:37:00.400 &    -29:51:50.07 &                        514.9 &                         15.74 &                           58 &                                               0.65 &             -149 &                             7.0 &                         320 &                           30 &                      0 \\
13:37:00.435 &    -29:51:49.76 &                        489.4 &                         13.66 &                           47 &                                               0.83 &             -167 &                             9.3 &                         300 &                           43 &                      0 \\
13:37:00.762 &    -29:51:43.94 &                        499.3 &                         13.43 &                           94 &                                               0.79 &             -150 &                              16 &                        1200 &                          240 &                      0 \\
13:37:00.610 &    -29:51:54.27 &                        481.9 &                         12.91 &                           57 &                                               0.52 &              -59 &                              11 &                         370 &                           75 &                      0 \\
13:37:00.334 &    -29:51:58.93 &                        564.5 &                         12.32 &                           74 &                                               0.68 &             -182 &                              11 &                         240 &                           91 &                      0 \\
$\vdots$ & $\vdots$ & $\vdots$ & $\vdots$ & $\vdots$ & $\vdots$ & $\vdots$ & $\vdots$ & $\vdots$ & $\vdots$ & $\vdots$ \\
\enddata
\tablecomments{Parameters of the identified cloud.
(1) Right ascension.
(2) Declination.
(3) Centroid velocity.
(4) Peak brightness temperature.
(5) Effective radius.
(6) Ratio between the minor and major axes lengths.
(7) Position angle of the major axis of the cloud, measured from the North.
(8) Velocity dispersion.
(9) CO (1--0) luminosity.
(10) Virial mass.
(11) Flag indicating whether the deconvolution was successfully made (=0) or not (=1).
}
\label{TableCatalog}
\end{deluxetable*}

\subsection{Fraction of CO luminosity sampled as GMC}
\label{SubsecCOLumFraction}

\begin{figure*}[ht!]
\plotone{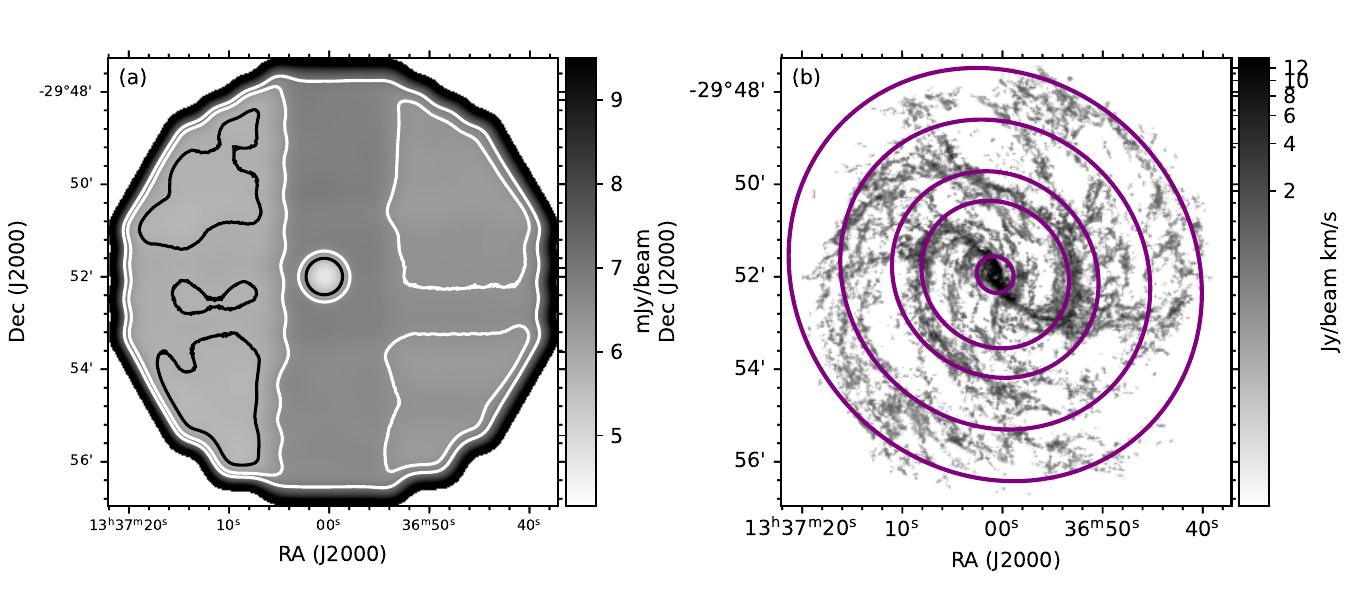}
\caption{
(a) RMS noise map of the $^{12}$CO (1--0) used here. The white lines indicate the RMS map with the contour levels of 0.9$\sigma$ and 1$\sigma$, where 1 $\sigma$ is the global median of the RMS noise of 6.5 mJy beam$^{-1}$ ($\sim$0.17 K), and the black line indicates the level of 1.1$\sigma$.
(b) Integrated intensity map of $^{12}$CO (1--) data used here.
The cyan lines indicate the galactocentric radii of 25$\arcsec$ ($\sim$0.55 kpc), 100$\arcsec$ ($\sim$2.2 kpc), 140$\arcsec$ ($\sim$3.1 kpc), 210$\arcsec$ ($\sim$4.6 kpc), and 280$\arcsec$ ($\sim$6.1 kpc), respectively.
}
\label{FigRMSMap}
\end{figure*}

\begin{figure}[ht!]
\plotone{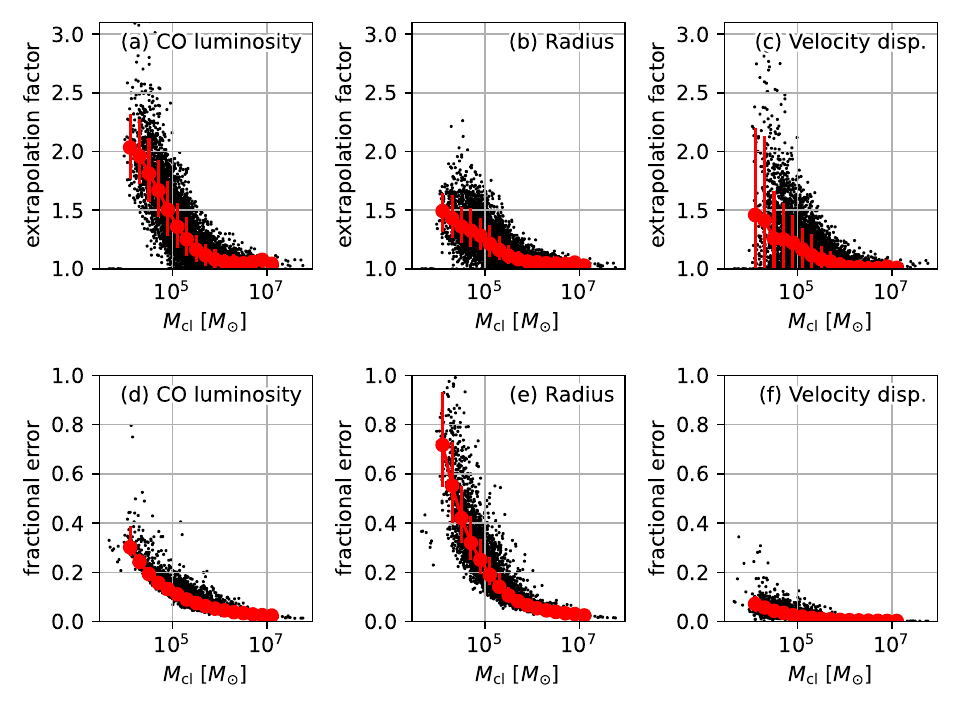}
\caption{
(a--c) Ratio of the extrapolated to un-extrapolated values for $\LCO$, $\Reff$, and $\sigmav$, respectively, are shown as a function of the cloud mass.
The red markers and error bars indicate the median and 16th to 84th percentile ranges of each quantity at the binning masses.
(d--f) Same as (a--c), but for fractional uncertainty for each quantity in (a-c) as a function of the cloud mass.
}
\label{FigCorrectionFactors}
\end{figure}

Table \ref{TblGlobalProps} lists the total CO luminosity $\LCO$ along with the fractional CO luminosities included within the signal mask ($f_{\mathrm{mask}}$) and sampled as molecular clouds ($f_{\mathrm{cl}}$).
To indicate their overall radial trends, we list the values for five radial bins separated at galactocentric radii shown in figure \ref{FigRMSMap}(b).
The $f_{\mathrm{mask}}$ is highest in the innermost radial bin ($\sim$97 percent) and decreases with increasing galactocentric radius to $\sim$68\% in the outermost radial bin.
This radially declining trend suggests that the contribution of lower mass clouds to total mass becomes increasingly significant at larger galactocentric radii.
\par

The extrapolation process recovers part of the emission removed by applying the signal mask.
Reflecting this, $f_{\mathrm{cl}}$ is higher than $f_{\mathrm{mask}}$ in all the radii.
However, in the central 550 pc, $f_{\mathrm{cl}}$ slightly exceeds 1 (1.04 $\pm$ 0.03), which indicates that the flux extrapolation is over-correcting the cloud flux by a few percent.
A radially declining trend of $f_{\mathrm{cl}}$ exists.
It is nearly 100 percent within the central 3.1 kpc.
It decreases radially down to $\sim$80\% in the outermost radial bin, suggesting an increased fraction of clouds below the detection limit.
\par

We estimate the detectable minimum cloud mass that is based on the observation and segmentation parameters.
The cloud identification process in (\S\ref{SubsecIdentify}) set the peak SNR threshold as 4, which corresponds to a brightness temperature limit of $\sim$0.5 K.
Assuming that the minimum cloud has an area equivalent to the beam area and that the cloud emission has a Gaussian profile with $\sigmav$ of 2 $\kmPerS$ and $\Tpeak$ of 0.5 K, the mass of the cloud is $\alphaCO \left(2 \pi \right)^{1/2} \Tpeak \sigmav \left(1.13 \theta_{\mathrm{b}}^2 \right)$. With the assumed $\alphaCO$ of 4.4 $\Msun$ / (K $\kmPerS$ pc$^2$) and $\theta_{\mathrm{b}}$ of 46 pc, the expected minimum cloud mass would be $\sim3 \times 10^{4}$ $\Msun$.
We note that this value is for clouds that exist in isolated areas, and the mass completeness limit is elevated in crowded regions.

\begin{deluxetable*}{lccc}
\tablecaption{Global CO luminosity and GMC mass in M83}
\tablewidth{0pt}
\tablehead{
\colhead{Radial range} & \colhead{$\LCO$} & \colhead{$f^{\mathrm{mask}}$} & \colhead{$f^{\mathrm{cl}}$} \\
\colhead{(kpc)} & \multicolumn{1}{c}{(10$^7$ K km s$^{-1}$ pc$^2$)}
}
\decimalcolnumbers
\startdata
All                 & 82.7 $\pm$ 0.17 & 0.85 & 0.93 $\pm$ 0.05 \\
0.00 - 0.55         & 12.6 $\pm$ 0.01 & 0.98 & 1.04 $\pm$ 0.03 \\
0.55 - 2.18         & 18.6 $\pm$ 0.05 & 0.90 & 1.00 $\pm$ 0.05 \\
2.18 - 3.05         & 17.7 $\pm$ 0.05 & 0.90 & 0.97 $\pm$ 0.05 \\
3.05 - 4.58         & 17.3 $\pm$ 0.08 & 0.79 & 0.86 $\pm$ 0.05 \\
4.58 - 6.11         & 14.4 $\pm$ 0.10 & 0.73 & 0.83 $\pm$ 0.06 \\
\enddata
\tablecomments{(1) Radial range. Locations of the boundary radii are shown in figure \ref{FigRMSMap}(b). (2) CO luminosity. (3) Fractional CO luminosity within the signal mask. (4) Fractional CO luminosity sampled as clouds.}
\label{TblGlobalProps}
\end{deluxetable*}

\section{Results}
\label{SecResults}

\begin{figure*}[ht!]
\plotone{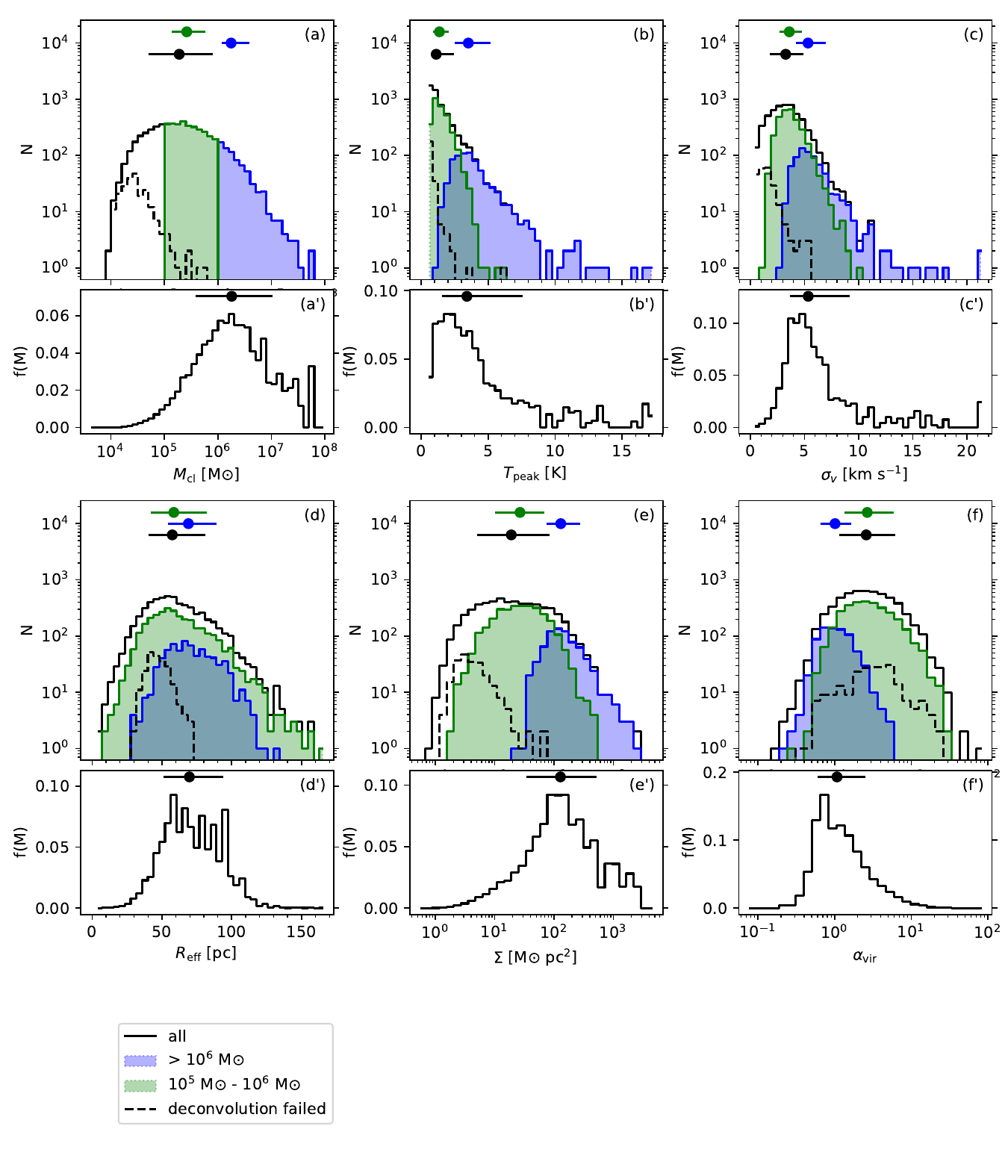}
\caption{
(a) Distribution of cloud mass $\Mcl$ in M83.
In the upper panel, the black histogram shows the overall distribution. The blue solid, green solid, and black dashed histograms indicate the sub-samples with $\Mcl$ $>$ 10$^6$ M$\odot$, $\Mcl$ in the range between 10$^5$ and 10$^6$ M$\odot$, and with sizes that failed in the deconvolution (\S\ref{SubsecDerivedProps}).
In the same panel, the marker and horizontal line indicate the median and 16th-to-84th percentile range for all clouds (black), clouds with $>10^6$ $\Msun$ (blue), and clouds with 10$^5$--10$^6$ $\Msun$ (green).
In the bottom panel, the mass-weighted histogram is shown. Note that as the constant CO-to-H$_2$ conversion factor is adopted, it is equivalent to the luminosity-weighted histogram. The marker and horizontal line indicate the median and percentiles as the upper panel.
(b to g) Same as (a), but for peak temperature $\Tpeak$, velocity dispersion $\sigmav$, effective radius $\Reff$, surface density $\SigmaGMC$, and virial parameter $\alphaVir$, respectively.
}
\label{FigCloudPropHistograms}
\end{figure*}

\subsection{Distribution of Cloud Properties}
\label{SubsecCloudPropDistribution}
Figure \ref{FigCloudPropHistograms} shows the distributions of the molecular cloud properties, including mass ($\Mcl$), peak temperature ($\Tpeak$), velocity dispersion $\sigmav$, radius ($\Reff$), surface density ($\SigmaGMC$), and virial parameter ($\alphaVir$).
Each cloud property is displayed in unweighted and mass-weighted histograms.
The top panels for each parameter show unweighted histograms for all the clouds (black line), clouds with masses above 10$^6$ $\Msun$ (blue line), those with masses between 10$^5$ and 10$^6$ $\Msun$ (green line), and those for which the deconvolution failed to determine their sizes (dashed line).
The median with 16-to-84 percentile ranges for the first three categories are displayed using markers and horizontal error bars.
The bottom panels for each parameter display the mass-weighted histogram with the mass-weighted medians and horizontal bars for the 16 and 84 percentiles.
Table \ref{TableClProps} lists the summary of the measurements.

\begin{deluxetable*}{lccccccc}
\tablecaption{Cloud property distribution summary}
\tablewidth{0pt}
\tablehead{
\colhead{} & \colhead{$T_{\mathrm{peak}}$} & \colhead{$\sigma_{\mathrm{v}}$} & \colhead{$R_{\mathrm{eff}}$} & \colhead{$M_{\mathrm{cl}}$} & \colhead{$\Sigma$} & \colhead{$\alpha_{\mathrm{vir}}$} & \colhead{$N_{\mathrm{cl}}$} \\
\colhead{} & \colhead{($K$)} & \colhead{(km s$^{-1}$)} & \colhead{(pc)} & \colhead{($10^5_{\odot}$)} & \colhead{($M_{\odot}$ pc$^{-2}$)} & \colhead{} & \colhead{}
}
\decimalcolnumbers
\startdata
\multicolumn4c{Un-weighted percentiles} \\
All                 & $1.1_{0.7}^{2.4}$ & $3.3_{1.8}^{4.9}$ & $59_{41}^{86}$ & $1.9_{0.51}^{8.1}$ & $22_{5.8}^{100}$ & $3.0_{1.0}^{7.0}$ & $5724$ \\
$>$10$^6$ M$\odot$  & $3.5_{2.5}^{5.2}$ & $5.3_{4.2}^{7.0}$ & $69_{54}^{89}$ & $18_{12}^{38}$ & $130_{75}^{280}$ & $1.0_{0.6}^{2.0}$ & $734$ \\
$>$10$^5$ M$\odot$  & $1.4_{0.91}^{2.1}$ & $3.6_{2.7}^{4.8}$ & $59_{42}^{83}$ & $2.6_{1.4}^{5.9}$ & $27_{10.0}^{68}$ & $3.0_{1.0}^{6.0}$ & $3124$ \\
\hline
\multicolumn4c{Weighted percentiles} \\
All                 & $3.4_{1.5}^{7.6}$ & $5.4_{3.6}^{9.2}$ & $70_{51}^{94}$ & $18_{3.8}^{110}$ & $130_{35}^{520}$ & $1.0_{0.6}^{3.0}$ & $5724$ \\
$>$10$^6$ M$\odot$  & $4.7_{3.0}^{9.3}$ & $6.5_{4.8}^{11}$ & $78_{58}^{96}$ & $37_{15}^{160}$ & $230_{99}^{870}$ & $0.8_{0.5}^{1.0}$ & $734$ \\
$>$10$^5$ M$\odot$  & $1.7_{1.1}^{2.5}$ & $3.9_{3.0}^{5.1}$ & $59_{43}^{81}$ & $4.6_{2.1}^{8.0}$ & $42_{16}^{88}$ & $2.0_{1.0}^{4.0}$ & $3124$ \\
\hline
\enddata
\tablecomments{Median and 16th-to-84th percentile range of the cloud properties.
(1) Radial range.
(2) Peak brightness temperature.
(3) Velocity dispersion.
(4) Effective radius.
(5) Mass.
(6) Surface density.
(7) Virial parameter.
(8) Number of clouds.
}
\label{TableClProps}
\end{deluxetable*}

\subsubsection{Mass}
\label{SubsubsecCloudMass}
Figure \ref{FigCloudPropHistograms}(a) shows the distribution of $\Mcl$.
The unweighted median of $\Mcl$ for the entire sample is 1.9 $\times$ 10$^5$ M$\odot$,
which is close to the median cloud mass found in the inner Galactic disk found by existing surveys \citep[e.g., $\sim$8 $\times$ 10$^4$ $\Msun$,][]{MivilleDeschenes2017MWGMC} and also close to the lower end of a conventional definition of GMC mass range \citep[around 10$^5$ $\Msun$, e.g., ][]{Sanders1985}.
The unweighted 16 percentile of cloud mass is 5 $\times$ 10$^4$ M$\odot$ and agrees with the roughly expected minimum cloud mass of $\sim3$ $\times$ 10$^{4}$ $\Msun$ (\S\ref{SubsecCOLumFraction}).
This suggests that the present catalog samples the full range of GMCs in M83's disk, at least when they are well isolated in space or velocity.
\par

Figure \ref{FigCloudPropHistograms}(a), when the top and bottom panels are compared, indicates that massive GMCs are a minority in number but occupy the dominant fraction of the total CO luminosity. About 50\% of the molecular gas mass (or total CO luminosity) resides in GMCs with $\Mcl>10^6$ M$\odot$, although such massive GMCs consist of less than 16\% in number. Therefore, the small number of massive clouds contain the major fraction of the total gas mass in M83. This trend is similar to the inner Galactic disk \citep{Solomon1987Larson, MivilleDeschenes2017MWGMC}. We will discuss the cloud mass spectrum in \S\ref{SecMF}.

\subsubsection{Peak temperature}
Figure \ref{FigCloudPropHistograms}(b) shows that $\Tpeak$ has a tail toward the higher end with a maximum value of $\sim$15 K.
The unweighted and mass-weighted median values among all the GMCs are 1.3 K and 3.4 K, respectively, having more than a factor of two difference. On the contrary, the difference is smaller for massive clouds with $\Mcl>10^6$ $\Msun$ (3.5 K and 4.7 K).
Again, this is because the small number of massive clouds determines the mass-weighted average properties of the GMC population in M83.
\par

We note that in previous extragalactic GMC studies, median $\Tpeak$ in the sampled clouds is often around 3 K, specifically 3.0 K in M51 \citep{Colombo2014Env} and 3.3 K in M83 \citep{Hirota2018M83}, and is higher than our median $\Tpeak$ of 1.3 K.
This is due to low sensitivities in the previous studies, which were made with similar spatial resolutions comparable to that in this study ($\sim46$ pc).
In the previous studies, the most frequent cloud masses were 10$^6$ $\Msun$ or higher.
This work detects abundant clouds with smaller masses by sampling clouds with the minimum $\Tpeak$ of $\sim0.5$ K.
\par

\subsubsection{Velocity dispersion and radius}
\label{SubsubsecVDispAndRadius}
Figure \ref{FigCloudPropHistograms}(c) and \ref{FigCloudPropHistograms}(d) show the distribution of $\sigmav$ and $\Reff$, respectively.
The two quantities show different tendencies concerning the classification by $\Mcl$.
The $\sigmav$ values tend to be higher for more massive clouds, but such a trend is less evident for $\Reff$.
\par

The distribution of $\sigmav$ has a skewed tail toward the higher end, extending over 20 $\kmPerS$.
The un-weighted and mass-weighted median values of $\sigmav$ are 3.3 $\kmPerS$ and 5.3 $\kmPerS$, respectively.
The 16 percentile of $\sigmav$ is 1.8 $\kmPerS$ and is well above the instrumental resolution of 0.3 $\kmPerS$ (\S\ref{SubsecDerivedProps}).
Hence, the measured $\sigmav$ should be almost unaffected by the instrumental resolution.
\par

In contrast to $\sigmav$, the mass dependence is not clear for $\Reff$.
The unweighted median $\Reff$ is 59 pc, and the maximum $\Reff$ is $\sim$100 pc.
There are 225 clouds with failure in the deconvolution for $\Reff$ since their pre-deconvolution sizes are less than the spatial resolution (\S\ref{SubsecDerivedProps}).
Figure \ref{FigCloudPropHistograms} shows that such deconvolution failed clouds mostly have $\Mcl$ smaller than 10$^5$ $\Msun$.
\par

In this study, the dynamic range in $\Reff$ is limited, with only a factor of two difference between the 16th and 84th percentiles (41 pc and 86 pc).
The 84th percentile of 86 pc is easily acceptable as the most significant cloud size should be limited by the thickness of the molecular gas disk, which is estimated to be of order 100 pc in local disk galaxies.
For example, the FWHM thickness of the molecular gas disk of the MW is $\sim$100 pc \citep{HeyerDame2015Review}.
On the other hand, the 16th percentile of 41 pc needs consideration because $\Reff$ is deconvolved, and typical GMCs in the MW have smaller sizes.
Most of the recent cloud identification algorithms have a known tendency to find structures with sizes a few times the instrumental resolution \citep{Verschuur1993BeamBias, Hughes2013GMCs}, and the $\Reff$ distribution suggests that the same tendency also applies here.
\par

The bootstrap error estimation made in \S\ref{SubsecBasicProps} indicates the measured $\Reff$ is more uncertain than $\sigmav$, which would indicate that the resolution is more marginal in the spatial directions than the velocity direction.
As we identified clouds from the masked data cube, in which all the valid voxels are positive ($>2$ $\sigma$), the cloud radii tend to be overestimated rather than underestimated due to the measurement uncertainty.
The relatively large 16th percentile of $\Reff$, 41 pc, could also be a consequence of the overestimation of radius in small clouds.
Therefore, even if the beam deconvolution is made to $\Reff$, there might still remain uncertainties on the beam filling factor. Further discussion about this point will be made in \S\ref{SecDiscussion}.

\subsubsection{Surface density}
Figure \ref{FigCloudPropHistograms}(e) shows the distribution of $\SigmaGMC$.
The unweighted and mass-weighted median values are 22 $\MsunPerSqPC$ and 128 $\MsunPerSqPC$, respectively; the mass-weighted median is skewed toward massive clouds.
This mass-weighted median is close to the median $\SigmaGMC$ for clouds with $M>10^6$ $\Msun$ (130 $\MsunPerSqPC$).
This alignment suggests a trend that massive GMCs account for the majority of the molecular gas mass in M83.
\par

We note that the previous extragalactic studies in molecular-rich gas disks with shallower mass sensitivities found median $\SigmaGMC$ similar to that for $\Mcl$ $>10^6$ $\Msun$ clouds in M83.
(e.g., \citealp[177 $\MsunPerSqPC$ in M51,][]{Colombo2014Env};
\citealp[106 $\MsunPerSqPC$ in M83,][]{Hirota2018M83};
\citealp[also 170 $\MsunPerSqPC$ in M83,][]{Freeman2017};
\citealp[100$^{+160}_{-60}$ $\MsunPerSqPC$ in 10 galaxies covered by the PHANGS program][]{Rosolowsky2021PhangsGMCs}
).
The unweighted median $\SigmaGMC$ of 22 $\MsunPerSqPC$ is below the median value by S87 and is close to that among the inner Galactic clouds by \cite{MivilleDeschenes2017MWGMC} ($31.6$ $\MsunPerSqPC$; see also Appendix \ref{SecAppendixMWCatalogComparison}).
However, we will discuss that our median aligns better with the high value of S87 after a beam dilution correction (see Section \S\ref{SecDiscussion}).
\par

\subsubsection{Virial parameter}
Figure \ref{FigCloudPropHistograms}(f) shows distribution of $\alphaVir$.
With an assumption of negligible support from magnetic fields and confinement by external pressure, $\alphaVir$ $\sim$ 1 indicates the virial equipartition and $\alphaVir$ $\sim$2 the equipartition between kinetic and gravitational energies (loosely gravitationally-bound).
The unweighted median $\alphaVir$ is $\sim$2.4, which is similar to the ones found in some of the recent extragalactic cloud surveys.
For example, \citet{Sun2020} found the median $\alphaVir$ of 2.7 from 70 galaxies included in the ALMA-PHANGS program.
Recent Galactic cloud catalogs also report higher $\alphaVir$ than $\sim$1, which is suggested from S87's data by assuming a constant $\alphaCO$ of 4.4 $\Msun$ / (K $\kmPerS$ pc$^2$).
Using the same $\alphaCO$, the clouds sampled by \citet{Rice2016} indicate a median $\alphaVir$ of 2.3 for the entire Galactic disk, although it lowers to $\sim$1.5 if the inner Galactic disk clouds are concerned (Appendix \ref{SecAppendixMWCatalogComparison}).
Even higher median $\alphaVir$ of $\sim7$ and $\sim11$ are found in the inner and outer Galactic disk by \citet[][see also \citealp{Evans2021BoundCloud}]{MivilleDeschenes2017MWGMC}.

\par

Mass-weighted mean $\alphaVir$ is close to 1; thus, massive GMCs appear more gravitationally bound than less massive ones at first glance.
We show in Appendix \ref{SecAppendixAlphaVirM} that $\alphaVir$ of the M83 clouds tends to vary as $\propto$ $\Mcl^{-0.5}$.
A similar tendency is also seen in previous cloud studies made in different environments \citep{Heyer2001OuterGalaxy, MivilleDeschenes2017MWGMC, Evans2021BoundCloud} \citep[also see figure 2 of][]{Chevance2023Review}. We argue in Appendix \ref{SecAppendixAlphaVirM} that the limited spatial resolution could be forming the trend in the sampled M83 clouds.

\subsection{Scaling relations}
\label{SubsecScalingRelations}

\begin{figure*}[htbp]
\plotone{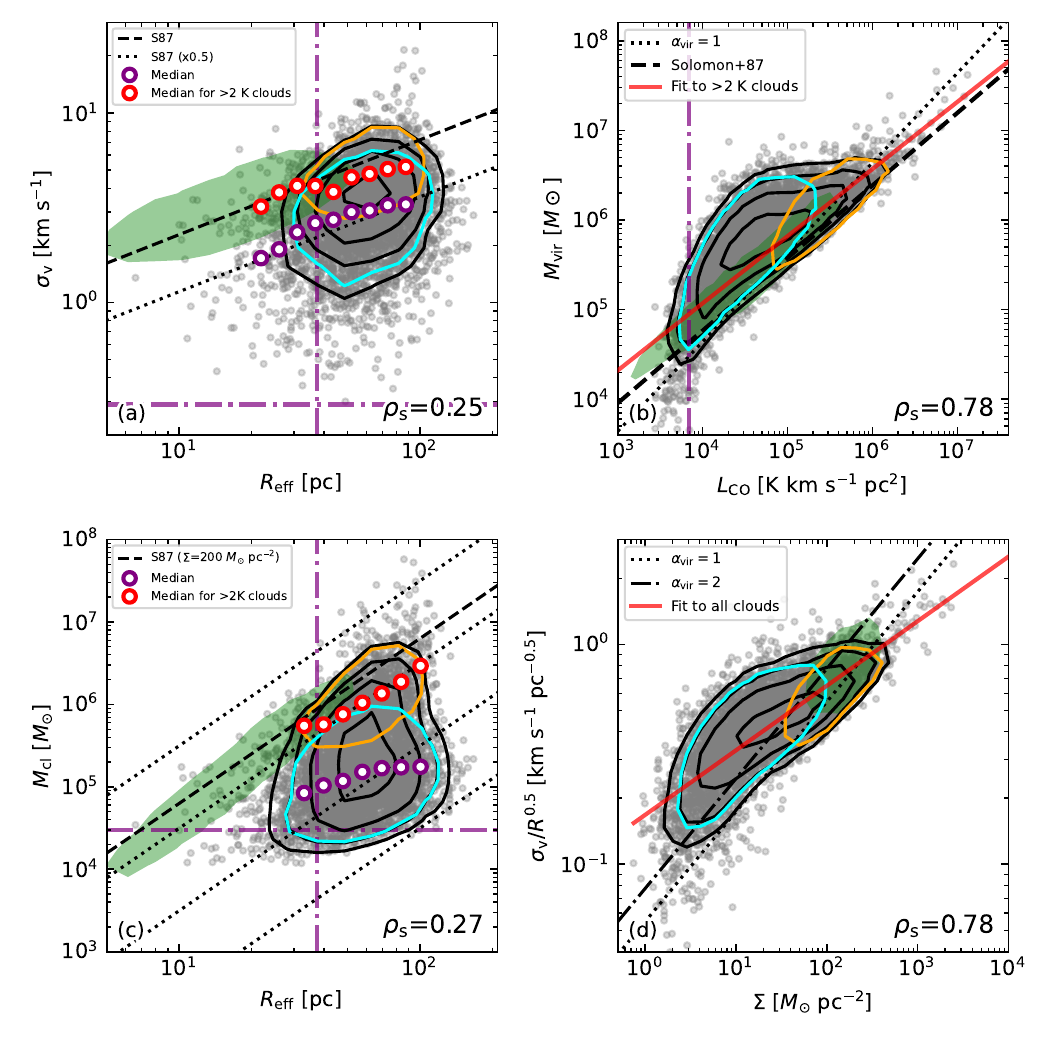}
\caption{
Scaling relations of cloud properties.
In each plot, the raw data points for the M83 clouds are shown with the gray markers, and the data density is displayed with the black contour lines at 10\%, 20\%, 40\%, and 80\% levels of the maximum density, respectively.
In the bottom right corner of the plot, Spearman's rank correlation coefficient is noted.
Orange and cyan contour lines indicate the distribution of $\Tpeak>2$ K and $<2$ K clouds, respectively, at the 20\% density level.
The green-shaded region indicates the distribution density of the Galactic clouds of S87 at a 20\% level.
Where appropriate, the resolution limits for $\sigmav$ of $1/\sqrt{12}$ $\kmPerS$ and $\Reff$ of $\sim37$ pc ($=1.91 \times \theta_{\mathrm{b}} / \sqrt{8 \log{2}})$) and the luminosity limit of $\sim6900$ K $\kmPerS$ pc$^2$, which corresponds to the expected minimum mass of $\sim3 \times 10^{4}$ $\Msun$ (\S\ref{SubsecCOLumFraction}), are shown with the purple dash-dotted line in each plot.
(a) $\sigmav$--$\Reff$ relationship.
The black dashed line indicates the $\sigmav$--$\Reff$ relation of S87, which is eq. (\ref{eq:Larson1}) with $C=0.72$ $\kmPerS$ pc$^{-1/2}$.
The dotted line is the one with $C$ half of S87's value.
The purple open markers indicate the median $\sigmav$ derived at the binned $\Reff$, and the red open marker running median for $\Tpeak>2$ K clouds.
(b) $\Mvir$--$\LCO$ relationship.
The dashed line indicates the relation of S87 ($\Mvir$ $\propto$ $\LCO^{0.81}$) and dotted line is $\alphaVir$=1.
The red line indicates the fit to the M83 clouds with $\Tpeak>2$ K (eq. \ref{EqMvirLCOFit}).
(c) $\Mcl$--$\Reff$ relationship. The dotted lines indicate the lines for $\SigmaGMC$ of 1, 10, 100, and 1000 $\MsunPerSqPC$, respectively.
The dashed line indicates $\SigmaGMC$=200 $\MsunPerSqPC$, implied from S87's scaling relations.
As in (a), the running median for all the samples and $>2$ K clouds are shown with red and purple markers.
(d) $\sigmav / \Reff^{1/2}$--$\SigmaGMC$ relationship.
The red line indicates the fit to all M83 clouds (eq. \ref{EqSizeLinewidthCoeffSigmaRelationFit}).
The lines of $\alphaVir$ of 1 and 2 are shown with the dotted and dash-dotted lines, respectively.
}
\label{FigSclRelations}
\end{figure*}

The scaling relations of cloud properties were empirically established by the early studies of Galactic clouds \citep{Larson1981, Solomon1987Larson}.
In particular, the following relations are collectively referred to as Larson's relationships and more qualitatively determined by S87:
molecular clouds
(i) have a power-law relation between their velocity dispersions $\sigmav$ and radius $\Reff$,
\begin{equation}
\sigmav = C \Reff^{0.5}, \label{eq:Larson1}
\end{equation}
with a scaling coefficient $C$ of 0.72 $\kmPerS \mathrm{pc}^{-1/2}$, (ii) have a near-linear scaling between virial mass and CO luminosity, and (iii) have a constant surface density of $\sim$200 $\MsunPerSqPC$ (S87).
\par

These three relationships are inter-dependent on each other \citep{Heyer2009, BallesterosParedes2011A}: by equating the two estimations of cloud mass based on luminosity ($\pi \Reff^2 \SigmaGMC$) and virial equations ($\Mvir/\alphaVir$), we have
\begin{equation}
C = {\sigmav}/\sqrt{\Reff}
= \left(\frac{2}{9} \alphaVir \pi G {\SigmaGMC} \right)^{1/2}.
\label{EqSizeLinewidthCoeff}
\end{equation}
Therefore, $\propto\Reff^{1/2}$ scaling of $\sigmav$ (1st relationship) is derived when $\alphaVir$ and $\SigmaGMC$ are constant, which are derived from the 2nd and 3rd relationships.
\cite{Heyer2009} also pointed out that the coefficient $C$ varies among clouds sampled in more diverse environments than S87 and it scales as $\SigmaGMC^{1/2}$ as expected by eq. (\ref{EqSizeLinewidthCoeff}). \cite{BallesterosParedes2011A} discussed the same dependence in the context of free-fall collapse \cite[see also][]{Larson1981}.
\par

Larson's relations (the S87's relations) are often taken as reference relations to compare cloud properties among various environments \citep[e.g.,][]{Bolatto2008GMCs, Hughes2013GMCs}.
In an inner portion of M83, \cite{Hirota2018M83} sampled clouds from $^{12}$CO (1--0) data observed with ALMA and found that GMCs in the disk region show similar characteristics as S87's clouds with median $\alphaVir$ and $\SigmaGMC$ of $\sim1$ and $\sim100$ $\MsunPerSqPC$, respectively. The new data surpass this previous study in sensitivity, although the spatial resolutions are similar: most clouds have $\Tpeak>2$~K in \citet{Hirota2018M83}, while median $\Tpeak$ is about 1 K in this study.
In what follows, we show clouds with $\Tpeak>$2~K and $<2$~K as separate subsets and show that the $\Tpeak>$2~K clouds mostly agree with S87's relations, while the $\Tpeak<$2~K clouds do not and are more aligned with more recent Galactic clouds surveys (see Appendix \ref{SecAppendixMWCatalogComparison}).
Later in \S\ref{SecDiscussion}, we will show that the majority of the $\Tpeak<$2~K clouds are likely affected by beam dilution and discuss that they could also follow S87's relations when the effect of dilution (i.e., beam filling factor) is corrected.
\par

We note one caveat about the scaling relations.
The near constant $\SigmaGMC$ of $\sim200$ $\MsunPerSqPC$ in S87's clouds is a consequence of the selection effect, resulting from the high-intensity thresholds adopted by S87 for identifying cloud peaks \citep{Lombardi2010}.
Recent studies of Galactic molecular clouds report a wider range of $\SigmaGMC$, in particular with a lower median $\SigmaGMC$ \citep[e.g.,][]{Rice2016, MivilleDeschenes2017MWGMC, Colombo2019},
which resembles low $\SigmaGMC$ values found for the $\Tpeak$ $<2$ K sample of clouds in M83.
Comparison of the M83 clouds with the catalogs of \cite{Rice2016} and \cite{MivilleDeschenes2017MWGMC} is presented in Appendix \ref{SecAppendixMWCatalogComparison}.

\subsubsection{Linewidth--size relation}
\label{SubsecLinewidthToSize}

Figure \ref{FigSclRelations}(a) shows the relationship between $\sigmav$ and $\Reff$ for the clouds in M83.
The linewidth--size relation of S87, eq. (\ref{eq:Larson1}), $C=0.72$ $\kmPerS$ pc$^{-1/2}$, is also shown with the black dashed line for reference.
\par

Any sign of correlation is hardly seen for $\sigmav$-$\Reff$ relation with the entire sample of M83 clouds with a low Spearman's rank correlation coefficient of $\sim$0.3.
The absence of the correlation is often seen in other extragalactic GMC studies \citep[e.g.,][]{Hughes2013GMCs}.
\par

Another point to note is that most samples have $\sigmav$ significantly lower than expected from S87's relation.
Clouds with low $\Tpeak$ mostly drive the downward deviation as the 20\% level contour of the distribution density of the clouds with $\Tpeak$ $<2$ K is located below the line of S87's relation.
On the other hand, at least some of the clouds with $\Tpeak$ $>$2 K are compatible with S87's relation, as indicated by the location of its distribution density contour.
\par

As further reference, we plot the median $\sigmav$ values derived for the binned cloud radii for the clouds with $\Tpeak$ $>2$ K and $<2$ K with red and purple markers, respectively.
The running-median values exhibit clear separation by $\Tpeak$.
While the median $\sigmav$ for $>2$ K clouds is located close to S87's relation of $\sigmav$ $\propto$ $C\Reff^{1/2}$ with $C$ of 0.72 $\kmPerS$ pc$^{-1/2}$, that for $<2$ K clouds follows the one with $C$ half of S87's value, i.e., 0.36 $\kmPerS$ pc$^{-1/2}$.
\par

The alignment of $>2$ K clouds with S87's relation suggests that luminous clouds in M83 might share the same characteristics as the Galactic clouds sampled by S87.
In the early study of S87, the Galactic clouds were identified by high-intensity detection thresholds, so the S87's relation would represent the properties of discrete GMCs with high $\SigmaGMC$.
Therefore, the luminous clouds in M83 might have properties similar to the dense GMCs in the MW. Again, we note that these clouds are the dominant contributor to the total CO luminosity of the galaxy.

\subsubsection{Virial mass--CO luminosity relation}
\label{SubsecMvirLCO}
Figure \ref{FigSclRelations}(b) shows the relationship between the $\Mvir$ and $\LCO$.
The dashed line on the figure indicates the case for the simple virial equilibrium, $\alphaVir$ = 1, with the `standard' CO-to-H$_2$ conversion factor of 2.0 $\times$ 10$^{20}$ cm$^{-2}$ / (K $\kmPerS$).
\par
As a whole, both quantities are well correlated to each other with Spearman's correlation coefficient of $\sim0.8$.
In particular, the $>2$ K clouds exhibit tight correlation, located within a factor of 3 variation to $\alphaVir$ of 1.
The $<2$K clouds also show correlation but are located upward of the line of $\alphaVir$ = 1.
Such upward deviation is often interpreted as the super-virial state of clouds or the variation in the CO-to-H$_2$ conversion factor.
\par

The Galactic clouds of S87 exhibit a narrow range of distribution on the plot, with the fitted result of $\Mvir$($\Msun$) = 43 $\left[ \LCO \left( \mathrm{K}\ \kmPerS\ \mathrm{pc}^2\right) \right]^{0.81}$ after adjusting for the adopted distance to the Galactic center \citep[S87; see also ][]{Bolatto2013ConversionFactor}.
The $\Tpeak$ $>2$K clouds mostly align with S87's reference relation.
The least-square fitting to $>2$ K clouds yield
\begin{equation}
\Mvir(\Msun) = (114 \pm 5) \left[ \LCO \left( \mathrm{K}\ \kmPerS\ \mathrm{pc}^2\right) \right]^{0.75\pm0.02},
\label{EqMvirLCOFit}
\end{equation}
which has an index similar to S87's relation.

\subsubsection{Mass--size relation}
\label{SubsecMassSize}
Figure \ref{FigSclRelations}(c) shows the relationship between $\Mcl$ and $\Reff$.
The lines for $\SigmaGMC$=1, 10, 100, and 1000 $\MsunPerSqPC$ are drawn for reference with dotted lines.
The Galactic clouds of S87 lie close to a constant $\SigmaGMC$ line of $200$ $\MsunPerSqPC$.
\par

In contrast to the narrow range of $\SigmaGMC$ seen in S87's MW clouds, the measured properties of the M83 clouds exhibit a wide range of $\SigmaGMC$, reaching about three orders of magnitudes on the plot.
The wide range is similar to the Galactic clouds sampled by \citet{MivilleDeschenes2017MWGMC} (see their figure 7a).
The correlation coefficient is low for the entire sample ($\sim0.3$).
Still, the $>$2 K clouds are centered around the constant $\SigmaGMC$ of 100 $\MsunPerSqPC$.
Taking the running median of $\Mcl$ for $>2$ K and $<2$ K clouds as was made in $\sigmav$--$\Reff$ relation, it is apparent that the $>$ 2K clouds are distributed close to the line of constant $\SigmaGMC$=100 $\MsunPerSqPC$.
For the $<2$ K clouds, the slope of the median mass values is shallower than those of the constant $\SigmaGMC$ lines, suggesting a trend that (apparently) larger clouds have smaller $\SigmaGMC$.

\subsubsection{Normalized velocity dispersion--surface density relationship}
\label{SubsecNormVelDispersionSigma}

Figure \ref{FigSclRelations}(d) shows the relationship between the normalization coefficient of the linewidth--size relation $C (=\sigmav/\Reff^{1/2})$ and $\SigmaGMC$.
As in eq. (\ref{EqSizeLinewidthCoeff}), $C$ is proportional to $\SigmaGMC^{1/2}$ for clouds dominated by self-gravity if $\alphaVir$ is approximately constant.
The dashed and dotted lines show the expected behaviors for $\alphaVir$ of 1 and 2, corresponding to the gravitational virial equilibrium and free-fall collapse, respectively.
\par

A high correlation coefficient of $\sim$0.8 is obtained, but this is trivially expected because the scaling between $C$ and $\SigmaGMC^{1/2}$ is mathematically equivalent to the $\Mvir$--$\LCO$ scaling (\S\ref{SubsecMvirLCO}).
\par

If all the clouds are strongly governed by self-gravity, we would expect all the data points to be clustered around the lines of $\alphaVir$=1 and 2.
However, we see upward deviations from these lines toward lower $\SigmaGMC$, mainly driven by the clouds with $\Tpeak$ $<2$ K.
To characterize the deviation, a least-square fitting to the entire sample is made,
\begin{equation}
\sigmav (\kmPerS) / \sqrt{\Reff (\mathrm{pc})} = (0.16 \pm 0.01) \left[ \SigmaGMC (\MsunPerSqPC) \right]^{0.30 \pm 0.01}.
\label{EqSizeLinewidthCoeffSigmaRelationFit}
\end{equation}
This index, $\sim$0.30, is shallower than the 0.5 for clouds governed by self-gravity (eq.\ref{EqSizeLinewidthCoeff}).
\par

The clouds with $\Tpeak>2$ K and $\Tpeak<2$ K make a clear separation on the plot.
The $>2$ K clouds are mostly clustered around the line of $\alphaVir$=1 and over a factor of a few range around $\SigmaGMC\sim 100$ $\MsunPerSqPC$.
On the other hand, the $<2$ K clouds are distributed systematically in the area of $\alphaVir>1$ (the 20\% density level barely crosses the line of $\alphaVir$=1).
Their $\SigmaGMC$ values are lower.
The shallow index of $\sim$0.3 of the $\sigmav/\sqrt{R}$--$\SigmaGMC$ relation is driven by the larger $\alphaVir$ in the clouds with $\Tpeak<2$ K.
We will discuss the impact of beam dilution on $\alphaVir$ and $\SigmaGMC$ in \S\ref{SecDiscussion} and also show that the beam dilution also explains the shallow index of $\sim$0.3 in Appendix \ref{SecAppendixSizelinewidthCoeffSigma}.

\subsection{Spatial distribution of the clouds}
\label{SubsecSpatialDistribution}

\begin{figure}[ht!]
\plotone{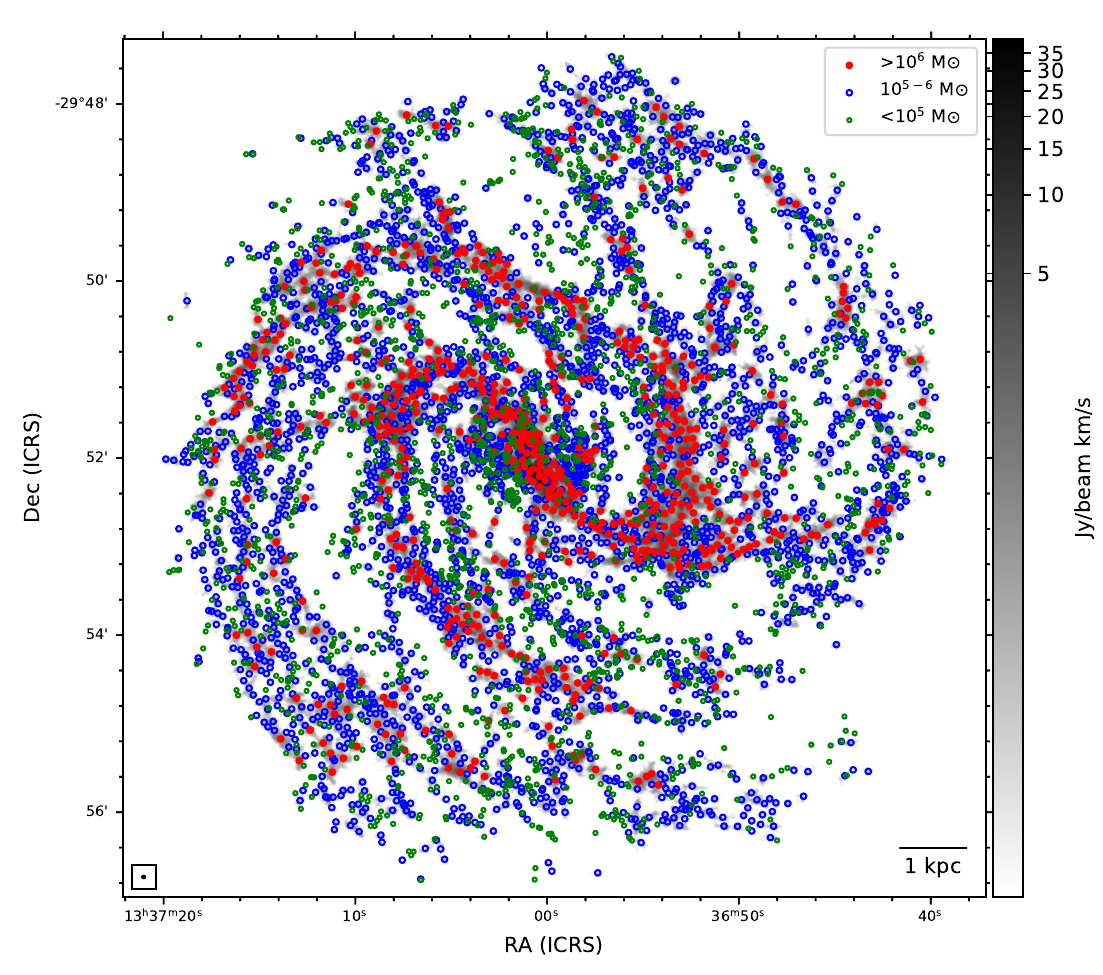}
\caption{
Distribution of the identified clouds overlaid on the CO map. Clouds with masses over 10$^6$ $\Msun$, between 10$^5$ $\Msun$ and 10$^6$ $\Msun$, and below 10$^5$ $\Msun$ are indicated with red-filled, blue open, and green open markers, respectively.
}
\label{FigM0ByMass}
\end{figure}

Figure \ref{FigM0ByMass} shows the distribution of the identified clouds, categorized into three groups by mass: $<$10$^{5}$, 10$^{5-6}$, and $>$10$^{6}$ $\Msun$, respectively. Massive clouds over 10$^6$ $\Msun$ are strongly concentrated around bright galactic structures, namely the galactic center, bar, and spiral arms. On the contrary, smaller clouds (below $10^6$ $\Msun$) prevail over the wider area,
residing both in arm and inter-arm regions.
This confirms the similar trends reported in other galaxies \citep[e.g., ][]{Koda2009, Colombo2014Env}.
\par

\begin{figure}[ht!]
\plotone{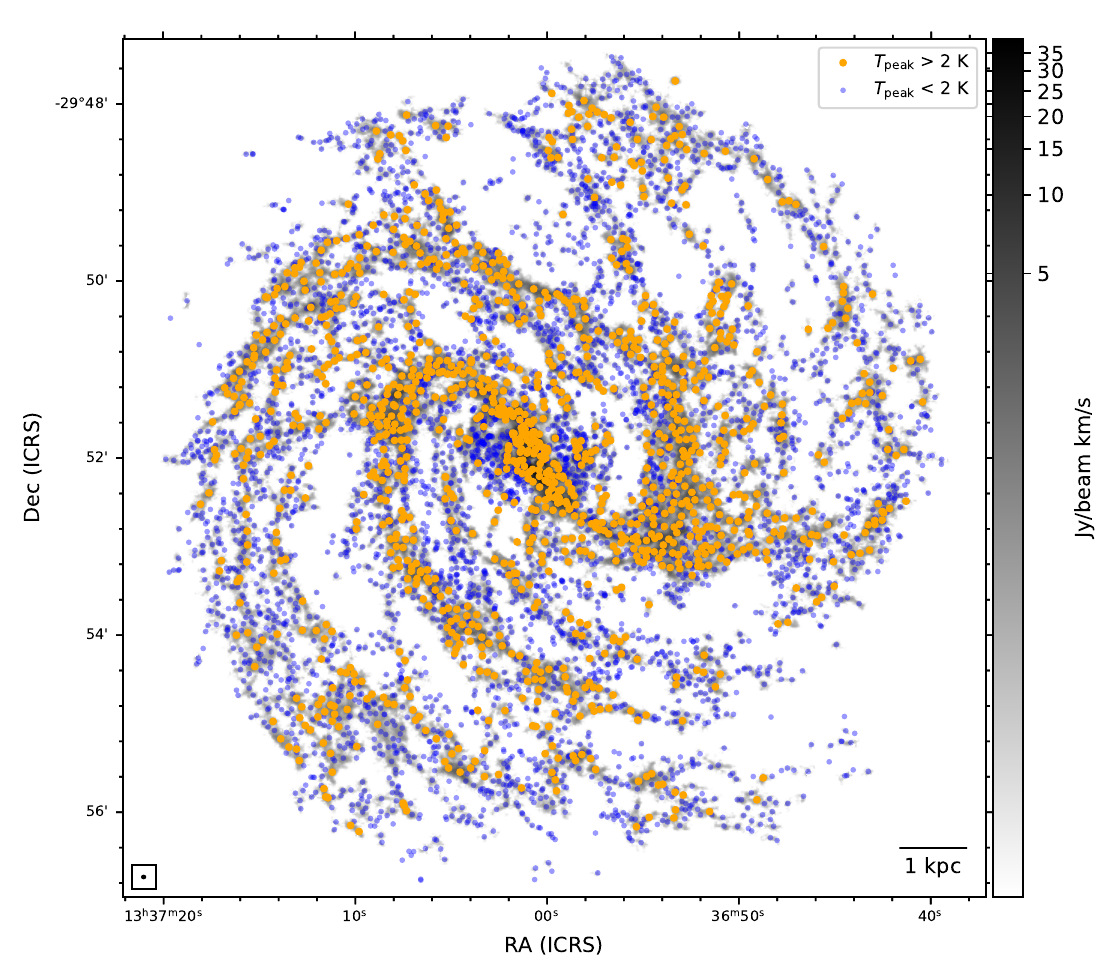}
\caption{
Same as figure \ref{FigM0ByMass}, but clouds are classified with a $\Tpeak$ threshold of 2 K. Clouds with $\Tpeak$ $>2$ K and $<2$ K are indicated with orange and blue markers, respectively.
}
\label{FigM0ByTpeak}
\end{figure}

Figure \ref{FigM0ByTpeak} also shows the distribution of the clouds but categorized with $\Tpeak>2$ K and $<2$ K.
As $\Mcl$ and $\Tpeak$ are correlated, figures \ref{FigM0ByTpeak} and \ref{FigM0ByMass} resemble each other.
High-$\Tpeak$ clouds ($>$2K) are often found within bright galactic structures, as were for the massive clouds.
In the previous subsection, we saw that the properties of the $>2$ K clouds mostly agree with the relations of S87.
The observation that $>2$ K clouds are found within or around larger-scale structures, such as bar, arm, and spurs, suggests that the galactic structures play an important role in assembling the massive, bright clouds that share similar properties as S87's sample.

\subsubsection{Spiral arm locations traced by molecular clouds}

\begin{figure*}[ht!]
\plotone{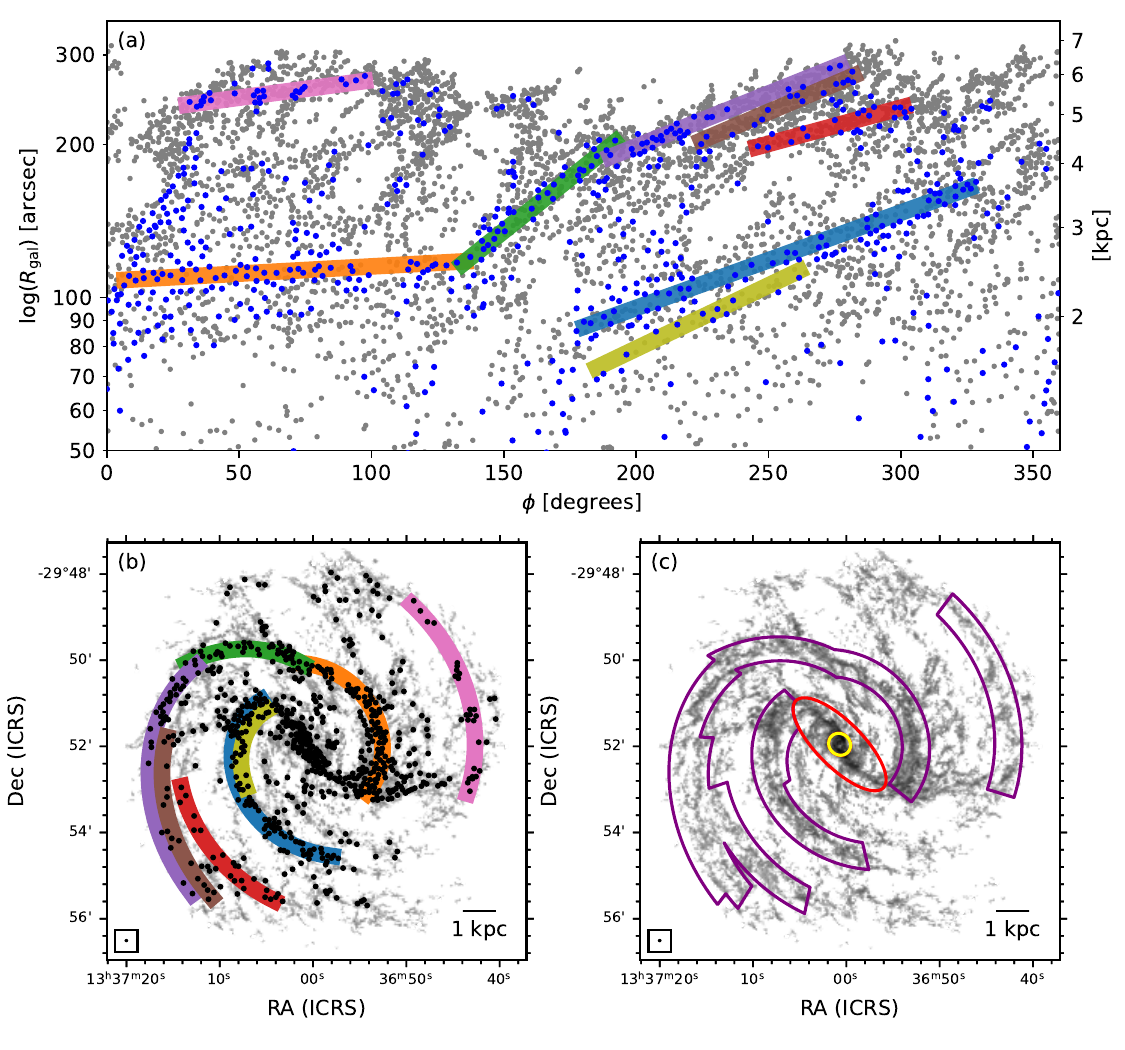}
\caption{(a) Spatial distribution of the clouds on $\log{R}$-$\phi$ plot.
Blue and gray markers indicate the locations of the clouds with masses above and below 10$^6$ $\Msun$.
The thick lines indicate the logarithmic spiral fitting to the selected clouds.
(b) CO map overlaid with the logarithmic spiral arms indicated in (a). Black markers indicate the clouds with masses above 10$^6$ $\Msun$.
(c) Same as (b), but overlaid with the regional mask created here.
The arms, the bar, and the central regions are indicated with purple, red, and yellow lines, respectively.
}
\label{FigArmLocs}
\end{figure*}

Figure \ref{FigArmLocs}(a) plots the clouds on the $\log{\Rgal}$ - $\phi$ plane.
As seen above, the massive clouds ($>$ 10$^6$ $\Msun$) highlight the locations of the spiral arms.
We identify sequences of the massive clouds on the $\log{\Rgal}$ - $\phi$ plane and linearly fit the sequences (i.e., logarithmic spiral arms with constant pitch angles).
The fit results are shown in the plot.
\par

Figure \ref{FigArmLocs}(b) displays the fitted logarithmic spiral structures in a two-dimensional map and the distribution of clouds with masses above 10$^6$ $\Msun$.
The identified structures are tied to the following knowledge.
In the inner part of M83, the two main spiral arms extend from both ends of the bar.
The eastern main arm is displayed with the blue line, and on the trailing side of it, a narrower molecular ridge runs parallel, shown as the yellow line.
This narrow ridge is seen as a dust lane at optical wavelengths \citep[see also][]{RandLordHidgon1999M83} but has weaker CO emission compared to the main arm \citet{Hirota2018M83}.
At first glance, The western spiral arm appears to have a long extension from the western end of the bar toward the eastern side, but it has two changes in pitch angle on the $\log{\Rgal}$ - $\phi$ plane.
Therefore, it is divided into three segments, shown with the orange, green, and purple lines.
In the outermost part, the eastern arm appears as "a bundle of narrower filaments or ripples" on the CO image \citep{Koda2023M83} -- the purple, brown, and red lines correspond to such molecular filaments.
Although the outermost western spiral segment is less defined than the inner main spirals, we tentatively identified a logarithmic spiral shown as the pink line.
Further discussion on the spiral structures in the CO image is also available in \citet{Koda2023M83} (see Fig. 11).
\par

We define a region mask using the fitted arm positions for the analyses in the following sections (figure \ref{FigArmLocs}c).
In this mask, each of the logarithmic spiral arm regions has a width of 40$\arcsec$ ($\sim$880 pc) and the bar region as an ellipse with major and minor axis lengths of 86$\arcsec$ and 32$\arcsec$ respectively, with a position angle of 225$\arcdeg$.
The central region is defined as $\Rgal$ $<16\arcsec$ ($\sim350$ pc).
The mask contains a majority of massive or bright clouds.
Out of 734 clouds with $>10^6$ $\Msun$ in M83, 498 clouds reside in the arm/bar/center mask.
The total mass of $>10^6$ $\Msun$ clouds is $\sim$2.2 $\times$ 10$^9$ $\Msun$ and about 80\% of the mass is included in the mask.

\section{Cloud mass function}
\label{SecMF}

\begin{figure*}[htbp]
\plotone{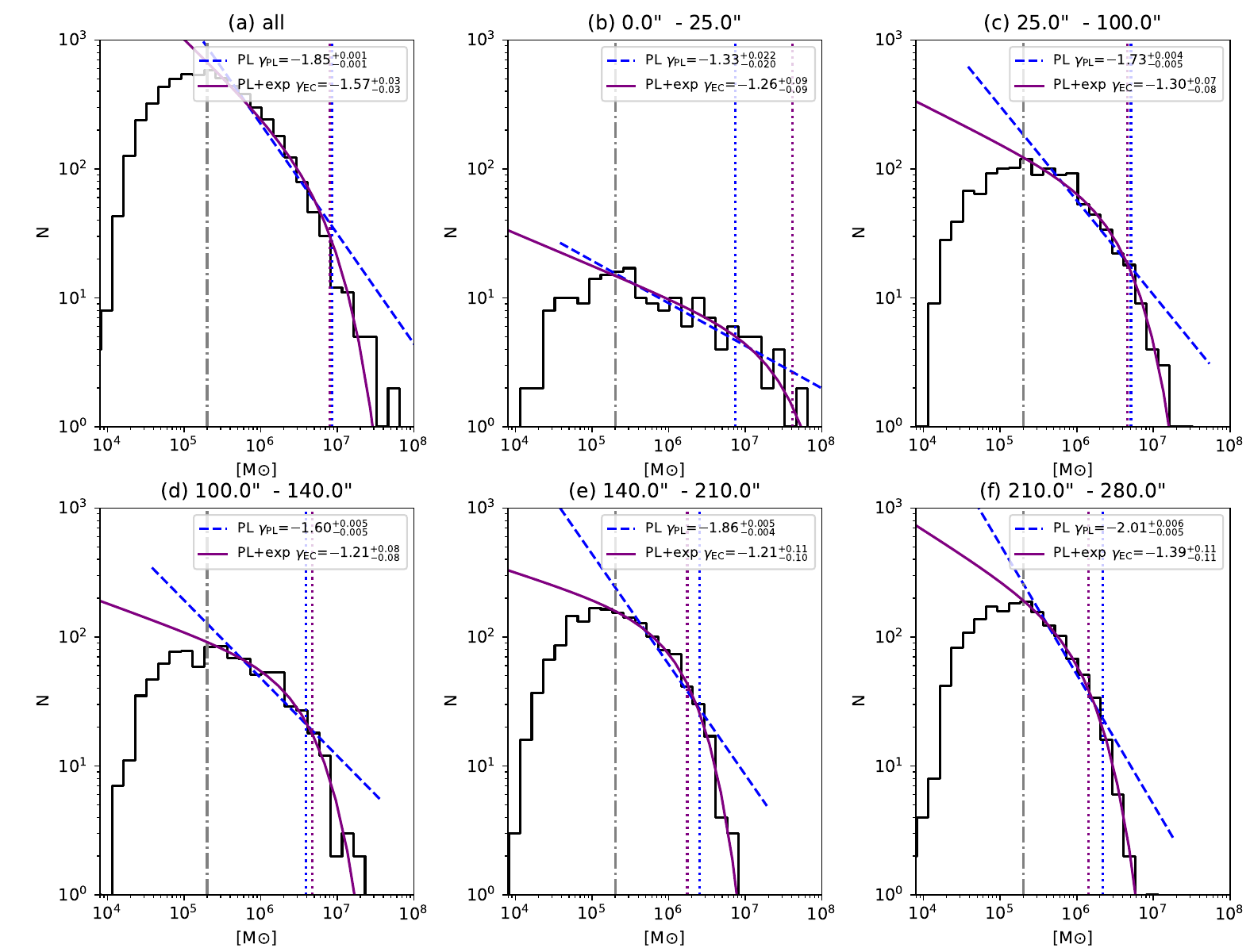}
\caption{(a) Mass distribution for the molecular clouds in M83 displayed in the differential.
The mass bins are configured with a width of 0.15 dex.
The blue dashed line the fit with a conventional form (eq. \ref{EqMFConventional}), and the purple solid line indicates the fit with a Schechter-like form (eq. \ref{EqMFExpCutoff}).
The vertical dash-dotted line indicates the lower-mass limit used for fitting the data (2 $\times$ 10$^5$ $\Msun$).
The blue and purple dotted lines indicate the upper cutoff masses, $\MuPLDeviate$ and $\MuSch$, respectively, determined by fitting the data.
(b--f) Same as (a), but for the radial range of
$R_{\mathrm{gal}}$ =
0$\arcsec$--$25\arcsec$,
$25\arcsec$-- $100\arcsec$,
$100\arcsec$-- $140\arcsec$,
$140\arcsec$-- $210\arcsec$, and
$210\arcsec$-- $280\arcsec$, respectively.
}
\label{FigMFs}
\end{figure*}

The mass distribution of GMCs is characterized by a mass function of a power-law form, i.e., ${dN}/{dM} \propto M^{\gamma}$.
The early GMC surveys in the inner Galactic disk consistently found the index $\gamma$ of around $-1.5$ \citep{Sanders1985, Solomon1987Larson, WilliamsMcKee1997, Rosolowsky2005MassSpectra}.
Subsequent studies in other environments also found $\gamma$ mainly in the range of -1.5 to -2.5 \citep{Heyer2001OuterGalaxy, Colombo2014Env}, but
even steeper relations
(more negative) values of $\gamma$ are also found in some environments \citep[e.g.,][]{Engargiola2003, Fukui2008}.
For a power-law distribution, $\gamma>-2$ means that a small number of massive clouds carry most of the total cloud mass, like the GMC mass distribution in the Galactic disk.
In contrast, $\gamma<-2$ means that most cloud mass resides in the smaller mass end.
\par

At the higher end of the mass distribution, the deficit of massive clouds is also often seen: in the Galactic disk, clouds more massive than a few 10$^6$ $\Msun$ are frequently under-populated with respect to the power-law relation \citep{WilliamsMcKee1997}.
Some previous studies reported $\gamma$ steeper (smaller) than -2, but this might be a consequence of fitting cloud mass distributions above such an upper limit mass \citep{Rosolowsky2005MassSpectra}.
\par

\subsection{Functional forms}
We fit the cloud mass distribution in two forms.
The first form is a power-law, restricted by the upper mass limit $\MuPL$ as,
\begin{equation}
\frac{dN}{dM} \propto \left(\frac{\Mcl}{\MuPL}\right)^{\gammaPL}, \ \ \Mcl < \MuPL.
\label{EqMFPowerLawDiff}
\end{equation}
The form is conventionally applied to the cloud mass distributions
\citep[e.g.,][]{WilliamsMcKee1997, Rosolowsky2005MassSpectra, Colombo2014Env}.
It can be written in the integral form as follows
\begin{equation}
N(>M) = \NuPL
\left [ \left(\frac{\Mcl}{\MuPL}\right)^{\gammaPL + 1} - 1  \right]
\label{EqMFConventional},
\end{equation}
where $\NuPL$ is the number of clouds more massive than $2^{1 / (\gammaPL + 1)} \MuPL$, which we denote as $\MuPLDeviate$.
We call $\MuPLDeviate$ the cutoff mass for the restricted power-law form because the deviation from a power-law becomes obvious above $\MuPLDeviate$.
\par

The second form combines the power-law distribution with an exponential cutoff, which is often called a Schechter-like function:
\begin{equation}
\frac{dN}{dM} \propto \left(\frac{M}{\MuSch}\right)^{\gammaSch} \exp{\left(-\frac{M}{\MuSch}\right)},
\label{EqMFExpCutoff}
\end{equation}
where $\gammaSch$ is the index of a power-law distribution, and $\MuSch$ is the cutoff mass limit above which the distribution is suppressed with respect to the power-law.
Fitting the cloud mass distribution with this Schechter-like form would help test the connection between the mass spectra of GMCs and stellar clusters as the cluster mass distributions are often characterized with this form \citep[e.g.,][]{Kruijssen2014Cluster, Adamo2015}.
However, the application of this form to molecular cloud distributions is limited in recent studies \citep{Freeman2017, Rosolowsky2021PhangsGMCs}.

\subsection{Cloud mass spectra}
Figure \ref{FigMFs} shows the cloud mass distributions for (a) all GMCs and (b-f) those in five radial bins. The latter is defined with the boundaries at $25\arcsec$ ($\sim0.55$ kpc), $100\arcsec$ ($\sim2.2$ kpc), $140\arcsec$ ($\sim3.1$ kpc), and $210\arcsec$ ($\sim4.6$ kpc), respectively, which are shown in figure \ref{FigRMSMap}(b).
They are to examine the radial and environmental variation of the cloud mass distribution:
The first three bins mainly cover the central region, bar, and two main spiral arms, respectively, and the last two bins divide the outer galactic disk into two.
\par

All the histograms show declines below around $2\times10^{5}$ $\Msun$, suggesting that cloud sampling is incomplete below that mass.
As discussed in \S\ref{SubsecCOLumFraction}, our mass sensitivity in cloud identification is $\sim3\times10^{4}$ $\Msun$ for a cloud in isolation.
The difference would be attributed to the blending of cloud emissions.
For a cloud to be detected above the high background surface density $\Sigma_\mathrm{bg}$ (e.g., in spiral arms), its mass has to exceed  $\SigmaBG$ A$_\mathrm{beam}$ with A$_\mathrm{beam}$ as beam area.
For example, in most spiral arms, the typical $\SigmaBG$ is 100$\MsunPerSqPC$.
Therefore, the minimum detectable mass in the arms cannot go below $\sim2 \times 10^{5}$ $\Msun$ with the current resolution of $\sim46$ pc.
Hereafter, we regard $2\times10^{5}$ $\Msun$ as the effective completeness limit.
This is a conservative limit in the disk region because the limit should be smaller in the inter-arm regions. However, the mass limit is higher in the central region because $\Sigma_{\mathrm{bg}}$ reaches nearly 1000 $\MsunPerSqPC$ or higher there \citep{Koda2023M83}.
\par

We fit only the clouds with $>$2$\times$ 10$^5$ $\Msun$ (above the completeness limit) with the two forms.
We fitted the power-law form using the cumulative mass distribution function and least-square minimization against the integral form (eq. \ref{EqMFConventional}).
For the Schechter-like form (eq. \ref{EqMFExpCutoff}), we sample the likelihood using the {\it emcee} software library, which implements a Markov chain Monte Carlo method \citep{ForemanMackey2013emcee}.
The prior distribution is configured such that the probability distribution of $\gammaSch$ is uniform between -5 and 0 and that $\MuSch$ is also uniform between 10$^5$ $\Msun$ and 10$^9$ $\Msun$ in the logarithmic space.
Table \ref{tableMFParams} reports the fitted parameters and $p$-values calculated with the two-sample K-S test for both forms with the null hypothesis that the fitted and observed cumulative mass distributions are drawn from the same distribution.
\par

Figure \ref{FigMFs}(a) shows the mass distribution for all the clouds.
The power-law fit gives $\gammaPL \sim -1.85$, which is comparable to the indices found in the inner Galactic disk \citep[e.g., -1.6 to -1.8,][]{WilliamsMcKee1997}.
The Schechter-like form yields an index $\gammaSch$ of $\sim-1.57$, which is shallower than $\gammaPL$ but is still comparable to the Galactic values.
Both forms suggest the existence of upper cutoff masses with similar values, namely $\MuPLDeviate$ of $\sim$8.6$\times$10$^6$ $\Msun$ and $\MuSch$ of $\sim$8.0$\times$10$^6$ $\Msun$, respectively.
\par

Figure \ref{FigMFs}(b) is for the innermost radial bin of $\Rgal < 25\arcsec$ ($\sim550$ pc).
The fitted indices are shallow: $\gammaPL\sim$-1.33 and $\gammaSch\sim$-1.26.
Due to the elevated background $\SigmaBG$, the shallow index is likely an artifact caused by the heavy blending of cloud emissions.
\par

Figure \ref{FigMFs}(c--f) shows the cloud mass distributions beyond $\Rgal\sim$550 pc.
Within the four radial bins, $\gammaPL$ varies within the range of $\sim$-1.7 to $\sim$-2.0 with a tendency of steeper index toward the outer radius.
On the contrary, the Schechter-like form results in more minor variations in $\gammaSch$, ranging from -1.2 to -1.4, with no notable radial trend.
In terms of the upper cutoff mass, both forms are again consistent.
As we will comment later in \S\ref{SubsecMFFormComparison}, the Schechter-like form tends to characterize the variation in mass spectra by a change in $\MuSch$ rather than in $\gammaSch$.
Both $\MuPL$ and $\MuSch$ exhibit a factor of three declines with increasing galactocentric radii.
\par

\begin{figure}[htbp]
\plotone{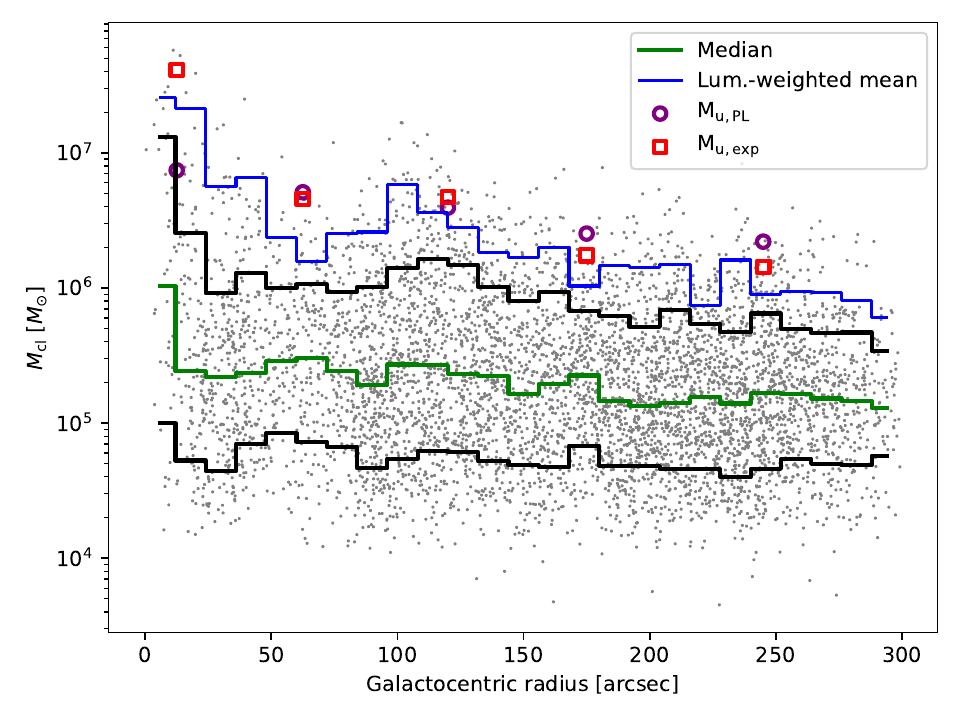}
\caption{
Cloud masses as a function of galactocentric radius, indicated with the gray dots.
The green line indicates the median values at the radial bins with 12$\arcsec$ averaging widths.
The blue line is the luminosity-weighted mean of the cloud masses calculated at the same bins.
Black lines indicate the 16 and 84 percentile values at the same bins.
Purple circle and red square markers indicate the estimated upper limits on cloud mass from the fittings of mass distributions, i.e., 2$^{1/\left(\gammaPL + 1\right)}$ $\MuPL$ and $\MuSch$, respectively, determined for five radial ranges listed in table \ref{tableMFParams}.
}
\label{FigMFParamRadial}
\end{figure}

Figure \ref{FigMFParamRadial} compares the upper-cutoff mass determined from the fittings, $\MuPLDeviate$ and $\MuSch$, with the radial distribution of the individual cloud masses.
This plot also shows the luminosity-weighted mean cloud mass in radial bins with a width of $12\arcsec$.
This luminosity-weighted mean mass declines radially, except for a slight secondary peak around $\Rgal = 100\arcsec$ where the main two spiral arms reside.
The radially declining trend of the luminosity-weighted mean generally agrees with the variation of $\MuPLDeviate$ and $\MuSch$ derived from the fittings.
The alignment would be reasonable as the slopes of the fitted mass functions in figure \ref{FigMFs} are mostly shallower than -2, and thus, most of the CO luminosity (and mass) should be concentrated around the upper cutoff mass.
\par

\subsection{Comparison of the fitted results between the two forms}
\label{SubsecMFFormComparison}
Because of the discrepancies between $\gammaPL$ and $\gammaSch$, one would wonder which of the two forms, namely the restricted power-law form or Schechter-like form, provides a better characterization of the cloud mass distributions.
The $p$-values of the K-S test in table \ref{tableMFParams} are inconclusive.
By taking the significance level of $p$-value as 0.05, the K-S test accepts almost all the cases except the restricted power-law form applied to all samples.
The values of $\gammaSch$ are shallow in most radial ranges with a slight variation (-1.2 to -1.4).
The shallow and limited variation in $\gammaSch$ might be due to the limited dynamic range in fitting -- there is only about an order of magnitude variation between $\MuSch$ and the adopted completeness limit of 2$\times$10$^5$ $\Msun$.
The exponential term induces a decline even below $\MuSch$ to some extent, further limiting the range available to fit as a power-law.
The determination of $\gammaSch$ might become reliable only if higher resolution data becomes available to overcome the crowding effect to lower the mass completeness limit \citep[see also][]{Rosolowsky2021PhangsGMCs}.
As the K-S test gives no clear preference and as we suspect the fitted values $\gammaSch$ might be biased, we consider that $\gammaPL$ to be preferable to $\gammaSch$, at least with the available data.

\subsection{Possible origins of the radial variation of mass spectra}
The power-law fitting shows a radially steepening of $\gammaPL$ from -1.3 to -2.0 and a decrease of $\MuPLDeviate$, indicating a relative deficit of massive clouds toward the outer disk.
The trends agree with other studies in M83 with shallower sensitivities \citep{Freeman2017, Hirota2018M83} and also in different environments \citep[e.g., ][]{Colombo2014Env}.
Several mechanisms might explain the radial trends in $\gammaPL$ and $\MuPLDeviate$.
First, given that M83 has many bubble structures in infrared, HI, and CO, which likely influence and drive the evolution of molecular clouds.
\citet{Inutsuka2015} and \citet{Kobayashi2017MF} provide semi-analytical calculations to describe the time evolution of molecular cloud mass functions regulated by multiple episodes of bubble expansions, which drive molecular cloud formation via shock compression of atomic gas. The idea would explain the steeper $\gammaPL$ in the outer disk as the slower cloud formation because the supernovae, which drive bubble expansions, are less frequent there. \citet{Kobayashi2017MF} also suggested the formation of massive clouds comparable to $\MuPLDeviate$ requires about 100 Myr by solving the time evolution, which agrees with the time scale predicted from the CO morphology of M83 \citep{Koda2023M83}.
Second, the under-population of massive clouds above $\MuPLDeviate$ might also be due to a lower molecular fraction in massive clouds --
\citet{Elmegreen1989MassMetalDependence} provides that self-shielding decreases approximately as $\propto$ $M^{-0.5}$ for virialized clouds, and therefore, CO line observations might trace only the core regions in most massive clouds.
As there is a radial decrease of per-cloud molecular fraction in the MW \citep{Elmegreen1987HISuperClouds}, it would be intriguing to compare molecular clouds with massive HI clouds in M83.

\begin{deluxetable*}{lccccccc}
\tablecaption{Mass function parameters}
\tablewidth{0pt}
\tablehead{
\colhead{Range} & \colhead{$\gamma_{\mathrm{PL}}$} & \colhead{$N_{\mathrm{u,PL}}$} & \colhead{$M_\mathrm{u,PL}$} & \colhead{$p$-value} & \colhead{$\gamma_{\mathrm{exp}}$} & \colhead{$M_\mathrm{u,exp}$} & \colhead{$p$-value} \\
\colhead{} & \colhead{} & \colhead{} & \colhead{($10^6 \Msun$)} & \colhead{} & \colhead{} & \colhead{($10^6 \Msun$)} & \colhead{}
}
\decimalcolnumbers
\startdata
0.0 kpc - 0.5 kpc   & $-1.33^{+0.02}_{-0.02}$ & $6.9^{+0.6}_{-0.7}$ & $7.41^{+0.14}_{-0.37}$ & 1.00 & $-1.26^{+0.09}_{-0.09}$ & $46.29^{+38.51}_{-15.64}$ & 1.00 \\
0.5 kpc - 2.2 kpc   & $-1.73^{+0.00}_{-0.00}$ & $25.3^{+0.5}_{-0.5}$ & $5.10^{+-0.06}_{--0.07}$ & 0.10 & $-1.30^{+0.07}_{-0.08}$ & $4.52^{+1.00}_{-0.73}$ & 0.44 \\
2.2 kpc - 3.1 kpc   & $-1.60^{+0.01}_{-0.01}$ & $30.8^{+0.6}_{-0.5}$ & $3.93^{+-0.02}_{--0.02}$ & 0.46 & $-1.21^{+0.08}_{-0.08}$ & $4.69^{+1.02}_{-0.81}$ & 1.00 \\
3.1 kpc - 4.6 kpc   & $-1.86^{+0.00}_{-0.01}$ & $40.4^{+0.7}_{-0.7}$ & $2.53^{+-0.02}_{--0.02}$ & 0.33 & $-1.21^{+0.11}_{-0.10}$ & $1.79^{+0.34}_{-0.26}$ & 0.99 \\
4.6 kpc - 6.1 kpc   & $-2.01^{+0.01}_{-0.01}$ & $33.8^{+0.8}_{-0.8}$ & $2.20^{+-0.03}_{--0.03}$ & 0.27 & $-1.39^{+0.11}_{-0.11}$ & $1.45^{+0.31}_{-0.23}$ & 1.00 \\
\hline
all                 & $-1.85^{+0.00}_{-0.00}$ & $52.3^{+0.4}_{-0.4}$ & $8.58^{+-0.06}_{--0.06}$ & 0.00 & $-1.57^{+0.03}_{-0.03}$ & $7.93^{+1.15}_{-0.85}$ & 0.21 \\
\enddata
\tablecomments{Mass function parameters. (1) Radial range. (2 to 4) Mass function parameters determined by the fitting with restricted power-law, namely power-law index $\gammaPL$, number of clouds at high mass end $\NuPL$, and upper-cutoff mass $\MuPLDeviate$. (5) K-S test $p$-value the restricted power-law fitting.  (6 and 7) Parameters determined by the fitting with the Schechter-like form, namely index $\gammaSch$ and upper-cutoff mass $\MuSch$. (8) K-S test $p$-value the Schechter-like form fitting.}
\label{tableMFParams}
\end{deluxetable*}

\section{Cloud-to-cloud velocity dispersion}
\label{SecVelDisp}
Vertical cloud-to-cloud velocity dispersion of a gaseous disk is a parameter that reflects the stability of the disk. Despite its importance, the measurements of the cloud-to-cloud velocity dispersion of molecular gas are still limited.
In the MW, \citet{Stark1984VelDispersion} found that the cloud-to-cloud velocity dispersion within the Galactic plane weakly depends on cloud mass, with 6.6 $\kmPerS$ and 9.0 $\kmPerS$ for cloud masses of 10$^{4-5.5}$ $\Msun$ and 10$^{2-4}$ $\Msun$, respectively. However, the estimation from the MW data involves an assumption of the isotropy of velocity dispersion.
\par

Extragalactic observations of nearly face-on galaxies permit estimation of the vertical cloud-to-cloud velocity dispersion.
Previous measurements were performed at sub-kpc scale resolutions, which blend several GMCs within resolution elements; they found dispersions of 6 to 9 $\kmPerS$ when an individual spectrum of each pointing is analyzed \citep{CombesBecquaert1997, Wilson2011JCMT, Mogotsi2016} and about 12 to 14 $\kmPerS$ when the spectra at multiple pointings are stacked \citep{CalduPrimo2013, CalduPrimo2016Andromeda}.
With the new cloud catalog and low inclination angle of M83, we can make a more direct estimation of the vertical cloud-to-cloud velocity dispersion.

\subsection{Method to estimate vertical velocity dispersion of clouds}
We adopt a method similar to the one used by \cite{Hernandez2019} to estimate the vertical velocity dispersion of stellar clusters.
The line-of-sight velocity $\vLOS$ of a cloud rotating in the disk
is written as
\begin{equation}
\vLOS = \vSys + (\vRot + \vPhi) \cos{\phi} \sin{i}
+ \vR             \sin{\phi} \sin{i}
+ \vZ                        \cos{i},
\end{equation}
where $\vSys$, $\vRot$, $\vPhi$, $\vR$, $\vZ$, $\phi$, and $i$ are the systemic recession velocity, the circular rotational velocity expected from the galactic rotation curve, in-plane azimuthal and radial non-circular velocities that are mostly governed by streaming motions, vertical motion velocity, the angle from the major axis, and the inclination angle.
We subtract the contributions of assumed $\vSys$ and $\vRot$ from $\vLOS$ for each cloud.
Then, the residual velocity $\vResid$ is left with the contributions from $\vZ$ and streaming motions as
\begin{equation}
\vResid = \vZ \cos{i} + \vPhi \cos{\phi} \sin{i} + \vR \sin{\phi} \sin{i}.
\label{EqVres}
\end{equation}
As long as the contributions from the in-plane non-circular motions, i.e., the last two terms in this equation, are negligible, we can estimate the cloud-to-cloud velocity dispersion, $\mathrm{<}\vZ\mathrm{>}$, as the dispersion of $\vResid/\cos{i}$ among a sample of clouds.
\par

We adopt $\vSys$ of 511 $\kmPerS$ and $i$ of 24$\arcdeg$ \citep{Koda2023M83}.
For $\vRot$, we use the rotation curve of \cite{Hirota2014}, which is obtained by subtracting the rotational velocities estimated from the stellar mass distribution using the 2MASS Large Galaxy Atlas image \citep{Jarrett20032MASS} with an assumption of the maximum disk contribution.

\subsection{Vertical cloud-to-cloud velocity dispersion}
Figure \ref{FigClClVelcityDispersion}(a) shows $\vResid$ as a function of galactocentric radius $\Rgal$ for the M83 clouds. The clouds with masses below and above 5 $\times$ 10$^5$ $\Msun$ are plotted separately with gray and cyan markers, respectively.
The radial bins are configured with a width and increment of 24$\arcsec$ and 12$\arcsec$, up to 288$\arcsec$. The 16th, 50th, and 84th percentiles of $\vResid$ are plotted for each radial bin.
\par

The absolute $|\vResid|$ values are elevated within the inner 100$\arcsec$ ($\sim$2.2 kpc) of M83, where the stellar bar resides.
Although $|\vResid|$ includes the contributions from the in-plane non-circular motions and vertical velocity dispersion (eq. \ref{EqVres}), the $\vPhi$ and $\vR$ terms appear to be negligible except in the bar, which induces large non-circular streaming motions.
Outside the bar, the 16th and 84th percentiles of $\vResid$ are within the range of $|\vResid|$ $<$ 10 $\kmPerS$, suggesting only a modest impact of the arm-induced streaming motions.
This suggestion agrees with \citet{TilanusAllen1993}, who found a modest streaming motion of about 15 $\kmPerS$ in the plane of M83.
\par

Figure \ref{FigClClVelcityDispersion}(b) shows the radial distribution of estimated cloud-to-cloud velocity dispersions, $\mathrm{<}\vZ\mathrm{>}$ for clouds above (cyan) and below (black) 5$\times$10$^5$ $\Msun$.
Except for the inner bar region ($\Rgal$ $<$ 100$\arcsec$; $\sim2.2$ kpc) where the impact of streaming motion is not negligible, we consider the data points to be the fair representatives of $\mathrm{<}\vZ\mathrm{>}$ based on the small variation in the scatters of $\vResid$ at $\Rgal$ $>$ 100$\arcsec$.
At $\Rgal$ $>$ 100$^{''}$,  $\mathrm{<}\vZ\mathrm{>}$ is found to be almost constant around 8.3 $\kmPerS$ for $>$5$\times$10$^5$ $\Msun$ clouds and 11 $\kmPerS$ for $<5$ $\times$ 10$^5$ $\Msun$ clouds.
\par

We performed Levene's test at each radial bin to see the statistical significances of the differences in $\vResid$ between $>5\times10^5$ $\Msun$ and $<5\times10^5$ $\Msun$ clouds (figure \ref{FigClClVelcityDispersion}c).
The test's null hypothesis is that the underlying $\vResid$ distribution is the same in both groups, i.e., the vertical velocity dispersion is the same.
We discard the null hypothesis if the $p$-value of the test is below 0.05.
The resultant $p$-values are far below the adopted significance level of 0.05 in most of the bins beyond $\Rgal$ $>$ 100$^{''}$, suggesting that the mass dependence of $\mathrm{<}\vZ\mathrm{>}$ is significant there.

\subsubsection{Assessment of the impact of streaming motions on the estimated vertical cloud-to-cloud velocity dispersion}
\begin{figure*}[htbp]
\plotone{
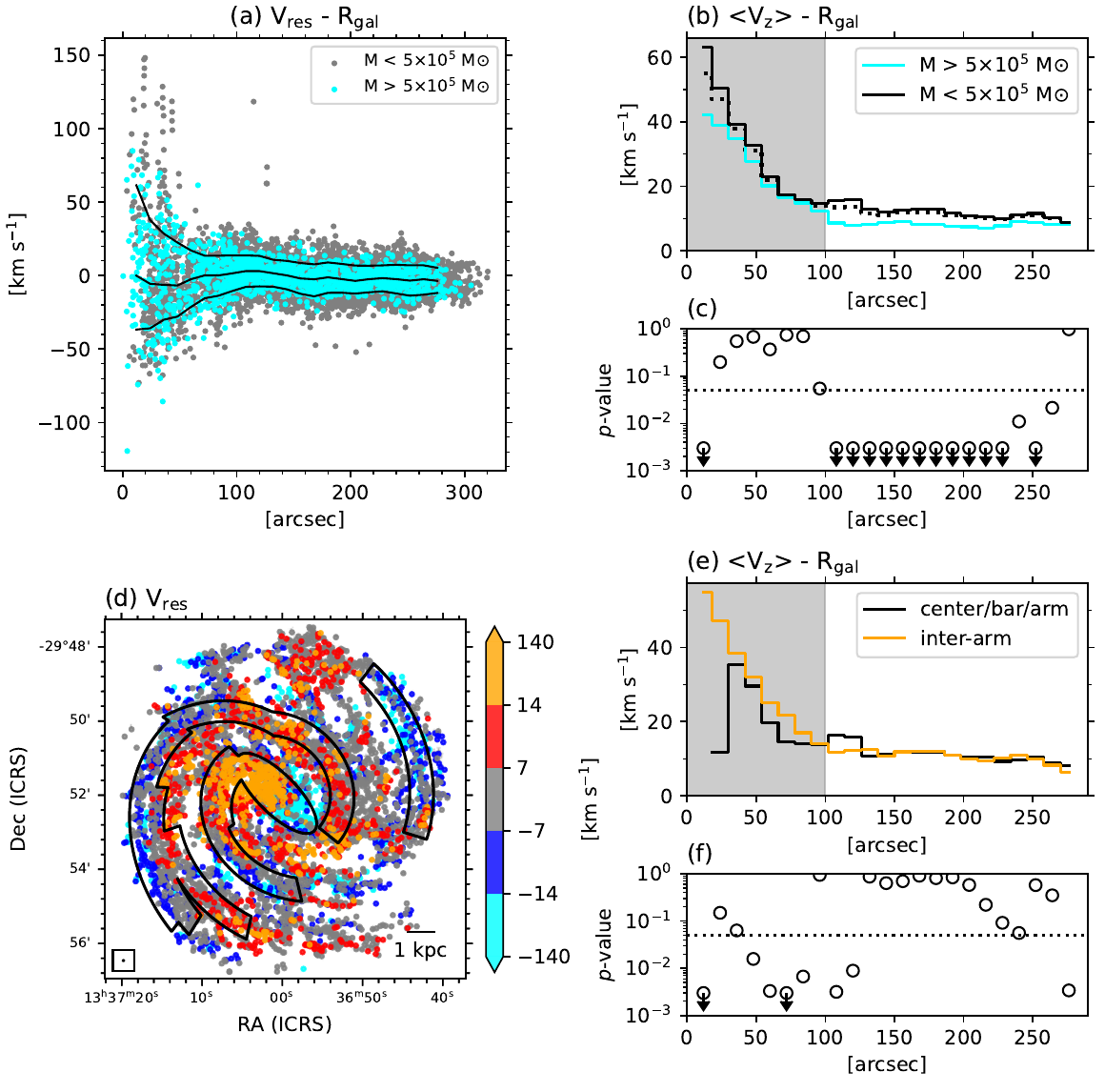
}
\caption{
(a) Cloud's residual velocity $\vResid$ as a function of galactocentric radius $\Rgal$. The gray and cyan markers indicate the clouds for mass below and above 5$\times$10$^5$ $\Msun$. The black lines indicate the 16th, 50th, and 84th percentiles of $\vResid$ derived within binned $\Rgal$ of 24$\arcsec$ width, spaced by every 12$\arcsec$.
(b) Estimated vertical velocity dispersion $\mathrm{<}\vZ\mathrm{>}$ derived within the same binned $\Rgal$ as (a).
Black and cyan solid lines indicate $\mathrm{<}\vZ\mathrm{>}$ for the clouds below and above 5$\times$10$^5$ $\Msun$.
The dashed line indicates $\mathrm{<}\vZ\mathrm{>}$ for the whole cloud masses.
As $\vResid$ is severely affected by non-circular motions in the inner disk fo M83, $\Rgal$ $<$ 100$\arcsec$ is shaded.
(c) $p$-values derived for the binned $\Rgal$ calculated with Levene's test, which indicate the statistical significance of the difference of $\mathrm{<}\vZ\mathrm{>}$ between clouds with mass below and above 5$\times$10$^5$ $\Msun$.
The $p$-values below 3$\times$10$^{-3}$ are indicated as upper limits with the arrow markers.
The dotted horizon line indicates the adopted threshold of 0.05.
(d) Spatial distribution of $\vResid$, compared with the regional mask.
(e) Same as (b), but comparing the difference between center/bar/arm and inter-arm regions shown with black and orange lines.
(f) Same as (c), but comparing the difference between center/bar/arm and inter-arm regions.
}
\label{FigClClVelcityDispersion}
\end{figure*}

A caveat of the $\mathrm{<}\vZ\mathrm{>}$ estimation made here is that the impact of in-plane non-circular motion is just neglected.
As massive clouds are concentrated toward spiral structures (\S\ref{SubsecSpatialDistribution}), one might suspect they are more susceptible to streaming motions.
To eliminate this possibility, we show the spatial distribution of $\vResid$ for individual clouds in figure \ref{FigClClVelcityDispersion}(d). The marker colors show the elevation of $|\vResid|$ around spiral arms.
The locations of the galactic structures identified in \S\ref{SubsecSpatialDistribution} are also shown for comparison.
\par

Streaming motions around the bar and spiral arms are coherent and should show ordered patterns in this figure. While such coherent motions appear limited, there are still some signatures. For example, across the eastern, outermost arm-like feature from the trailing to leading side, $\vResid$ changes from positive (red markers) to negative (blue markers).
\par

Figure \ref{FigClClVelcityDispersion}(e) is similar to figure \ref{FigClClVelcityDispersion}(b), but plots the radial variation of $\mathrm{<}\vZ\mathrm{>}$ for clouds in the center/bar/arm region and the inter-arm regions using the spatial mask shown in figure \ref{FigClClVelcityDispersion}(d).
Except for the innermost two radial bins, the dispersions within the bar/spiral arms and inter-arm regions are similar to each other. The $p$-values obtained with Levene's test (figure \ref{FigClClVelcityDispersion}f) also support their similarity.
We here conclude that the arm-streaming motions have less impact on $\vResid$ than the mass-dependence of the vertical cloud-to-cloud velocity dispersion at $\Rgal$ $>2.2$ kpc.

\subsubsection{Comparision with the previous studies}
We identify a weak mass dependence of $\mathrm{<}\vZ\mathrm{>}$, with $\sim$8.3 $\kmPerS$ and $\sim$11 $\kmPerS$ for clouds above and below 5$\times$10$^5$ $\Msun$ at $\Rgal$ beyond 2.2 kpc.
The former is close to the values found in other extragalactic studies at kpc-scale resolutions \citep{CombesBecquaert1997, Mogotsi2016}.
The consistency with our study at $\sim$46 pc resolution may be because massive clouds occupy the dominant fraction of CO luminosity and also because our radial bin has a wide width of 24$\arcsec$ ($\sim520$ pc).
\par

The smaller vertical velocity dispersion in more massive clouds ($>$ 5 $\times$ 10$^5$ $\Msun$) is similar to the tendency for clouds in the Solar neighborhood \citep{Stark1984VelDispersion}.
Massive clouds are scarce in the inter-arm regions, and thus, the reduced $\vResid$ dispersion in massive clouds suggests that clouds in spiral arms are likely more settled near the galactic plane than inter-arm regions.
Such a trend is also observed in the Galactic plane \citep{StarkLee2005} as the reduced scale height of clouds near the spiral arm.

\section{Discussion}
\label{SecDiscussion}

We identified 5724 molecular clouds from the $^{12}$CO (1--0) data of M83 with a spatial resolution of $\sim$46 pc, and found a median $\alphaVir$ of 2.7 (\S\ref{SubsecCloudPropDistribution}) and a median $\SigmaGMC$ of $\sim$20 $\MsunPerSqPC$ assuming a Galactic CO-to-H2 conversion factor of 2.0 $\times$ 10$^{20}$ cm$^{-2}$ (K $\kmPerS$)$^{-1}$.
Some of the recent cloud studies suggest that most molecular clouds are at most loosely bound by self-gravity, with $\alphaVir$ a factor of a few higher than 1 \citep{Heyer2001OuterGalaxy, Dobbs2011NotBound, MivilleDeschenes2017MWGMC, Sun2020, Evans2021BoundCloud} and also have a wide range of variation in $\SigmaGMC$ \citep{Heyer2001OuterGalaxy, MivilleDeschenes2017MWGMC, Lada2020MassSizeRelation, Sun2020}.
At first glance, the median $\alphaVir$ and $\SigmaGMC$ agree with these recent results and suggestions.
However, we take a step back to consider the possible impact of insufficient spatial resolution on the measured cloud properties.
This assessment is needed because while the median of the sampled cloud masses is $\sim$1.9 $\times$ 10$^5$ $\Msun$, the utilized $\sim$46 pc spatial resolution is larger than the sizes of the Galactic GMCs at a similar mass \citep[e.g., $\sim$20 pc, ][]{Sanders1985}.
\par

We organize the discussion as follows.
In \S\ref{SubsecDiscussionBeamDilutionFormulate}, we discuss how the $\sim46$ pc beam affects cloud properties, in particular $\alphaVir$ and $\SigmaGMC$, via a variation in the area-filling factor of clouds.
Next, in \S\ref{SubsecDiscussionFillingFactorCorrectedProps}, we argue the possible range of the area-filling factor and intrinsic cloud properties considering the sensitivity of the observational data.
In \S\ref{SubsecDiscussionImplication}, we discuss the implications of this section.

\subsection{Impact of sampling clouds with a relatively large beam}
\label{SubsecDiscussionBeamDilutionFormulate}

We consider the relations between intrinsic and observed parameters of the clouds affected by the effects of a limited resolution.
When clouds are sampled with marginal resolution and SNR, measured cloud sizes could be overestimated even when the beam deconvolution (\S\ref{SubsecBasicProps}) is made \cite{Rosolowsky2006CPROPS}.
This over-estimation would arise as a consequence of the SNR thresholding imposed during the cloud identification process, as only positive value pixels are included in cloud masks, and thus, the presence of noise tends to bias the measured cloud sizes toward larger values.
For a cloud located in a crowded region, blending of emissions from the surrounding material also leads to the overestimation of cloud sizes.
The observed distribution of $\Reff$ appears to imply the sign of the over-estimation of radii, as over 80 percent of the clouds have $\Reff$ larger than 40 pc \S\ref{SubsubsecVDispAndRadius}, which is approximately the resolution limit
\footnote{As $\Reff$ relates to root-mean-square cloud size $\sigma_r$ as $\Reff=1.91\sigma_r$, the beam size $\theta_{\mathrm{b}}$ of 46 pc corresponds to $\sim$37 pc (=1.91 $\times$ $\theta_{\mathrm{b}}$ / $\sqrt{8\log{2}}$)}.
The velocity dispersion $\sigmav$ could be similarly affected by the resolution bias, but the observed values of $\sigmav$ are well above the resolution limit \S\ref{SubsubsecVDispAndRadius}, suggesting $\sigmav$ is less affected.
Therefore, we consider that the resolution bias mostly affects $\Reff$ for the M83 catalog.
This is also supported by the bootstrap error estimation made in \S\ref{SubsecBasicProps}, which indicates significantly larger fractional uncertainties on $\Reff$ than $\sigmav$.
\par
From the nature of the data considered here, we assume that the beam dilution affects the observed cloud radius $\Reff$ such that it overestimates the intrinsic radius $\RCh$, but does not affect the velocity dispersion $\sigmav$.
We also assume that cloud mass $\Mcl$ is unaffected.
When observing a cloud with a larger beam, the cloud emission is diluted, and thus, the observed brightness temperature is lowered, but the luminosity, and hence cloud mass $\Mcl$, should be preserved to the first order.
\par
Under the assumed simplified configuration, the relation between the observed and intrinsic radii is
\begin{equation}
\Reff = \fFill^{-1/2} \RCh,
\label{EqRFilling}
\end{equation}
where $\fFill$ is the area-filling factor, the intrinsic-to-observed area ratio.
By definition, $\fFill$ is $\le$ 1.
\par
The cloud surface mass density $\SigmaGMC$ is affected by the beam dilution as it is dependent on $\Reff$.
The relation between the intrinsic $\SigmaCh$ and the observed $\SigmaGMC$ is
\begin{equation}
\SigmaGMC = \fFill \SigmaCh.
\label{EqSigmaFilling}
\end{equation}
\par

The virial parameter $\alphaVir$ depends on $\Reff$ as $\alphaVir$ $\propto$ $\Reff\sigmav^2$.
Denoting the intrinsic value of $\alphaVir$ as $\alphaVirCh$, from eq. (\ref{EqMVirDef}), eq. (\ref{EqAlphaVirDef}), and eq. (\ref{EqRFilling}), the relation between $\alphaVir$ and $\alphaVirCh$ is
\begin{equation}
\alphaVir = \alphaVirCh \fFill^{-1/2}.
\label{EqAlphaVirWithFillingFactor}
\end{equation}
As $\fFill$ is $\le$ 1 by definition, $\alphaVir$ is elevated as $\fFill^{-1/2}$.
For example, suppose a bound cloud with $\RCh=10$~pc and $\alphaVirCh$ = 1. With a spatial resolution of $\sim46$ pc, $\fFill$ is $\sim$0.1 and thus, the observed $\alphaVir$ is elevated to $\sim$3.1, which would apparently indicate that the cloud is unbound even though its intrinsic $\alphaVirCh$ is 1.
\par

\begin{figure}[htbp]
\plotone{
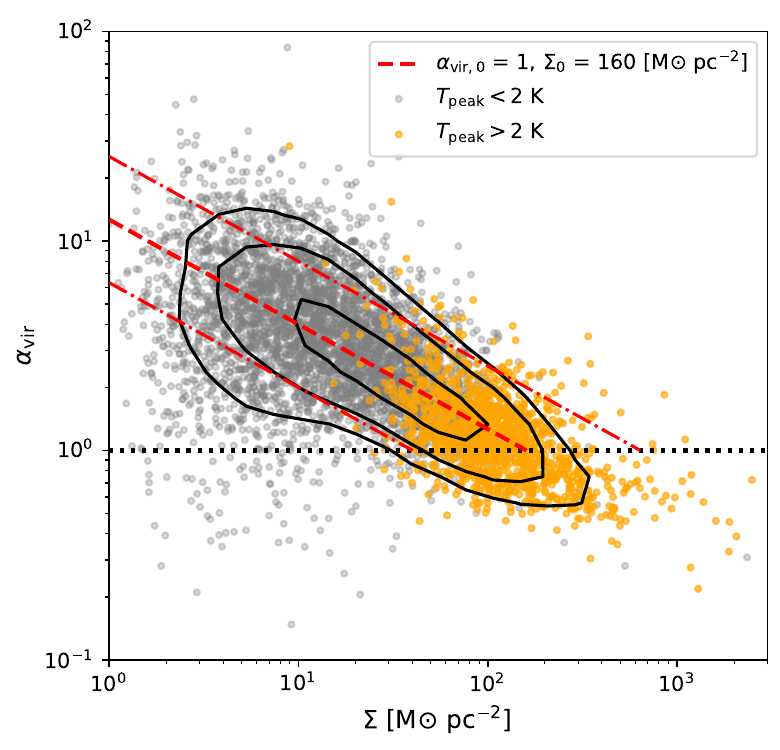
}
\caption{
$\alphaVir$--$\SigmaGMC$ relationship for the clouds in M83.
The contour lines indicate the density distribution of the data points at the levels of 0.2, 0.4, and 0.8, respectively.
As in \S\ref{SubsecScalingRelations}, the clouds with $\Tpeak$ $>2$ K and $<2$ K are indicated with orange and gray markers, respectively.
The black horizontal dotted line indicates $\alphaVir$ of 1.
The red dashed line indicates the case in which the clouds are affected by the beam dilution, and they uniformly have the intrinsic $\alphaVirCh$ of 1 and $\SigmaCh$ of 160 $\MsunPerSqPC$, and the two red dot-dashed lines indicate the same but for a factor of 4 variations around it.
}
\label{FigAlphaToSigma}
\end{figure}

Figure \ref{FigAlphaToSigma} shows the relationship between the observed $\alphaVir$ and $\SigmaGMC$ for the M83 clouds.
The plot exhibits a solid declining trend of $\alphaVir$ toward higher $\SigmaGMC$.
This decline is consistent with the trend expected by a variation in $\fFill$ made by the beam dilution.
We re-write eq. (\ref{EqAlphaVirWithFillingFactor}) by casting eq. (\ref{EqSigmaFilling}) to eliminate $\fFill$:
\begin{equation}
\alphaVir = \alphaVirCh \left(\SigmaCh / \Sigma\right)^{1/2},
\label{EqAlphaSigma}
\end{equation}
where $\SigmaGMC$ $\le$ $\SigmaCh$.
If clouds share a combination $\SigmaCh$ and $\alphaVirCh$ similar to each other, their observed $\alphaVir$ should decrease as $\propto$ $\SigmaGMC^{-1/2}$.
The dashed lines in figured \ref{FigAlphaToSigma} show the expected variation of the observed $\alphaVir$ for clouds in a simple virial equilibrium ($\alphaVirCh$=1) with three different $\SigmaCh$ of 40, 160, and 640 $\MsunPerSqPC$.
Most clouds are distributed along these lines, and moreover, over half of the clouds are located within a factor of 4 variations around $\SigmaCh=$160 $\MsunPerSqPC$.
Notably, 160 $\MsunPerSqPC$ is close to the average surface density of 200 $\MsunPerSqPC$ implied from Larson's relations \citep[S87; ][]{HeyerDame2015Review}.
\par

A caveat is that the formulation made here is based on a crude approximation that the resolution effect mostly affects $\Reff$, but not $\sigmav$ and $\Mcl$. In Appendix \ref{SimSourceInjection}, we verify this approximation with a source injection test.
\par

The cloud distribution on the $\alphaVir$-$\SigmaGMC$ plane suggests that the intrinsic virial parameter and surface density may be more uniform than the observed $\alphaVir$ and $\SigmaGMC$ apparently tell.
This also suggests the need to consider the effect of the beam dilution when interpreting the observed variations in $\alphaVir$ and $\SigmaGMC$.

\subsection{Distribution of filling factor and intrinsic cloud properties}
\label{SubsecDiscussionFillingFactorCorrectedProps}
In the previous subsection, we showed that the observed spread of the virial parameter and surface mass density can be explained, at least partly, by the beam dilution.
Their intrinsic $\alphaVirCh$ and $\SigmaCh$ are potentially more uniform than they appear from $\alphaVir$ and $\SigmaGMC$.
Here, we discuss the potential fraction of the clouds compatible with this possibility and estimate the distribution of intrinsic cloud properties.
We address these questions by first defining a criterion to select clouds for which the discussion on the beam-dilution effect is applicable in \S\ref{SubsubsecCriterion}, and following
in \S\ref{SubsubsecCorrectedProperties}, we discuss the intrinsic cloud properties after the correction for $\fFill$.
\par

\subsubsection{Criterion to select clouds for filling factor estimation}
\label{SubsubsecCriterion}

In the previous sub-section, we discussed the $\fFill$-dependence of the observed cloud properties.
Here, we crudely estimate $\fFill$ for each cloud by comparing the observed $\alphaVir$ with an assumed value of $\alphaVirCh$ using eq. (\ref{EqAlphaVirWithFillingFactor}).
This procedure should apply only to clouds with a simple/single structure that can be approximated with a single temperature when observed at the current resolution and thus are describable with a single $\fFill$.
\par

To select such clouds, we define a peak-equivalent temperature, or simply equivalent temperature, $\Teq$.
By using the cloud's CO luminosity $\LCO$ and assuming a Gaussian spectral profile,
\begin{equation}
\Teq = \LCO / \left[ \sqrt{2 \pi} \sigmav \left( \pi \Reff^2 \right ) \right].
\label{EqEquivalentPeakTemperature}
\end{equation}
Note that if the line profile of a cloud is a single Gaussian (i.e., if the cloud's observed emission has a simple structure characterized by a single temperature), then $\Teq$ $\sim$ $\Tpeak$.
As we adopted the 4 $\sigma$ detection threshold of $\Tpeak$ $\gtrsim$ 0.5 K (\S\ref{SubsecIdentify}), clouds that agree with the single-component approximation should have $\Teq$ $>$ 0.5 K.
Therefore, we adopt $\Teq$ $>$ 0.5 K as a criterion to select clouds to estimate $\fFill$.
\par

\begin{figure}[htbp]
\plotone{
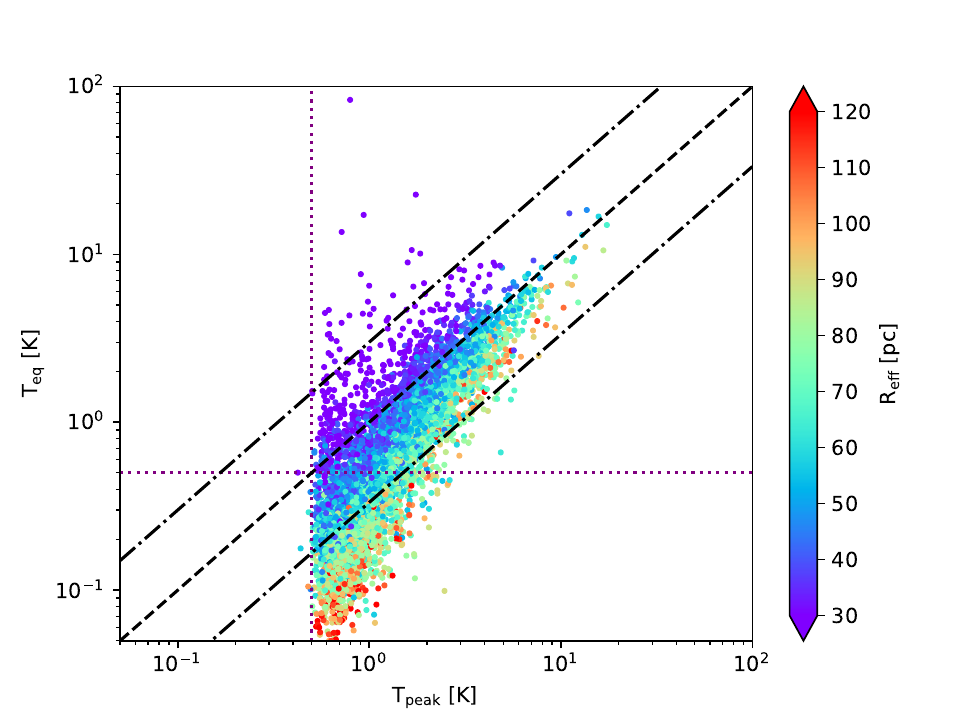
}
\caption{
Equivalent peak temperature $\Teq$, derived with equation (\ref{EqEquivalentPeakTemperature}) as a function of peak temperature $\Tpeak$ for the M83 clouds.
Marker colors are coded to represent $\Reff$.
The dashed line indicates $\Teq$ = $\Tpeak$, and the dash-dotted lines indicate a factor of 3 variations.
The vertical dotted line indicates $\Tpeak$ $=$ 0.5 K, roughly the detection threshold of the clouds.
The horizontal dotted line indicates the $\Teq$ threshold of 0.5 K, adopted for selecting clouds for $\fFill$ estimation.
}
\label{FigTavgTpeak}
\end{figure}

We plot $\Teq$ vs. $\Tpeak$ in figure \ref{FigTavgTpeak} with $\Reff$ color-coded.
Smaller clouds are closer to the line of $\Teq$ = $\Tpeak$, and larger clouds deviate downward toward $\Teq$ $<$ $\Tpeak$.
This trend is expected because smaller clouds are more likely to have a simple appearance when convolved by the beam
The plot indicates that the clouds with $\Teq$ $>$ 0.5 K mostly have $\Teq$ $\sim$ $\Tpeak$, although with modest deviations of up to a factor of 3, hinting that our criterion indeed selects the clouds that can be approximated with a single temperature under the current resolution.
\par

\subsubsection{Estimation of intrinsic cloud properties}
\label{SubsubsecCorrectedProperties}

\begin{figure}[htbp]
\plotone{
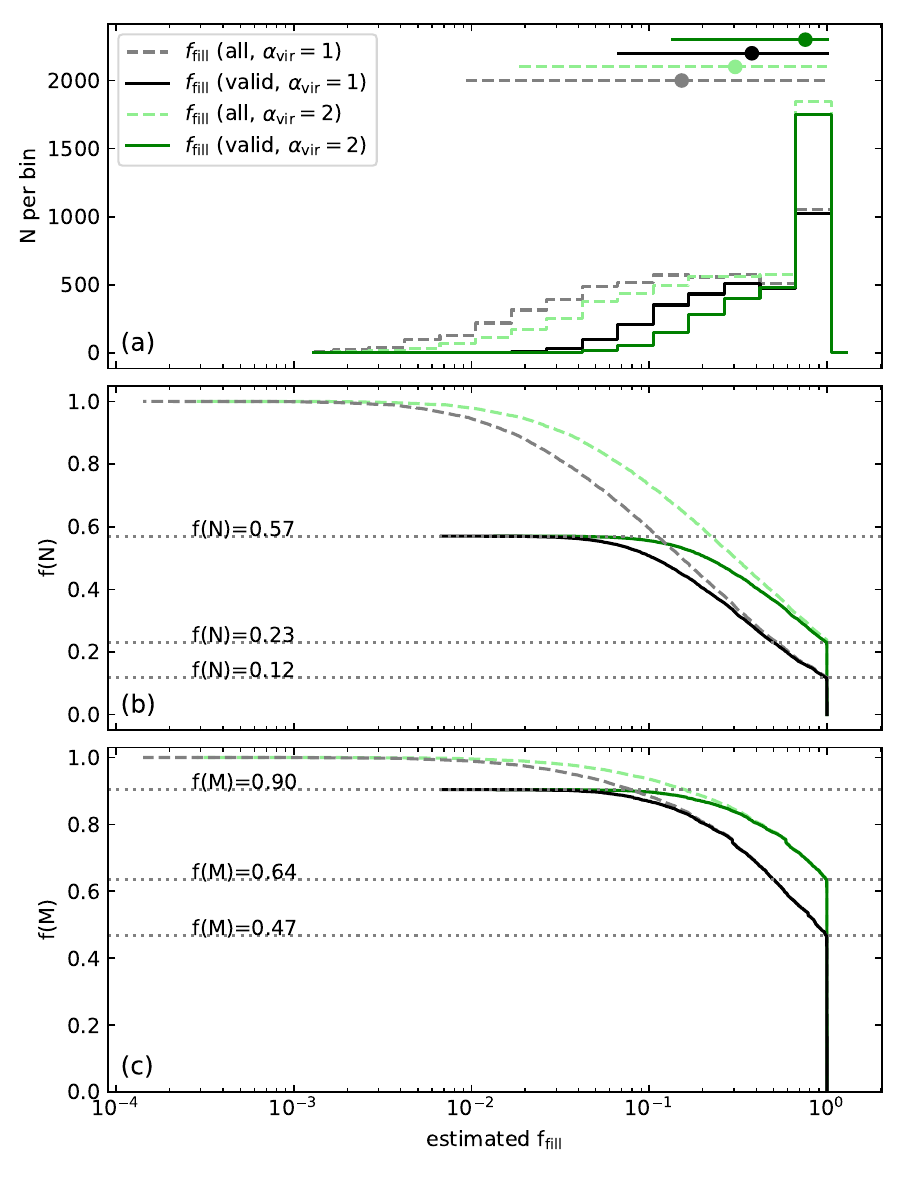
}
\caption{
(a) Number distribution of the estimated area-filling factor $\fFill$ for the clouds in M83.
The black and green lines show the distribution of $\fFill$ by assuming $\alphaVirCh$ of 1 and 2 for clouds that provide 'valid' estimates, i.e., the clouds that satisfy the $\Teq$ $>$ 0.5 K condition.
The dashed gray and light green lines are the same as the black and green lines but for all the cloud samples.
For each category shown here, the 16th--84th percentile range and 50th percentile are indicated with the horizontal line and the filled-circle marker, respectively.
(b)
Cumulative number distribution of the estimated $\fFill$ for the clouds.
The colors and styles of the lines are the same as in (a).
The dotted horizontal lines indicate the fraction number of clouds with 'valid' estimates of $\fFill$; about 12 (23) percent of the clouds have $\fFill=1$ for the case of $\alphaVirCh$ of 1 (2), and about 56 percent of the clouds have 'valid' estimate of $\fFill$.
(c) Same as (b), but showing as a cumulative fractional mass distribution.
}
\label{FigEstimatedF}
\end{figure}

\begin{figure}[htbp]
\plotone{
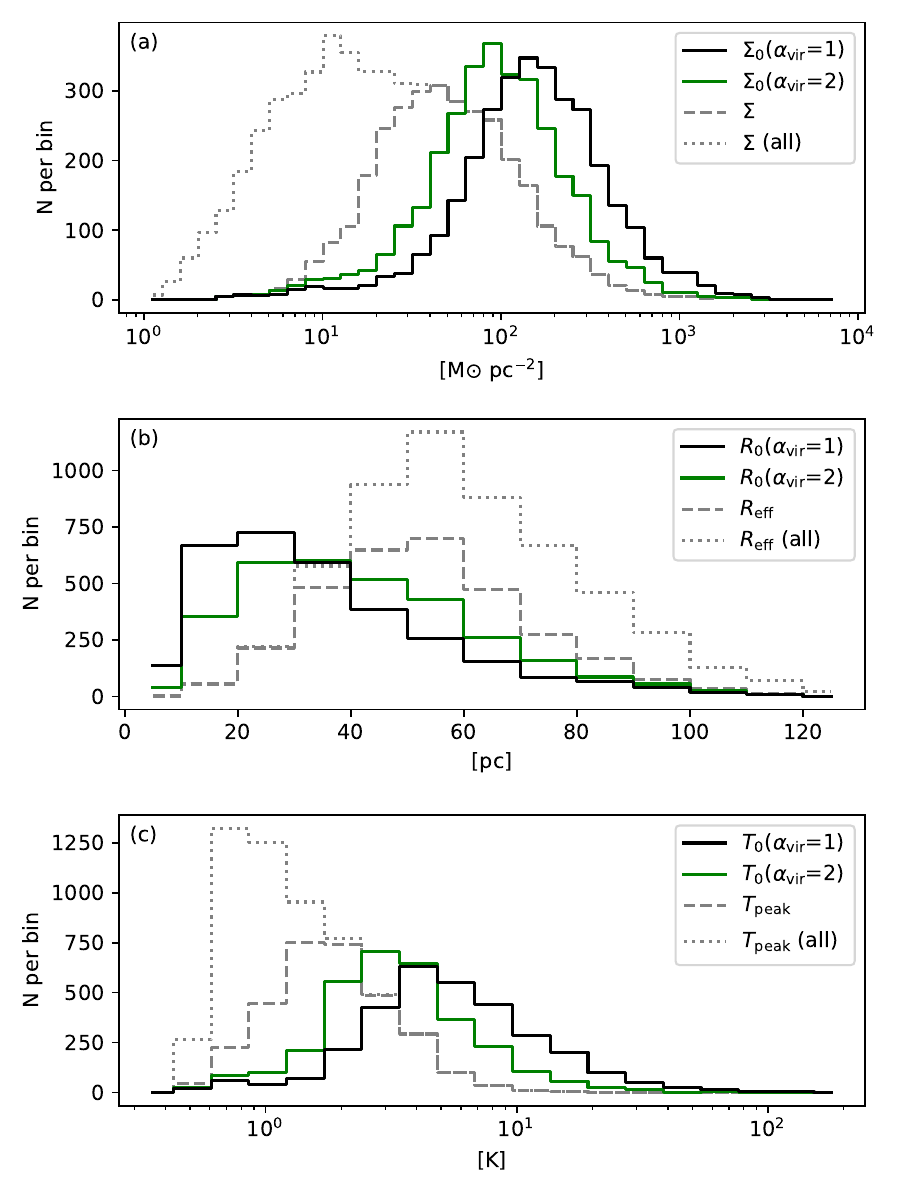
}
\caption{
(a)
Number distribution of cloud surface densities.
The black and green lines indicate the estimated intrinsic $\SigmaCh$ by assuming $\alphaVirCh$ of 1 and 2 for clouds that provide `valid' estimates, i.e., the clouds that satisfy the $\Teq$ $>$ 0.5 K condition.
The dashed line indicates the observed (uncorrected) $\SigmaGMC$ for clouds that satisfy the condition of $\Teq$ $>$ 0.5 K, and the dotted line is the same for all the clouds.
(b) Same as (a), but for cloud radii, showing the intrinsic $\RCh$ and the observed $\Reff$.
(c) Same as (a), but for cloud peak temperatures, showing the estimated intrinsic $\TpeakCh$ and the observed $\Tpeak$.
}
\label{FigEstimatedIntProps}
\end{figure}

We estimate $\fFill$ for each cloud by casting an assumed value of $\alphaVirCh$ into eq. (\ref{EqAlphaVirWithFillingFactor}). We adopt two values of $\alphaVirCh$, 1 (virialized clouds) and 2 (loosely gravitationally-bound clouds). If a cloud's $\alphaVir$ is less than the assumed $\alphaVirCh$, then $\fFill$ is capped at 1.
Although this procedure allows us to derive $\fFill$ for all the clouds in M83, we regard the derived values of $\fFill$ to be ``valid" only for clouds with $\Teq$ $>$ 0.5 K, i.e., clouds that are compatible with the single-component approximation.
The $\fFill$ values for clouds with $\Teq$ $<$ 0.5 K are presented just for reference.
We also note the derivation of $\fFill$ involves the assumptions of a constant CO-to-H$_{2}$ conversion factor and dominance of H$_{2}$ in the cloud masses.
\par

Figure \ref{FigEstimatedF}(a) shows the distribution of $\fFill$.
The solid lines indicate the ``valid" $\fFill$ values derived from the clouds compatible with the single-component approximation by assuming $\alphaVirCh=$1 (black) and 2 (green), respectively.
As for reference, the dashed lines indicate $\fFill$ for all clouds, which include ``invalid" values derived from the clouds incompatible with the single-component approximation.
Although the distribution of ``valid" $\fFill$ extends down to $\sim$0.01, it is chiefly between 0.1 and 1, as the 16th percentile is $\sim$0.07 ($\sim0.13$) for $\alphaVirCh$ of 1 (2).
\par

The minimum value of $\fFill\sim$0.1 is seemingly consistent with the detection limit of the clouds.
For example, if we suppose a typical cloud with an intrinsic brightness temperature of 8 K \footnote{In most parts within Taurus molecular cloud, which is one of the most well-studied molecular clouds, the intrinsic temperature is estimated to be between 6 and 12 K \cite{Goldsmith2008Taurus}}, the observed peak brightness temperature would be 0.5 K if $\fFill$ is 0.1.
As our cloud-finding procedure adopted 0.5 K as the detection limit, it is consistent that the lower limit of the estimated $\fFill$ is around 0.1.
\par

Figure \ref{FigEstimatedF}(b) is the same as figure \ref{FigEstimatedF}(a) but shows in the cumulative number form.
On the plot, the fractional number of clouds with $\fFill=1$ for a particular value of $\alphaVirCh$ indicates the fractional number of clouds with their observed $\alphaVir$ less than the adopted $\alphaVirCh$; about 12 (24) percent of the clouds indicate $\fFill=$1 with $\alphaVirCh$ of 1 (2).
Therefore, if the area-filling factor were not considered, we would be led to the conclusion that only a small number of clouds, less than about 20 percent, are gravitationally bound.
However, contrary to this small fractional number, a majority of clouds ($\sim$57 percent) are compatible with the single-component approximation as $\Teq$ $>$ 0.5 K.
We saw that the $\fFill$ values estimated for such clouds by assuming $\alphaVirCh$ of 1 (or 2) are consistent with the sensitivity limit of the observation; therefore, we deduce that more than half of the clouds in number are potentially bound, i.e., $\alphaVirCh < 2$ if we accept a variation in $\fFill$.
\par

The increase of the fraction of bound cloud by taking $\fFill$ into account is also prominent on a mass basis.
Figure \ref{FigEstimatedF}(c) shows the fractional cumulative mass distribution of $\fFill$.
About 47 (64) percent of the clouds have the observed $\alphaVir$ less than 1 (2), as seen from the cloud numbers with $\fFill=1$.
Allowing a change in $\fFill$, most of the cloud mass ($\sim90$ percent) is explainable as being bound.
\par

The derived $\fFill$, if they are correct, provide the intrinsic cloud properties.
We derive $\SigmaCh$ and $\RCh$ from $\SigmaGMC$ and $\Reff$ using eq. (\ref{EqRFilling}) and eq. (\ref{EqSigmaFilling}).
Also, we derive the intrinsic peak brightness temperature of cloud $\TpeakCh$ as
\begin{equation}
\label{EqTFilling}
\TpeakCh = \Tpeak \fFill^{-1}.
\end{equation}
Figure \ref{FigEstimatedIntProps}(a--c) show the distributions of $\SigmaCh$, $\RCh$, and $\TpeakCh$.
In each panel, the solid line indicates the $\fFill$-corrected estimates derived assuming $\alphaVirCh=$1 (black) and 2 (green), respectively, for the nearly ``single-component" clouds with $\Teq$ $>$ 0.5 K.
The uncorrected observed values, namely $\SigmaGMC$, $\Reff$, and $\Tpeak$, are also shown for the nearly ``single-component" clouds and whole cloud samples with gray dashed and dotted lines, respectively.
\par

Figure \ref{FigEstimatedIntProps}(a) shows that $\SigmaCh$ is in a narrow range.
The median $\SigmaCh$ is $\sim$150 ($\sim90$) for $\alphaVirCh$ of 1 (2), with a spread of $\sim0.3$ dex ($1\sigma$).
These median and narrow ranges in $\SigmaCh$ are similar to the values derived by S87 for the Galactic GMCs.
The estimation of the surface density is vastly different before the filling factor correction: the uncorrected $\SigmaGMC\sim40$ $\MsunPerSqPC$ with a much wider spread.
\par

Figure \ref{FigEstimatedIntProps} (b)(c) show $\RCh$ and $\TpeakCh$.
The medians are $\RCh \sim30$ pc ($\sim40$ pc) and $\TpeakCh \sim3.7$ K ($\sim2.3$ K) and for the assumed $\alphaVirCh$ of 1 (2).
These values of $\RCh$ and $\TpeakCh$ are similar to the typical radii and kinetic temperature (peak brightness temperature) of Galactic GMCs \citep{Sanders1985}, supporting the consistency of the filling-factor estimation.
\par

\begin{figure}[htbp]
\plotone{
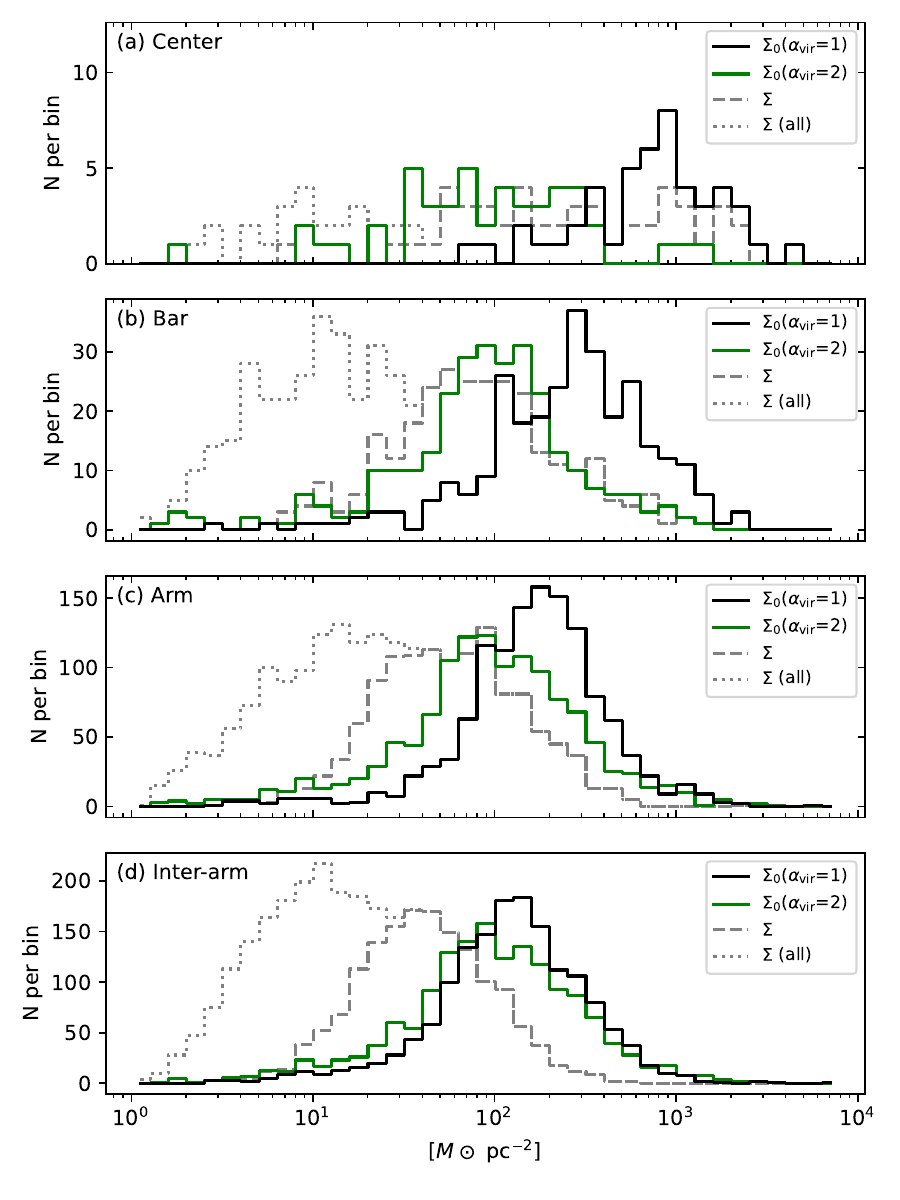
}
\caption{
(a--d) Same as figure \ref{FigEstimatedIntProps}(a), but showing for the central region, bar, arms, and inter-arm regions, respectively.
The regional mask shown in figure \ref{FigArmLocs}(c) is used for the classification.
}
\label{FigEstimatedIntPropsByRegion}
\end{figure}

We also show the environmental variation of $\SigmaCh$ in figure (\ref{FigEstimatedIntPropsByRegion}).
The environmental mask defined in \S\ref{SubsecSpatialDistribution} is used here.
The medians in the arm and inter-arm regions are $\sim$170 ($\sim96$) $\MsunPerSqPC$ and $\sim$120 ($\sim95$) $\MsunPerSqPC$, respectively, for the assumed $\alphaVirCh$ of 1 (2).
The 1 $\sigma$ scatter is only about 0.3 dex for each.
In the bar and center regions, $\SigmaCh$ is elevated compared to the arm/inter-arm regions with medians of $\sim270$ $\MsunPerSqPC$ and $\sim770$ $\MsunPerSqPC$, respectively.
Thus, the cloud properties vary depending on the environments; the inter-arm clouds have lower $\SigmaCh$, or higher $\alphaVirCh$,  compared to the arm clouds, and the disk clouds have lower $\SigmaCh$ than the clouds in the central region.
\par

\begin{figure*}[htbp]
\plotone{
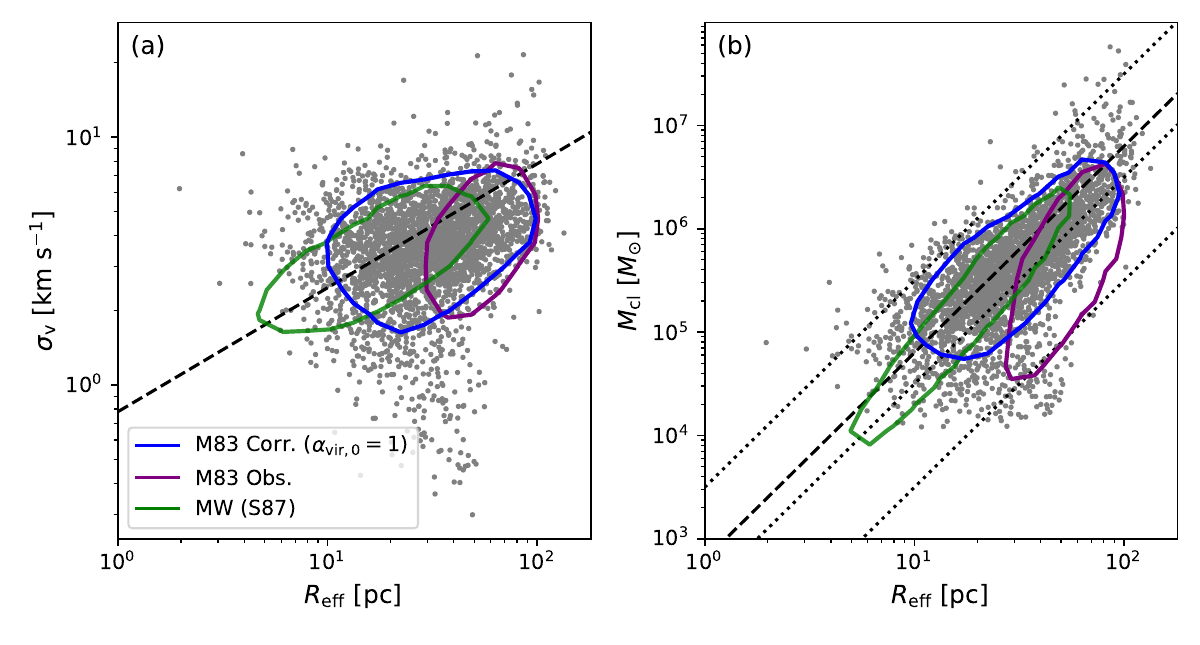
}
\caption{
Scaling relations of cloud properties.
In each plot, the filling-factor corrected data points for the M83 clouds are shown with the gray markers.
The blue and purple contour lines indicate the data densities for the corrected and uncorrected cloud properties in M83 at the level of 20\% of the maximum density for each. The green contour is the same for S87's Galactic samples.
The correction for $\fFill$ is made by assuming $\alphaVirCh$ of 1, and only the clouds with $\Teq$ $>$ 0.5 K are shown for M83 clouds.
(a) Velocity dispersion to radius relationship.
The dashed line indicates the $\sigmav$--$\Reff$ relation of S87.
(b) Mass to radius relationship.
The dashed line indicates $\SigmaGMC$=200 $\MsunPerSqPC$, the constant $\SigmaGMC$ implied from S87's scaling relations.
Also, the dotted lines indicate the lines for $\SigmaGMC$ of 10, 100, and 1000 $\MsunPerSqPC$, respectively.
}
\label{FigSclRelationIntProps}
\end{figure*}

Finally, we compare the filling-factor corrected intrinsic cloud properties with the uncorrected plain values on Larson's scaling relations.
Figure \ref{FigSclRelationIntProps}(a) compares the corrected relationship, $\sigmav$--$\RCh$, with the uncorrected $\sigmav$-$\Reff$, and figure \ref{FigSclRelationIntProps}(b) compares $\Mcl$--$\RCh$ with $\Mcl$--$\Reff$.
The clouds with $\Teq$ $>$ 0.5 K are shown, and we here assumed $\alphaVirCh$ of 1.
From the near constancy of $\SigmaCh$ obtained by assuming a constant $\alphaVirCh$ (figure \ref{FigEstimatedIntProps}), we would expect that the filling-corrected intrinsic cloud properties follow Larson's scaling relations, and both plots show that the expectation is the case.
Figure \ref{FigSclRelationIntProps} also plots the distribution density of S87's data points.
M83's data points have a larger scatter than S87's points, for which the environmental variation of $\SigmaCh$ seen in figure \ref{FigEstimatedIntPropsByRegion} should be at least partly responsible.

\subsection{Implications}
\label{SubsecDiscussionImplication}

The observed $\alphaVir$ and $\SigmaGMC$ vary over wide ranges in the M83 clouds (\S\ref{SubsecCloudPropDistribution}).
The wide range of variations in $\alphaVir$ and $\SigmaGMC$ agree with some recent GMC studies (see references at the beginning of this section).
However, we saw that the $\alphaVir$ and $\SigmaGMC$ in the M83 clouds are distributed approximately along the $\alphaVir$ $\propto$ $\SigmaGMC^{-1/2}$ lines on the $\alphaVir$-$\SigmaGMC$ plane (figure \ref{FigAlphaToSigma}), which are exactly in the direction of the beam dilution (along the change of $\fFill$).
Within the limitations of the current data, it seems conceivable that a significant fraction of the clouds in M83 is consistent with those studied by S87, having $\alphaVir\sim$1-2 and $\SigmaGMC\sim$200 $\MsunPerSqPC$, but being beam-diluted with $\fFill\sim$0.1-1.0.
Specifically, we found that a cloud population of up to $\sim$57\% in number and $\sim$90\% in mass are potentially in agreement with the intrinsic $\alphaVir$ of $1-2$ (figure \ref{FigEstimatedF}), and the intrinsic $\SigmaCh$ is near-constant around the median of $\sim150$ ($\sim90$) $\MsunPerSqPC$ with $\alphaVirCh=1$ (2) for such clouds.
Despite relatively large variations in $\SigmaGMC$ and $\alphaVir$ suggested in some of the recent studies, discrete and bound clouds with a high and near constant $\SigmaGMC$ as suggested by Larson's relations might still be the dominant component in molecular disks.
\par

The dependence of $\alphaVir$ and $\SigmaGMC$ on $\fFill$ gives a caveat when interpreting these observed parameters.
The relation between the size-normalized velocity dispersion $\sigmav/\sqrt{\Reff}$ and $\SigmaGMC$ is often used to visualize the dynamical stabilities of clouds, including the impact of the confinement by external pressure \citep[e.g.,][]{Field2011, Heyer2001OuterGalaxy, Sun2020DE}.
For clouds with a constant $\alphaVir$, $\sigmav/\sqrt{\Reff}$ should scale as $\propto \SigmaGMC^{1/2}$, but in \S\ref{SubsecScalingRelations}, we saw that it scaled as with a shallow index, $\propto \SigmaGMC^{\sim0.3}$.
The shallower index is explainable as the variation in $\fFill$ (Appendix \ref{SecAppendixSizelinewidthCoeffSigma}).
In addition, some studies note the strong mass-dependence of $\alphaVir$, which is mostly $\alphaVir$ $\propto$ $\Mcl^{-0.5}$ \citep[][see references therein]{Chevance2023Review}.
At least for the M83 clouds presented here, the $\alphaVir$-$\Mcl$ scaling is also explainable by the beam-dilution effect (Appendix \ref{SecAppendixAlphaVirM}).
\par

The fact that the distribution of the $\fFill$-corrected $\SigmaCh$ has a narrow range around 100-200 $\MsunPerSqPC$ by assuming $\alphaVirCh$ of 1 (or 2) (\S\ref{SubsecDiscussionFillingFactorCorrectedProps}) would imply that in many, but not all, molecular clouds, the CO luminosity and mass are potentially concentrated at a bound component with high surface densities of around 100 $\MsunPerSqPC$.

We note that the concentration of cloud mass to $\SigmaGMC$ of around 100 $\MsunPerSqPC$ and $\alphaVir$ of $<2$ is also similarly seen in the distribution of the cloud properties inspected in \S\ref{SubsecCloudPropDistribution}: despite that the inter-cloud median of $\SigmaGMC$ is as low as $\sim20$ $\MsunPerSqPC$ and also that median $\alphaVir$ is 2.7, the mass-weighted average of $\SigmaGMC$ of $\alphaVir$ are $\sim130$ $\MsunPerSqPC$ and $\sim$1.1, respectively (\S\ref{SubsecCloudPropDistribution}).
This similarity would reinforce the implication that molecular gas material with high $\SigmaGMC$, which is also strongly influenced by self-gravity, {could be} the dominant component of the molecular gas disk.
It would also be of interest to investigate the internal distribution of $\SigmaGMC$ and $\alphaVir$ within individual clouds to see whether the concentration of CO luminosity to possibly gravitationally bound and high $\SigmaGMC$ components are universal property of molecular clouds not only on inter-cloud scale scales but also on sub-cloud scales.

\section{Summary}
\label{SecSummary}
We identified molecular clouds over the whole disk of M83, with the $^{12}$CO (1--0) data from ALMA \citep{Koda2023M83}.
M83 is one of the closest molecular gas-rich spiral galaxies seen face-on, and it resembles the Milky Way in many aspects, including the existence of the stellar bar, multiple spiral arms, and its Solar-like metallicity. At a resolution of $\sim$46 pc and high mass sensitivity of $\sim$10$^4$ $\Msun$, we sampled 5724 clouds.
By assuming a uniform CO-to-H$_2$ conversion factor of 2.0 $\times$ 10$^{20}$ cm$^{-2}$ (K $\kmPerS$)$^{-1}$, the median cloud mass is $\sim$2$\times$10$^5$ $\Msun$ (\S\ref{SubsecCloudPropDistribution}),
which is close to the median cloud mass found in the inner Galactic disk found by existing surveys \citep[e.g., $\sim$8 $\times$ 10$^4$ $\Msun$,][]{MivilleDeschenes2017MWGMC} and also close to the lower end of a conventional definition of GMC mass range \citep[around 10$^5$ $\Msun$, e.g., ][]{Sanders1985}.
This is a significant improvement because even up to recently, molecular cloud studies that sample clouds in extragalactic disks beyond the Local Group were often made with a spatial resolution of 50-100pc with a shallower sensitivity, with the median mass of sampled clouds being $\sim10^6$ $\Msun$ or higher
\citep[e.g.,][]{Colombo2014Env, Hirota2018M83, Rosolowsky2021PhangsGMCs}.
\par

The overall distributions of the cloud properties are examined in \S\ref{SubsecCloudPropDistribution}.
\begin{itemize}
\item{
The median cloud mass is $\sim$1.9$\times$10$^5$ $\Msun$, and the mass-weighted mean cloud mass is $\sim$1.8 $\times$ 10$^6$ $\Msun$.
The difference indicates that the CO luminosity and molecular gas mass of M83 are concentrated predominantly in massive clouds, similar to the MW disk's cloud population.
Specifically, clouds more massive than $>10^6$ $\Msun$ are only $\sim13\%$ in cloud number but account for $\sim65\%$ of the total CO luminosity.
}

\item{
For a subset of clouds with a mass threshold of $10^6$ $\Msun$, which roughly corresponds to the ones sampled by the previous clouds studies in extragalactic molecular-rich disks, the median surface mass density $\SigmaGas$ and virial parameter $\alphaVir$ are about 130 $\MsunPerSqPC$ and 1.0, respectively.
}

\item{
Without any consideration of beam dilution, the observed median $\SigmaGas$ and $\alphaVir$ are 22 $\MsunPerSqPC$ and 2.7, respectively, for M83 clouds.
The observed median $\SigmaGas$ is much lower than the that of the clouds of S87 \citep[200 $\MsunPerSqPC$, ][]{Solomon1987Larson, HeyerDame2015Review}, but is comparable to those found in the recent Galactic studies, which extensively sampled clouds down to low-intensity threshold \citep{MivilleDeschenes2017MWGMC, Lada2020MassSizeRelation}.
Also, the observed median $\alphaVir$ is similar to the values found in recent extragalactic studies, including PHANGS \citep{Sun2020, Rosolowsky2021PhangsGMCs}.
However, we caution that the observed median values of $\alphaVir$ and $\SigmaGMC$ in M83 are potentially subject to the beam filling factor as discussed in \S\ref{SecDiscussion}.
}

\end{itemize}
The scaling relations of cloud properties (\S\ref{SubsecScalingRelations}), spatial distributions (\S\ref{SubsecSpatialDistribution}), mass distributions (\S\ref{SecMF}), and cloud-to-cloud velocity dispersion (\S\ref{SecVelDisp}) are examined.

\begin{itemize}
\item{{\it Scaling relations}: The observed cloud properties are compared with the scaling relations of \cite{Solomon1987Larson}, namely the velocity dispersion to size, virial mass to CO luminosity, and mass to size relationships.
Although the $\Mvir$-$\LCO$ relationship shows an apparent correlation with a correlation coefficient of 0.78, the signs of correlation are hardly seen for the other two relations.
Still, bright clouds selected with $\Tpeak>2$ K, which comprise about 20 percent of the clouds, are more aligned with Larson's relations.
}
\item{{\it Normalized cloud velocity to surface mass density relation}: We see a correlation between the size-normalized velocity dispersion $\sigmav/\sqrt{\Reff}$ and $\SigmaGMC$, which is expected for clouds influenced by self-gravity \citep{Heyer2009, BallesterosParedes2011A}. However, clouds with low $\SigmaGMC$ exhibit upward deviations and fitting of the data yielded $\sigmav/\sqrt{\Reff} \propto \SigmaGMC^{\sim0.3}$, which is shallower than $\propto$ $\SigmaGMC^{1/2}$ expected for bound clouds with a uniform $\alphaVir$.
}

\item{{\it Spatial distribution of clouds}:
Clouds more massive than 10$^6$ $\Msun$ are strongly concentrated around bright galactic structures, such as the spiral arms, bar, and center.
Smaller clouds are more prevalent in the inter-arm regions.
We fitted the distributions of the massive clouds as logarithmic arms and produced a regional mask.
}

\item{{\it Mass function}: Mass distributions of the identified clouds are examined and fitted with two functional forms: the power-law form, widely used in cloud studies, and a modified power-law with exponential cutoff (Schechter-like form). The fit with the power-law form yields an index $\gammaPL$ of $\sim$-1.9, while the fit with the Schechter-like form yields $\gammaSch$ of $\sim$-1.5. Both forms provide the upper cutoff mass at around $\sim$8.0 $\times$ 10$^6$ $\Msun$. Radial variations of the mass distributions are also examined by dividing the galactic disks into five radial bins. Steepening of the slopes of mass functions is implied from the radial variations of $\gammaPL$, but not with $\gammaSch$.
}

\item{{\it Cloud-to-cloud velocity dispersion}:
Taking advantage of the face-on inclination of M83, the vertical cloud-to-cloud velocity dispersion is estimated.
Outside of the central 2.2 kpc, the estimated vertical dispersion is $\sim$8.3 $\kmPerS$ and $\sim$11 $\kmPerS$ for clouds above and below 5$\times$10$^5$ $\Msun$, respectively.
}

\end{itemize}

In \S\ref{SecDiscussion}, we discussed the possible impact of beam dilution on the cloud properties.
In particular, the dependence of the observed values of virial parameter and surface density ($\alphaVir$ and $\SigmaGMC$) on area-filling factor $\fFill$ was discussed.

\begin{itemize}
\item{
The observed $\alphaVir$ and $\SigmaGMC$ are distributed approximately along $\alphaVir$ $\propto$ $\SigmaGMC^{-1/2}$.
As a hypothetical consideration, we suggest this tendency is explainable as the effect of beam dilution.
If a cloud's beam-convolved emission distribution is simple enough to be approximated as a single component, the relations between the observed and intrinsic values are $\SigmaGMC$ = $\fFill$ $\SigmaCh$ and $\alphaVir$ = $\alphaVirCh$ $\fFill^{-1/2}$.
}

\item{
We devised a temperature criterion to select clouds compatible with the single-component approximation, which is the prerequisite to assume the $\fFill$ dependence of $\alphaVir$.
For the selected clouds with the criterion, we estimated the values of $\fFill$ for each cloud by assuming that they are intrinsically bound with $\alphaVirCh$ of $\le2$ and found that $\fFill$ of the clouds is mostly between 0.1 and 1.
In this way, up to $\sim57$ percent of the clouds in number, or $\sim90$ percent of the cloud masses in M83's disk, are compatible with the assumed $\alphaVirCh$.
Also, the estimated intrinsic $\SigmaCh$ for such clouds are found to concentrate around $150$ $\MsunPerSqPC$ if the intrinsic $\alphaVirCh$ of 1 is assumed, or 90 $\MsunPerSqPC$ if the intrinsic $\alphaVirCh$ of 2 is assumed.
The combination of $\alphaVirCh$ and $\SigmaCh$ resemble the clouds found by S87, suggesting that such bound and discrete clouds are potentially the dominant component in a MW-type disk of M83.
}
\end{itemize}

\vspace{5mm}
\begin{acknowledgments}
This paper makes use of the following ALMA data: ADS/JAO. ALMA\#2017.1.00079.S. ALMA is a partnership of ESO (representing its member states), NSF (USA) and NINS (Japan), together with NRC (Canada), MOST and ASIAA (Taiwan), and KASI (Republic of Korea), in cooperation with the Republic of Chile. The Joint ALMA Observatory is operated by ESO, AUI/NRAO, and NAOJ. The National Radio Astronomy Observatory is a facility of the National Science Foundation operated under cooperative agreement by Associated Universities, Inc.
This research made use of astrodendro, a Python package to compute dendrograms of Astronomical data (http://www.dendrograms.org/).  This research made use of APLpy, an open-source plotting package for Python hosted at http://aplpy.github.com.
LCH was supported by the National Science Foundation of China (11991052, 12011540375, 12233001), the National Key R\&D Program of China (2022YFF0503401), and the China Manned Space Project (CMS-CSST-2021-A04, CMS-CSST-2021-A06).
\end{acknowledgments}

\facilities{ALMA}
\software{
astrodendro \citep{Robitaille2019}, scikit-image \citep{vanDerWalt2014ScikitImage}, matplotlib \citep{Hunter2007Matplotlib}, emcee \citep{ForemanMackey2013emcee}
}

\appendix
\section{Source injection test}
\label{SimSourceInjection}

The observed virial parameter $\alphaVir$ of the clouds in M83 exhibits a trend, which is approximately $\alphaVir$ $\propto$ $\SigmaGMC^{-0.5}$ (figure \ref{FigAlphaToSigma}).
In \S\ref{SubsecDiscussionBeamDilutionFormulate}, we argued that this trend possibly owes to a variation in the area-filling factor, driven by the overestimation of cloud radius $\Reff$ for clouds smaller than the beam.
We assumed there that (1) the resolution effect overestimates $\Reff$ but does not affect the velocity dispersion $\sigmav$ and mass $\Mcl$ of clouds and that (2) the $\Reff$ overestimation is coupled with the decrease of the area-filling factor of the beam (eq. \ref{EqRFilling}).
We test these assumptions with a simulation test, which injects model sources with known properties into a portion of the data cube and samples the injected sources via the same identification procedure utilized in \S\ref{SecIdentify}.
\par

\subsection{Procedure}
The test is made with two setups.
A $\sim0.8$ kpc$^2$ segment of M83 is used as the target field in both setups.
Within the field, CO emission from M83 mostly resides within the LSR velocity range between 410 $\kmPerS$ and 455 $\kmPerS$.
The two setups are configured with different velocity ranges as follows.
\begin{itemize}
\item{
\emph{Noise-only} setup:
A CO emission-free velocity range of from 340 $\kmPerS$ to 409$\kmPerS$ is used.
This setup is used to see the impact of the noise on the cloud property measurements.
}
\item{
\emph{Blended} setup:
The velocity range between 401 $\kmPerS$ and 470 $\kmPerS$ is used to assess the combined impact of the noise and blending of CO emission surrounding the input sources.
}
\end{itemize}
\par

Input model clouds are generated by assuming a Gaussian profile in both spatial and velocity directions.
For each model cloud, $\Mcl$ is given as a parameter and $\Reff$ and $\sigmav$ are determined by setting $\alphaVir$ as 1.5 and $\SigmaGMC$ as 100 $\MsunPerSqPC$, respectively.
Five model clouds with $\Mcl$ of 0.5, 1, 2, 4, and 8 $\times$ 10$^5$ $\Msun$ are generated and convolved to emulate the spatial resolution of $\sim$46 pc and velocity channel of 1 $\kmPerS$.
In each round of the test, the model clouds are injected into the random positions within the data cube with an imposed minimum spacing of 80 pc.
For the \emph{noise-only} setup, a constant receding velocity of 375 $\kmPerS$ is assigned to the clouds.
For the \emph{blended} setup, the receding velocity of each cloud is determined such that it is offset by 12 $\kmPerS$ from the median velocity of the existing CO emission around the cloud. The offset velocity of 12 $\kmPerS$ is comparable to the RMS cloud-to-cloud velocity dispersion in the M83 clouds (\S\ref{SecVelDisp}).
From the source-added data cube, clouds are identified using the procedure described in \S\ref{SecIdentify} and cross-matched with the input model clouds with a matching tolerance of half the FWHM cloud size in both spatial and velocity directions.
The injection-identification sequence is repeated 40 times for each setup.
\par

\subsection{Simulation results}
\begin{figure*}[htbp]
\plotone{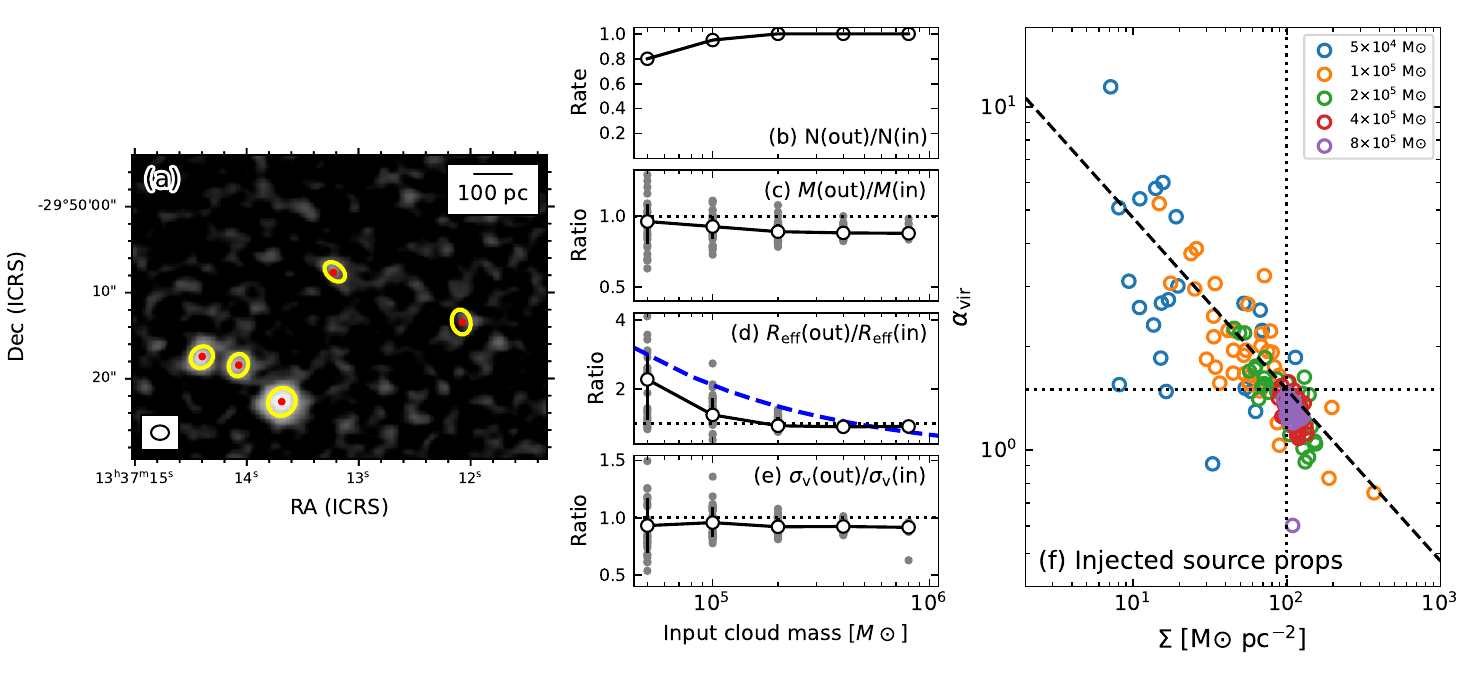}
\caption{
Simulation results of the source injection test made with the \emph{noise-only} setup, which uses a CO emission-free portion of the data cube.
(a) Integrated CO intensity image of the data used in one of the 40 rounds of the test.
In the image, five sources with $\Mcl$ of 0.5, 1, 2, 4, and 8 $\times$ $10^5$ $\Msun$ are injected.
The red dotted markers indicate the positions of the input sources.
Yellow ellipses indicate the output FWHM sizes of the detected sources.
(b) Detection rate as a function of the input $\Mcl$.
(c) Output-to-input ratios of $\Mcl$ for the detected sources as a function of the input $\Mcl$.
The black dotted horizontal line indicates the output-to-input ratio of 1.
Gray dots indicate each of the detected clouds in each round.
Open markers indicate the median output-to-input ratios for each input $\Mcl$ with error bars showing the 16th to 84th percentiles range.
(d) Same as (c), but for $\Reff$.
The blue dashed line indicates the line of $\theta_{\mathrm{b}}/\theta_{\mathrm{source}}$, where $\theta_{\mathrm{b}}$ and $\theta_{\mathrm{source}}$ are the FWHM sizes of the beam ($\sim46$ pc) and the input model sources, respectively.
(e) Same as (c), but for $\sigmav$.
(f) $\alphaVir$-$\SigmaGMC$ relationship of the detected sources.
The vertical and horizontal dotted lines indicate $\SigmaGMC$ of 100 $\MsunPerSqPC$ and $\alphaVir$ of 1.5 uniformly assigned to all the input sources.
The dashed line indicates the relation of $\alphaVir$ $\propto$ $\SigmaGMC^{-0.5}$.
}
\label{FigSimAlphaVirOff}
\end{figure*}

\begin{figure*}[htbp]
\plotone{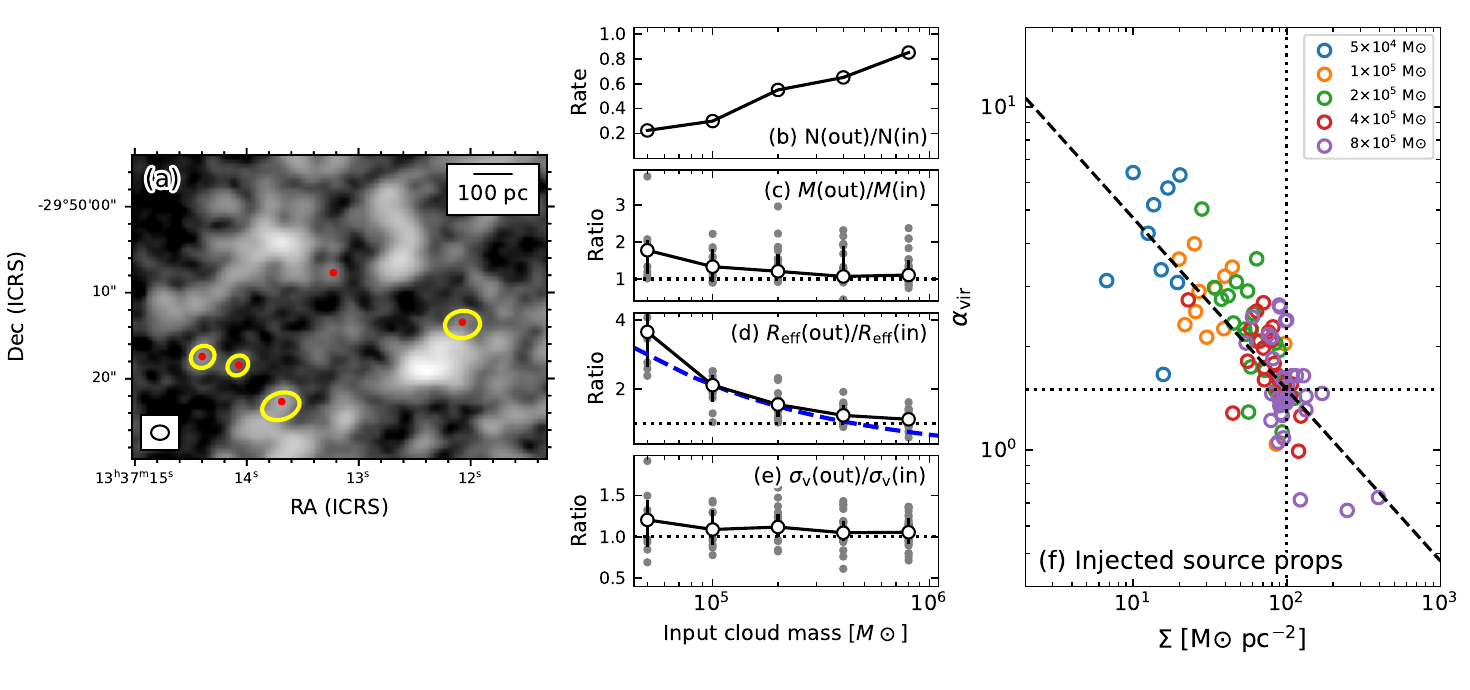}
\caption{
Same as figure \ref{FigSimAlphaVirOff}, but for the \emph{blended} setup.
}
\label{FigSimAlphaVirOn}
\end{figure*}

Figure \ref{FigSimAlphaVirOff} and figure \ref{FigSimAlphaVirOn} show the test results for \emph{noise-only} and \emph{blended} setups, respectively.
In the \emph{noise-only} setup, the input sources are detected with high detection rates.
The rate is 100 percent for the input $\Mcl$ of $>10^5$ $\Msun$ and slightly lowers to $\sim80$ percent at 5$\times$10$^4$ $\Msun$.
These high recovery rates are in accordance with the minimum cloud mass of $\sim3$ $\times$ $10^4$ $\Msun$, estimated for uncrowded regions in \S\ref{SubsecCOLumFraction}.
Figure \ref{FigSimAlphaVirOff}(c), \ref{FigSimAlphaVirOff}(d), and \ref{FigSimAlphaVirOff}(e) show the ratios between the output-to-input values of $\Mcl$, $\Reff$, and $\sigmav$.
At the lower mass range, $\Reff$ shows a significant increase in the output-to-input ratio while the ratios for $\Mcl$ and $\sigmav$ are more stable around 1.
Figure \ref{FigSimAlphaVirOff}(f) shows the $\alphaVir$-$\SigmaGMC$ plot derived from the output properties of the detected sources.
As a consequence of $\Reff$ overestimation at lower $\Mcl$, a trend of $\alphaVir$ $\propto$ $\SigmaGMC^{-0.5}$ appears despite the input sources uniformly have $\alphaVir$ of 1.5 and $\SigmaGMC$ of 100 $\MsunPerSqPC$.
\par

In the \emph{blended} setup, the recovery rates are lower than the \emph{noise-only} setup in all the mass ranges, with larger drops in smaller cloud masses.
The rates are above 60 percent with input $\Mcl$ of $>2$ $\times$ $10^5$ $\Msun$, roughly in agreement with the effective completeness limit suggested from the cloud mass spectra \S\ref{SecMF}.
The trend of $\Reff$ overestimation is more prominent in the \emph{blended} setup (figure \ref{FigSimAlphaVirOn}d).
In particular, the median output-to-input ratios of $\Reff$ approximately follow the line of $\theta_{\mathrm{b}}/\theta_{\mathrm{source}}$ at lower $\Mcl$ of $<2\times10^5$ $\Msun$.
The $\alphaVir$-$\SigmaGMC$ plot for the \emph{blended} setup shown in figure \ref{FigSimAlphaVirOn}(f) again shows a trend in agreement with $\alphaVir$ $\propto$ $\SigmaGMC^{-0.5}$ as in the \emph{noise-only} setup.
\par

As the CO emission widely prevails over the disk of M83, the \emph{blended} setup would be more representative of the uncertainty trends in the M83 catalog than the \emph{noise-only} setup.
The blended setup indicates that the $\Reff$ measurement tends to derive the beam size as cloud size in crowded environments despite the beam size deconvolution.
As the overestimated cloud size is comparable to beam size, the intrinsic and observed cloud radii are associated with each other by the beam filling factor, as in eq. \ref{EqRFilling}.
In addition, the trend of $\alphaVir$ $\propto$ $\SigmaGMC^{-0.5}$ seen with the model clouds with a uniform combination of $\alphaVir$ and $\SigmaGMC$ suggests that, to the first order, the impact of beam dilution on $\sigmav$ and $\Mcl$ are more moderate compared to $\Reff$.

\section{Filling factor dependence seen in size-normalized velocity dispersion to surface density relation}
\label{SecAppendixSizelinewidthCoeffSigma}

The relation between the size-normalized velocity dispersion $C$ ($\equiv$ $\sigmav/\sqrt{R}$) and surface mass density $\SigmaGMC$ is often used to gauge the dynamical state of clouds \citep{Heyer2009, BallesterosParedes2011A}.
As $\Mvir$ $\left( \propto R \sigmav^2 \right)$ = $\alphaVir$ $\Mcl$ is mathematically equivalent to $\sigmav^2/R$ $\propto$ $\alphaVir \SigmaGMC$, clouds with a constant $\alphaVir$ should follow a line on the $\sigmav/\sqrt{R} - \SigmaGMC$ plane following equation (\ref{EqSizeLinewidthCoeff}).
\par

We saw in \S\ref{SubsecScalingRelations} that a majority of the M83 clouds are located upward of the line expected for $\alphaVir$ = 1 on the $\sigmav/\sqrt{R}$--$\Sigma$ plot.
The clouds with such upward deviations are often considered to be confined by external pressure \citep{Heyer2009, Field2011, Sun2020DE}, but we also saw that the deviations are rather systematic; the fitted relation indicated that $\sigmav/\sqrt{R}$ varies as $\propto$ $\SigmaGMC^{\sim0.3}$ (eq. \ref{EqSizeLinewidthCoeffSigmaRelationFit}), although $\SigmaGMC^{1/2}$ is expected for a constant value of $\alphaVir$ (eq. \ref{EqSizeLinewidthCoeff}).
\par

The observed index of $\sim0.3$ for the $\sigmav/\sqrt{R}$--$\Sigma$ relation is explainable with the beam-dilution effect if a significant fraction of the clouds share a similar set of combinations of the intrinsic virial parameter and surface mass density.
Consider a cloud with intrinsic radius $\RCh$, surface mass density $\SigmaCh$, virial parameter $\alphaVirCh$, and velocity dispersion $\sigmav$. As we assume the beam dilution does not affect the velocity dispersion, $\sigmav$ does not have a subscription.
If the cloud follows Larson's relations, the following applies to the cloud,

\begin{equation}
\sigmav / \sqrt{\RCh} = \left( \frac{2}{9} {\alphaVirCh} \pi G \SigmaCh \right)^{1/2}.
\label{EqSizeLinewidthCoeffCh}
\end{equation}
The observed cloud radius $\Reff$ and surface mass density $\SigmaGMC$ are related to $\RCh$ and $\SigmaCh$ with equation (\ref{EqRFilling}) and equation (\ref{EqSigmaFilling}), respectively.
Substituting them into equation (\ref{EqSizeLinewidthCoeffCh}), we obtain
\begin{equation}
\sigmav / \sqrt{\Reff} = \left( \frac{2}{9} {\alphaVirCh} \pi G \right)^{1/2} \left( \SigmaCh \SigmaGMC \right)^{1/4}.
\label{EqSizeLinewidthCoeffWithChObs}
\end{equation}
The equation postulates that if $\alphaVirCh$ and $\SigmaCh$ are constant, the scaling between the observed coefficient, $\sigmav/ \sqrt{\Reff}$ to be proportional to $\SigmaGMC^{1/4}$, which is similar to the observed relation of $\sigmav/\sqrt{\Reff}$ $\propto$ $\SigmaGMC^{\sim0.3}$.
\par

Figure \ref{FigAppendix}(a) shows the $\sigmav/\sqrt{\Reff}$--$\SigmaGMC$ relation for the M83 clouds. The red line on the plot is the expected behavior of equation (\ref{EqSizeLinewidthCoeffCh}) in the case of $\alphaVirCh$ of 1 and $\SigmaCh$ of 160 $\MsunPerSqPC$ for a range of filling factor between 0.1 and 1, showing an alignment with the running median of the observed $\sigmav/\Reff$.

\section{Filling factor dependence seen in virial parameter to mass relation}
\label{SecAppendixAlphaVirM}

\begin{figure*}[htbp]
\plotone{
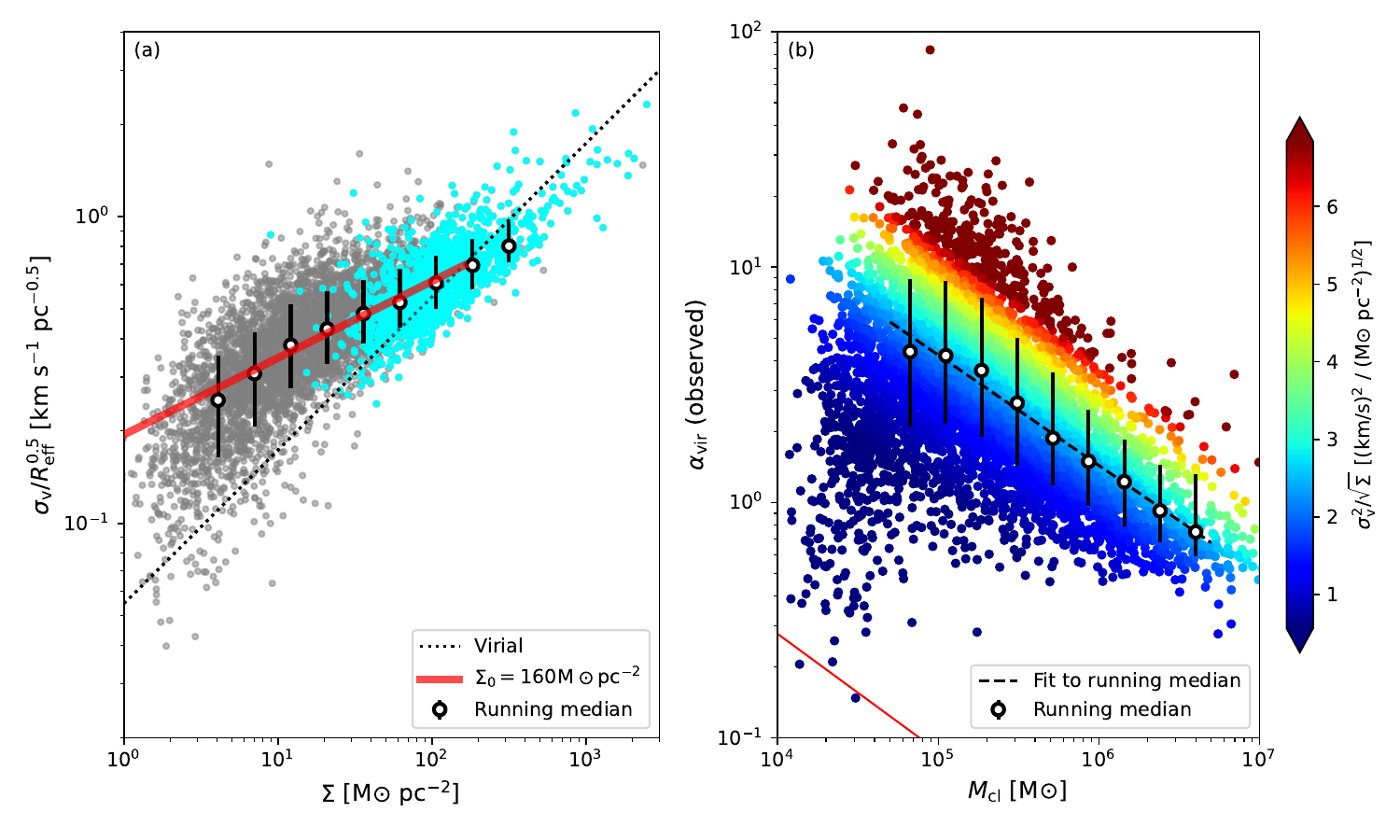
}
\caption{
(a) $\sigmav/\sqrt{\Reff}$--$\SigmaGMC$ relationship for clouds in M83.
Cyan and gray markers indicate the clouds with $\Tpeak$ above and below 2 K, respectively.
White circle markers with black outline indicate the running median of $\sigmav/\sqrt{\Reff}$--$\SigmaGMC$.
The dotted line indicates a state of virial equilibrium, $\alphaVir$=1.
The red thick line shows the expected range of variation for clouds that have $\alphaVir$ = 1 and a constant intrinsic surface density $\SigmaCh$ of 160 $\MsunPerSqPC$ but are affected by the beam dilution with $\fFill$ of $< 1$.
(b) $\alphaVir$--$\Mcl$ relationship for clouds in M83.
The running median of $\alphaVir$ and fitted to the median values are displayed similarly to (a).
The red line indicates the minimum $\alphaVir$ expected from equation (\ref{EqAlphaLimit}) by setting $\sigmav$ to be 1 / $\sqrt{12}$ $\kmPerS$, which is the instrumental limit of the data, and $\Tpeak$ of 0.5 K, which is the near the detection limit of the clouds.
Most clouds are located well above the line, indicating that the velocity resolution is not driving the observed $\Mcl$-dependence of $\alphaVir$.
The marker colors are coded to express the variation in $\sigmav^2/\SigmaGMC^{1/2}$.
}
\label{FigAppendix}
\end{figure*}

Some studies of molecular clouds, including this one, noted that $\alphaVir$ varies as $\propto$ $\Mcl$$^{\sim-0.5}$ in their samples \citep{Heyer2001OuterGalaxy, MivilleDeschenes2017MWGMC}
\citep[also see references in][]{Chevance2023Review}.
Figure \ref{FigAppendix}(b) shows the $\alphaVir$-$\Mcl$ relation for the M83 clouds.
The running median values of $\alphaVir$ derived within the binned $\Mcl$ values are also shown, and fit the running median values yield that $\alphaVir$ $\propto$ $\Mcl$$^{-0.45}$, which is close to the $\alphaVir$ $\propto$ $\Mcl$$^{-\sim0.5}$ seen in other studies quoted here.
\par

\citet{Heyer2001OuterGalaxy} pointed out that a relation of $\alphaVir$ $\propto$ $\Mcl^{-0.5}$ could arise due to a limited velocity resolution; the argument of \citet{Heyer2001OuterGalaxy} is as follows.
From eq. (\ref{EqMVirDef}) and eq. (\ref{EqAlphaVirDef}), $\alphaVir =\ (9/2)\Reff\sigmav^2 (G\alpha_{\mathrm{CO}}\LCO)^{-1}$, where $\alpha_{\mathrm{CO}}$ is the mass-to-luminosity conversion factor for the CO line.
Combining this with $\LCO$ $\simeq$ $\sqrt{2 \pi} \Tpeak \sigmav \pi \Reff^2$ to eliminate $\Reff$, one obtains
\begin{equation}
\alphaVir = 178
\left( \frac{\alpha_{\mathrm{CO}}}{4.4 \Msun / (\mathrm{K}\ \kmPerS \mathrm{pc}^2)} \right)^{-1/2}
\left( \frac{\sigmav}{\kmPerS} \right)^{3/2}
\left( \frac{\Tpeak}{\mathrm{K}} \right)^{-1/2}
\left( \frac{\Mcl}{\Msun} \right)^{-1/2}.
\label{EqAlphaLimit}
\end{equation}
Many clouds in the samples of \citet{Heyer2001OuterGalaxy} have $\sigmav$ close to the instrumental resolution limit. Substituting the instrumental limit of $\sigmav$ and the typical temperature of the clouds observed, \citet{Heyer2001OuterGalaxy} demonstrated that the above relation explains the lower envelope of their samples on the $\alphaVir$-$\Mcl$ plot.
\par

An argument similar to \citet{Heyer2001OuterGalaxy} would hold for the M83 cloud samples, but not in exactly the same way because $\sigmav$ of the clouds are well above the velocity resolution of the data (\S\ref{SubsecCloudPropDistribution}).
The red line in figure \ref{FigAppendix}(b) shows that the minimum $\alphaVir$ expected by equation (\ref{EqAlphaLimit}) are far below the observed data points.
In turn, we consider that the $\alphaVir$ $\propto$ $\Mcl^{\sim1/2}$ relation is related to the limitation in the spatial resolution.
\par

The marker colors in figure \ref{FigAppendix}(b) are coded to represent the value of $\sigmav^2/\sqrt{\SigmaGMC}$, which is dimensionally equal to $\sigmav^{3/2} \Tpeak^{-1/2}$ as $\SigmaGMC$ $\propto$ $\Tpeak \sigmav$.
From the distribution of the data points, it is apparent $\sigmav^2/\SigmaGMC^{1/2}$ is almost constant along the direction of $\alphaVir$ $\propto$ $\Mcl^{-1/2}$.
We may rewrite eq. (\ref{EqSizeLinewidthCoeffWithChObs}) as
\begin{equation}
\sigmav^{2} / \sqrt{\SigmaGMC} \propto \sqrt{\SigmaCh} \Reff.
\label{EqSizeLinewidthCoeffWithChObsRewrite}
\end{equation}
As seen in \S\ref{SubsecCloudPropDistribution}, the observed $\Reff$ indeed has a limited range of variation, with 16th and 84th percentiles of 40 and 80 pc, respectively.
We argued in \S\ref{SecDiscussion} that once the beam dilution is taken into account, a majority of the clouds agree with $\alphaVir$ of 1 (or 2) with near uniform $\SigmaCh$.
Therefore, when we assume that near-virial clouds with almost constant $\SigmaCh$ account for the dominant fraction of the M83 clouds as suggested in \S\ref{SecDiscussion}, equation (\ref{EqSizeLinewidthCoeffWithChObsRewrite}) leads that ${\sigmav}^2 / \sqrt{\SigmaGMC}$, which has the same dimension as $\sigmav^{3/2} \Tpeak^{-1/2}$, would also be constant and thus, $\Mvir$ $\propto$ $\Mcl^{\sim0.5}$ emerges from equation (\ref{EqAlphaLimit}).

\section{Comparison of Galactic cloud catalogs}
\label{SecAppendixMWCatalogComparison}

In \S\ref{SubsecScalingRelations}, the clouds in M83 are compared with the scaling relations of S87, which have been used as the reference lines up to now.
The scaling relations of S87 are determined from 273 clouds sampled from the inner Galactic disk with relatively high-intensity thresholds of a few to several Kelvin, but subsequent studies made a more extensive sampling of clouds down to lower intensity levels.
In particular, \cite{Rice2016} (hereafter R16) and \cite{MivilleDeschenes2017MWGMC} (hereafter M17) sampled clouds from the Galactic plane data of \cite{Dame2001}, which observed most part of the thin Galactic molecular disk in $^{12}$CO (1--0) with the 1.2m telescopes.
Here, we extend the examination made in \S\ref{SubsecScalingRelations} by including clouds in the R16 and M17 catalogs.
\par

\subsection{Catalogs for comparison}

The R16 catalog contains 1064 clouds with a total cloud mass of 2.5 $\times$ 10$^8$ $\Msun$ and the M17 catalog lists 8107 clouds with a total mass of 1.2 $\times$ 10$^9$ $\Msun$. The M17 catalog sampled 98 percent of the total CO luminosity within the mapped area.
To compare the M83 clouds with the Galactic clouds at similar environmental conditions, we use the clouds with the Galactocentric radius within 8.5 kpc for the Galactic catalogs.
Both M83 and Milky Way have radially decreasing trend of molecular fraction $f_\mathrm{mol}$ ($\equiv$ $\Sigma_{\mathrm{mol}}$ / ($\Sigma_{\mathrm{mol}}$ + $\Sigma_{\mathrm{atom}}$)). In M83, $f_{\mathrm{mol}}$ is about 0.5 at the edge of the mapped area \citep{Crosthwaite2002M83, Lee2024M83}. In the Milky Way, it is about 0.1 at $\Rgal$ of 8.5 kpc \citep{NakanishiSofue2016}.
\par

For the R16 catalog, we take the radius $R$, velocity dispersion $\sigmav$, and cloud mass $\Mcl$ from the catalog.
The cloud radius in R16 follows the same definition as S87.
Therefore, $R$ in the R16 catalog can be directly compared with $\Reff$ of the M83 clouds and S87 clouds.
CO luminosity $\LCO$ of the R16 clouds is computed from $\Mcl$ assuming the same conversion factor as in \S\ref{SubsecCOLumFraction}.
The surface density $\SigmaGMC$ and virial parameter $\alphaVir$ are derived following eq. (\ref{EqSigmaDef}) and eq. (\ref{EqAlphaVirDef}).
\par

For the M17 catalog, the velocity dispersion $\sigma_v$, surface density $\SigmaGMC$, and cloud mass $\Mcl$ are taken from the catalog.
We compute the effective radius of M17 clouds as $\Reff$ = $\left[M (\pi \SigmaGMC)^{-1}\right]^{1/2}$.
For $\LCO$ and $\alphaVir$, they are derived in the same way as the R16 clouds.
\par

\subsection{Comparison}
Figure \ref{FigSclRelationsExtended} compares the scaling relations of cloud properties among the M83 catalog and the Galactic cloud catalogs of S87, R16, and M17.
For the MW catalogs, we divide the clouds into two groups with a kinematic distance $\Dkin$ threshold of 8.5 kpc.
The upper and bottom rows of figure \ref{FigSclRelationsExtended} show the Galactic clouds at $\Dkin$ $<$ 8.5 kpc and $>$ 8.5 kpc, respectively.
In both rows, the MW clouds with $\Rgal$ $>8.5$ kpc are excluded, as mentioned above.
In addition, we excluded the clouds with Galactic longitude $l$ within $|l|$ $<$ 5 $\degr$ to exclude the clouds located in the directions close to the Galactic Center (GC) because the GC clouds have properties distinct from the disk clouds \citep[e.g.,][]{Oka2001} and such clouds are contained only in the M17 catalog among the three catalogs.
Table \ref{TablePropsMW} lists the 16, 50, and 84 percentiles of the cloud properties for the three Galactic catalogs classified with the same criteria as figure \ref{FigSclRelationsExtended}.
We note that the R16 and M17 catalogs contain clouds in the outer Galactic disk, which have lower $\SigmaGMC$ and higher $\alphaVir$ than the inner Galactic clouds \citep{Heyer2001OuterGalaxy}.
The last three rows in Table \ref{TablePropsMW} list the cloud properties in the three catalogs without any filtering.
\par

\subsubsection{R16 catalog}
Comparing the R16 and S87 clouds, the R16 clouds mostly follow scaling relations similar to that of S87, with a slight modification in coefficients.
The coefficients of the cloud scaling relations are determined by $\SigmaGMC$ and $\alphaVir$.
As can be seen from table \ref{TablePropsMW}, the R16 clouds have a median $\SigmaGMC$, a factor of three lower than that of S87, but the median $\alphaVir$ is comparable.
Because of the comparable $\alphaVir$, clouds in both catalogs show a similar scaling on the $\Mvir$-$\LCO$ plot and $\sigmav/\sqrt{\Reff}$-$\SigmaGMC$ plots.
On the $\sigmav$-$\Reff$ plot, the R16 clouds indicate a smaller $\sigmav$ at given $\Reff$ compared to S87.
R16 determined the $\sigmav$-$\Reff$ scaling in the inner Galactic disk to be $\sigmav$ = 0.50 $\Reff$$^{0.52}$.
The coefficient value of 0.50 $\kmPerS$ pc$^{-1/2}$ is slightly smaller than for S87, 0.72 $\kmPerS$ pc$^{-1/2}$ \citep[S87,][]{Heyer2009}.
Considering the factor of two differences in the median $\SigmaGMC$, this difference in the $\sigmav$-$\Reff$ coefficient value agrees with the $\SigmaGMC$-dependence of the $\sigmav$-$\Reff$ coefficient for bound clouds (eq. \ref{EqSizeLinewidthCoeff}).
\par

In \S\ref{SubsecScalingRelations}, the M83 clouds with $\Tpeak$ $>2$ K mostly follow the scaling relations of S87. As R16 and S87 clouds follow similar scaling relationships, the argument in \S\ref{SubsecScalingRelations} would not change if the R16 clouds were taken as reference samples instead of the S87 clouds.

\subsubsection{M17 catalog}
\begin{figure*}[htbp]
\plotone{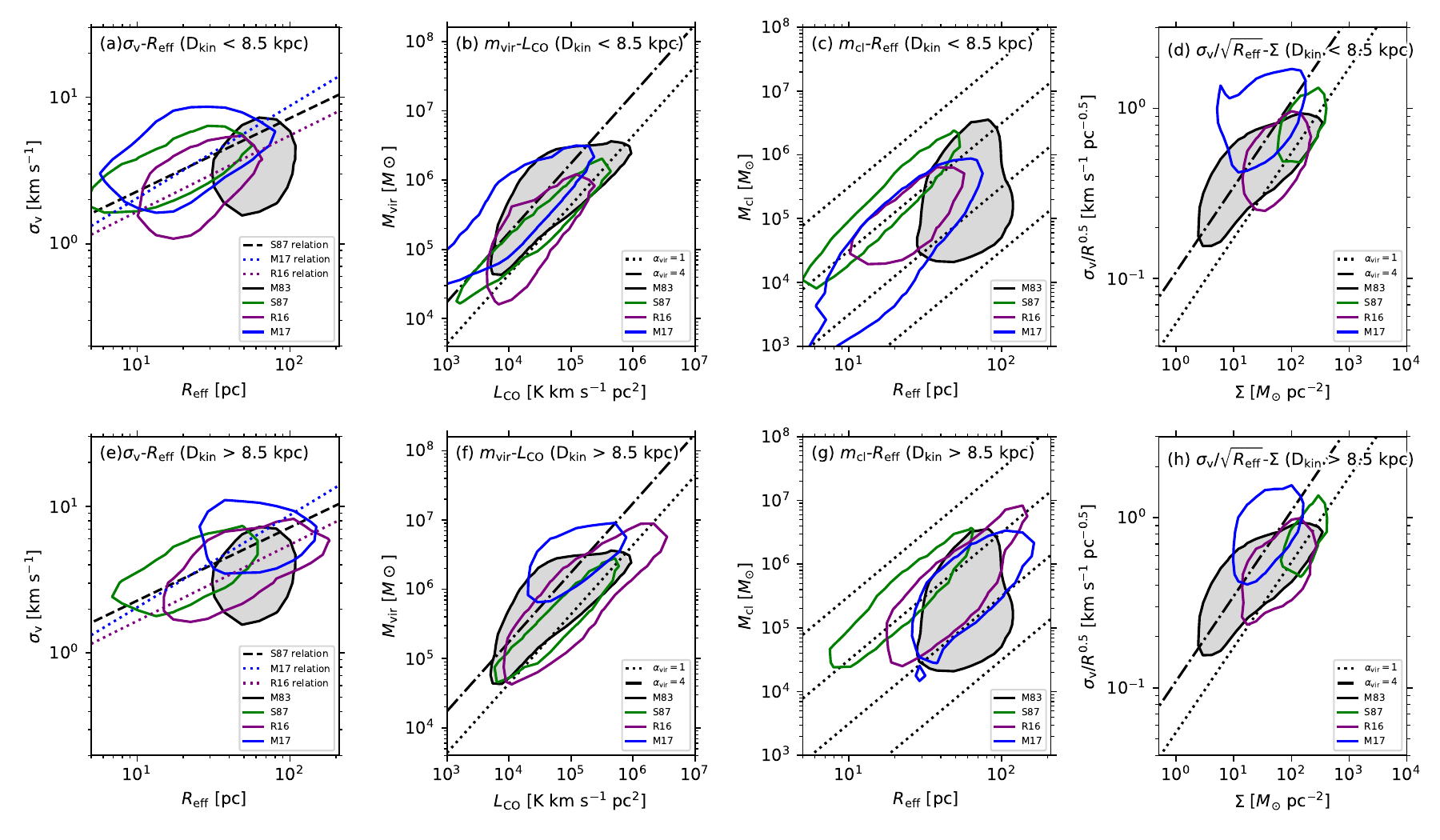}
\caption{
Scaling relations of cloud properties.
In each plot, the density distribution of the clouds at the 20\% level of the maximum density is shown for M83 clouds with the black contour and for the Galactic clouds of S87, R16, and M17 with green, purple, and blue contours, respectively.
For the Galactic catalogs of S87, R16, and M17, we excluded the clouds with the Galactocentric of $\Rgal$ $>8.5$ kpc and the Galactic longitude $l$ within $|l|$ $<$ 5 $\degr$.
The former filtering is made to compare clouds among molecular-dominated environments, as in the disk of M83.
The latter is to exclude the Galactic Center clouds.
In the top and bottom rows, the Galactic clouds with the kinematic distance at $\Dkin$ $<8.5$ kpc and $\Dkin$ $>8.5$ kpc are shown, respectively.
(a, e) $\sigmav$--$\Reff$ relationship.
The black dashed line indicates the $\sigmav$--$\Reff$ relation of S87, which is eq. (\ref{eq:Larson1}) with $C=0.72$ $\kmPerS$ pc$^{-1/2}$.
The purple and blue dotted lines indicate the relations reported in R16 and M17, respectively.
(b, f) $\Mvir$--$\LCO$ relationship.
The dotted and dash-dotted lines indicate $\alphaVir$ of 1 and 4, respectively.
(c, g) $\Mcl$--$\Reff$ relationship. The dotted lines indicate the lines for $\SigmaGMC$ of 1, 10, 100, and 1000 $\MsunPerSqPC$, respectively.
(d, h) $\sigmav / \Reff^{1/2}$--$\SigmaGMC$ relationship.
The dotted and dash-dotted lines indicate $\alphaVir$ of 1 and 4, respectively.
}
\label{FigSclRelationsExtended}
\end{figure*}

At $\Dkin$ $>$ 8.5 kpc, the M17 clouds cover a range of $\Reff$ similar to that of the M83 clouds as median $\Reff$ is similarly around 50 pc.
However, despite the similar radii, there is quite a small overlap with that of M83 clouds on the $\sigmav$-$\Reff$, $\Mvir$-$\LCO$ and $\sigmav/\sqrt{\Reff}$-$\SigmaGMC$ plots (figure \ref{FigSclRelationsExtended}e, \ref{FigSclRelationsExtended}f, \ref{FigSclRelationsExtended}h).
The small overlap in these plots suggests that the M83 clouds cannot be deemed as the simple counterparts of the M17 clouds sampled in another galactic disk.
\par

The small overlap between the M17 and M83 clouds owes to higher $\alphaVir$ in the M17 clouds compared to other Galactic catalogs.
In particular, even though M17 and R16 sampled clouds from similar data sets of \cite{Dame2001}, M17 clouds have about a factor 4 higher $\alphaVir$ compared to the R16 clouds.
The high $\alphaVir$ in the M17 clouds is linked to the higher $\sigma_v$ at a given $\Reff$ compared to the R16 clouds, seen on the $\sigma_v$-$\Reff$ relation (figure \ref{FigSclRelationsExtended}a, \ref{FigSclRelationsExtended}e).
The reason for this systematically higher $\sigmav$ in the M17 clouds compared to the R16 clouds is not completely clear, but it might reflect the difference in the ways clouds are identified, as also was suggested by \citet{Evans2021BoundCloud}.
M17 identified clouds by decomposing each spectrum of the data cube into a set of Gaussian components and then clustering the Gaussian components into clouds, while the methods used by S87, R16, and here in M83 are similar in that cloud peaks are identified by setting intensity thresholds.
The clustering process of M17 accepts the joining of two clusters of Gaussian components at different velocities if the velocity difference between the two clusters is less than twice the velocity dispersion of individual clusters.

\begin{deluxetable}{lcccccc}
\tablecaption{Median cloud properties in the MW catalogs}
\tablewidth{0pt}
\tablehead{
\colhead{Catalog name} & \colhead{$R_{\mathrm{eff}}$} & \colhead{$M_{\mathrm{cl}}$} & \colhead{$\Sigma$} & \colhead{$\sigma_{\mathrm{v}}$} & \colhead{$\alpha_{\mathrm{vir}}$} & \colhead{$N$} \\
\colhead{} & \colhead{(pc)} & \colhead{($10^{5}$ $M_{\odot}$)} & \colhead{($M_{\odot}$ pc$^{-2}$)} & \colhead{(km s$^{-1}$)} & \colhead{} & \colhead{}
}
\decimalcolnumbers
\startdata
S87 ($D<8.5$ kpc, $R_{\mathrm{gal}}<8.5$ kpc, $|l|>5\degr$) & $14.3_{8.2}^{27.9}$ & $1.0_{0.2}^{4.7}$ & $152.2_{100.2}^{236.8}$ & $3.1_{2.3}^{4.3}$ & $1.4_{0.9}^{2.1}$ & 166 \\
S87 ($D>8.5$ kpc, $R_{\mathrm{gal}}<8.5$ kpc, $|l|>5\degr$) & $26.8_{17.3}^{45.0}$ & $4.8_{1.9}^{11.1}$ & $194.5_{142.6}^{288.2}$ & $3.9_{3.0}^{5.0}$ & $1.0_{0.7}^{1.4}$ & 62 \\
R16 ($D<8.5$ kpc, $R_{\mathrm{gal}}<8.5$ kpc, $|l|>5\degr$) & $24.0_{15.5}^{40.6}$ & $0.8_{0.4}^{3.0}$ & $53.8_{25.9}^{102.9}$ & $2.5_{1.7}^{3.7}$ & $1.5_{0.8}^{3.0}$ & 324 \\
R16 ($D>8.5$ kpc, $R_{\mathrm{gal}}<8.5$ kpc, $|l|>5\degr$) & $47.4_{26.5}^{78.8}$ & $4.2_{1.0}^{13.1}$ & $67.8_{25.8}^{120.4}$ & $3.3_{2.3}^{5.1}$ & $1.3_{0.7}^{2.9}$ & 118 \\
M17 ($D<8.5$ kpc, $R_{\mathrm{gal}}<8.5$ kpc, $|l|>5\degr$) & $19.8_{9.4}^{37.0}$ & $0.4_{0.0}^{1.9}$ & $28.8_{7.6}^{75.2}$ & $4.0_{2.3}^{5.9}$ & $6.8_{3.2}^{35.0}$ & 2893 \\
M17 ($D>8.5$ kpc, $R_{\mathrm{gal}}<8.5$ kpc, $|l|>5\degr$) & $60.3_{36.7}^{95.7}$ & $4.1_{1.3}^{12.4}$ & $32.7_{15.4}^{75.9}$ & $6.4_{4.5}^{8.2}$ & $5.4_{2.9}^{14.9}$ & 1177 \\
\hline
S87 (Full) & $17.0_{8.0}^{31.9}$ & $1.3_{0.2}^{6.3}$ & $152.6_{92.8}^{250.5}$ & $3.1_{2.2}^{4.4}$ & $1.3_{0.8}^{2.1}$ & 273 \\
R16 (Full) & $27.6_{15.4}^{53.6}$ & $0.5_{0.1}^{3.0}$ & $22.1_{8.6}^{76.7}$ & $2.1_{1.3}^{3.5}$ & $2.3_{1.1}^{5.2}$ & 1064 \\
M17 (Full) & $26.4_{9.0}^{60.4}$ & $0.4_{0.0}^{3.2}$ & $16.5_{4.9}^{60.0}$ & $3.6_{1.8}^{6.5}$ & $8.0_{3.2}^{34.0}$ & 8107 \\
\enddata
\tablecomments{Median and 16th-to-84th percentile range of the cloud properties for the Galactic cloud catalogs of \citet{Solomon1987Larson}, \citet{Rice2016}, and \citet{MivilleDeschenes2017MWGMC}.
In the first six rows, the clouds with the Galactocentric of $\Rgal$ $>8.5$ kpc and the Galactic longitude $l$ within $|l|$ $<$ 5 $\degr$ are excluded, and the remaining clouds are classified by the line-of-sight distance threshold of 8.5 kpc.
In the last three rows, the full sample summaries for the catalogs are listed.
(1) Catalog name and $\Dkin$ range.
(2) Effective radius.
(3) Cloud mass.
(4) Surface density.
(5) Velocity dispersion.
(6) Virial parameter.
(7) Number of clouds.
}
\label{TablePropsMW}
\end{deluxetable}

\end{document}